\documentclass[]{aa}
\usepackage{longtable}
\usepackage{natbib,afterpage}
\usepackage{graphicx,color}
\usepackage{amssymb,times,graphics, subfigure}
\usepackage[english]{babel}
\usepackage{subfigure}
\usepackage{caption}
\usepackage{clearpage}
\usepackage{wrapfig}
\usepackage[rightcaption]{sidecap}

\def\Msun{\hbox{M$_{\odot}$}}               
\def\Rstar{\hbox{R$_{\star}$}}              
\def\Mdot{\hbox{$\dot{M}$}}               
\def\arcsec{\hbox{$^{\prime\prime}$}}
\def\arcmin{\hbox{$^{\prime}$}}

\def\deg{\hbox{$^\circ$}}       

\bibpunct{(}{)}{;}{a}{}{,}

\begin{document}
\selectlanguage{english}
\newcommand{\red}{\textcolor[rgb]{1,0,0}}
  \title{The enigmatic nature of the circumstellar envelope and bow shock surrounding Betelgeuse as revealed by Herschel. 
\thanks{Herschel is an ESA space observatory with science instruments provided by European-led Principal Investigator consortia and with important participation from NASA. 
}}
\subtitle{I. Evidence of clumps, multiple arcs, and a linear bar-like structure.}
 \author{L. Decin\inst{1,2}
  \and
  N.L.J. Cox\inst{1}
  \and
  P. Royer\inst{1}
  \and
  A.J. Van Marle\inst{1}
  \and
  B. Vandenbussche\inst{1}
  \and
  D. Ladjal\inst{3}
  \and
  F. Kerschbaum\inst{4}
  \and
  R. Ottensamer\inst{4}
  \and	
  M.J. Barlow\inst{5}
 \and	
  J.A.D.L. Blommaert\inst{1}
  \and
  H.L. Gomez\inst{8}
  \and
  M.A.T. Groenewegen\inst{6}  
  \and
  T. Lim\inst{7}
  \and	
  B.M. Swinyard\inst{9}
  \and
  C. Waelkens\inst{1}
  \and
  A.G.G.M. Tielens\inst{10}
}

  \offprints{Leen.Decin@ster.kuleuven.be}

  \institute{
Instituut voor Sterrenkunde,
             Katholieke Universiteit Leuven, Celestijnenlaan 200D, 3001 Leuven, Belgium
 		\and
	Sterrenkundig Instituut Anton Pannekoek, University of Amsterdam, Science Park 904, NL-1098 Amsterdam, The Netherlands
\and
University of Denver, 2112 E.\ Wesley Ave.\ Denver CO 80208, US
    \and
 University of Vienna, Department of Astrophysics, T{\"u}rkenschanzstra\ss{}e 17, A-1180 Vienna, Austria
    \and
 Dept.\ of Physics \& Astronomy, University College London, Gower St, London WC1E 6BT, UK 
  \and
Koninklijke Sterrenwacht van Belgi\"e, Ringlaan 3, B-1180 Brussel, Belgium 
  \and
  Space Science and Technology Department, Rutherford Appleton Laboratory, Oxfordshire, OX11 0QX, UK
\and
  School of Physics and Astronomy, Cardiff University, Queens Buildings, The Parade, Cardiff CF24 3AA, UK
\and
Space Science and Technology Department, Rutherford Appleton Laboratory, Didcot, Oxfordshire, OX11 0QX, UK
\and
Leiden Observatory, Leiden University, P.O.\ Box 9513,
NL- 2300 RA Leiden, The Netherlands
}


  \date{Received / accepted }

 \abstract{The interaction between stellar winds and the interstellar medium (ISM) can create complex bow shocks. The photometers on board the Herschel Space Observatory are ideally suited to studying the morphologies of these bow shocks.}
{We aim to study the circumstellar environment and wind-ISM interaction of the nearest red supergiant, Betelgeuse.}
{Herschel PACS images at 70, 100, and 160\,$\mu$m and SPIRE images at 250, 350, and 500\,$\mu$m were obtained by scanning the region around Betelgeuse. These data were complemented with ultraviolet GALEX data, near-infrared WISE data, and radio 21\,cm GALFA-HI data. The observational properties of the bow shock structure were deduced from the data and compared with hydrodynamical simulations.}
{The infrared Herschel images of the environment around Betelgeuse are spectacular, showing the occurrence of multiple arcs at $\sim$6--7\arcmin\ from the central target and the presence of a linear bar at $\sim$9\arcmin. Remarkably, no large-scale instabilities are seen in the outer arcs and linear bar. The dust temperature in the outer arcs varies between 40 and 140\,K, with the linear bar having the same colour temperature as the arcs.  The inner envelope shows clear evidence of a non-homogeneous clumpy structure (beyond 15\arcsec), probably related to the giant convection cells of the outer atmosphere. The non-homogeneous distribution of the material even persists until the collision with the ISM. A strong variation in brightness of the inner clumps at a radius of $\sim$2\arcmin\ suggests a drastic change in mean gas and dust density $\sim$32\,000\,yr ago.  Using hydrodynamical simulations, we try to explain the observed morphology of the bow shock around Betelgeuse. }
{Different hypotheses, based on observational and theoretical constraints, are formulated to explain the origin of the multiple arcs and the linear bar and the fact that no large-scale instabilities are visible in the bow shock region. We infer that the two main ingredients for explaining these phenomena are a non-homogeneous mass-loss process and the influence of the Galactic magnetic field. {The hydrodynamical simulations show that a warm interstellar medium, reflecting a warm neutral or partially ionized medium, or a higher temperature in the shocked wind also prevent the growth of strong instabilities.} The linear bar is probably an interstellar structure illuminated by Betelgeuse itself. }

  \keywords{Stars: AGB and post-AGB, Stars: mass loss, Stars: circumstellar matter, Stars: individual: Betelgeuse}
\titlerunning{The enigmatic envelope of Betelgeuse}
  \maketitle
%

\section{Introduction} \label{SECT:Introduction}

For decades, it has been understood that the mass lost by red giants and supergiants dominates the mass return to the Galaxy \citep{Maeder1992A&A...264..105M}, yet the actual mechanisms of such stellar mass loss have remained a mystery. The generally accepted idea on the mass-loss mechanism for asymptotic giant branch (AGB) stars is based on pulsations and radiation pressure on newly formed dust grains. However, oxygen-rich AGB stars suffer from the so-called `acceleration deficit' dilemma, which   states that mass-loss rates due to the formation of silicate dust alone are orders of magnitude less than  observed ones \citep{Woitke2006A&A...460L...9W}. The formation of both carbon and silicate grains  \citep{Hoefner2007A&A...465L..39H}   or micron-sized Fe-free silicates  \citep{Hoefner2008A&A...491L...1H, Norris2012Natur.484..220N} have been proposed as possible solutions to this dilemma. The proposed models for the mass loss of AGB stars are very unlikely to be applicable to red supergiants (RSGs) since 
they are irregular variables with small amplitudes. 

Processes linked to convection, chromospheric activity, or rotation might play an important role as triggers for the mass loss in RSGs. \citet{Josselin2007A&A...469..671J} proposes that turbulent pressure generated by convective motions, combined with radiative pressure on molecular lines, might initiate mass loss, but Alfv\'en waves generated by a magnetic field might also contribute \citep{Hartmann1984ApJ...284..238H}. The last few years have provided more and more evidence that, at least for a fraction of evolved stars, the mass is not lost in a homogeneous way, but irregular and clumpy structures prevail \citep[e.g.][]{Weigelt2002A&A...392..131W, Kervella2009A&A...504..115K, Decin2011A&A...534A...1D}. This might have serious consequences for the total amount of mass lost during these late evolutionary stages. Solving the riddle of the mass loss of these (higher mass) targets is important in the framework of stellar evolution as a key to estimating when a supernova explosion may occur. 

decade (as CRIRES, IRAM, APEX etc.) made it possible to study these chemical processes in detail, and to refine our estimates of the total chemical enrichment of the ISM by these evolved stars.

When these evolved stars with their surrounding winds move through the interstellar medium (ISM), a wind-ISM interface structure is expected in the form of a bow shock, a kind of  cometary like structure pointing in the direction of motion. Photometric observations performed with the IRAS telescope \citep{Noriega1997AJ....114..837N} and later confirmed with the Spitzer Space Telescope \citep{Ueta2006ApJ...648L..39U, Ueta2008PASJ...60S.407U} and the \textit{Galaxy Evolution Explorer} satellite {\citep[GALEX][]{Martin2007Natur.448..780M,  Sahai2010ApJ...711L..53S}} indeed showed the presence of  bow shock structures around evolved stars. However, these early observations of the far-IR and UV-bright  emission structures lacked any spectral resolution, and since they only had poor spatial resolution, allowed only basic morphological studies.

Thanks to the capabilities of the instruments on board the Herschel Space Observatory \citep{Pilbratt2010A&A...518L...1P}, one is now able to image the circumstellar envelopes (CSEs) and the CSM-ISM interaction regions with unprecedented detail. In the MESS GTKP \citep{Groenewegen2011A&A...526A.162G} several known and new interaction objects have been imaged with the PACS and SPIRE instruments \citep[e.g. CW\,Leo,][]{Ladjal2010A&A...518L.141L, Decin2011A&A...534A...1D, Cox2012A&A...537A..35C}, among which several supergiants. Betelgeuse ($\alpha$~Ori, M2~Iab) is the closest supergiant, and it serves as the prototype of the red oxygen-rich supergiants. It has been an important astrophysical laboratory for many decades and its envelope has been studied with a variety of observational techniques. 
Betelgeuse is, therefore, a prime target for Herschel to study the still unknown
mechanisms that lead to mass loss and eventually result in an energetic bow shock when the wind collides with the ISM. 

\begin{figure*}[!htbp]
  \includegraphics[width=\textwidth]{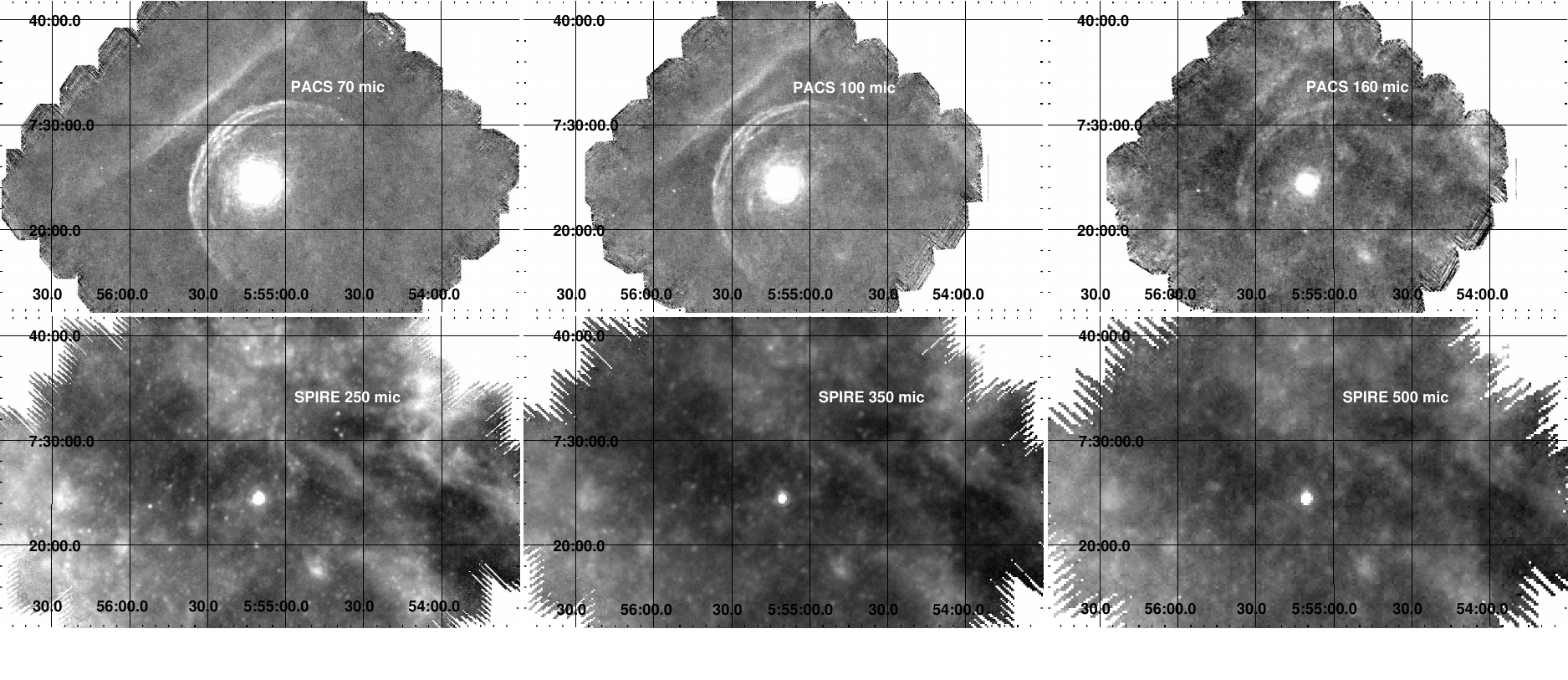}
\caption{Herschel PACS and SPIRE images of Betelgeuse. North is up, east to the left. The field-of-view (FOV)  is $\sim$49\arcmin$\times$29\arcmin.}
\label{FIG:ALL_HERSCHEL}
\end{figure*}

\begin{figure*}[!htbp]
        \centerline{\resizebox{\textwidth}{!}
{\includegraphics{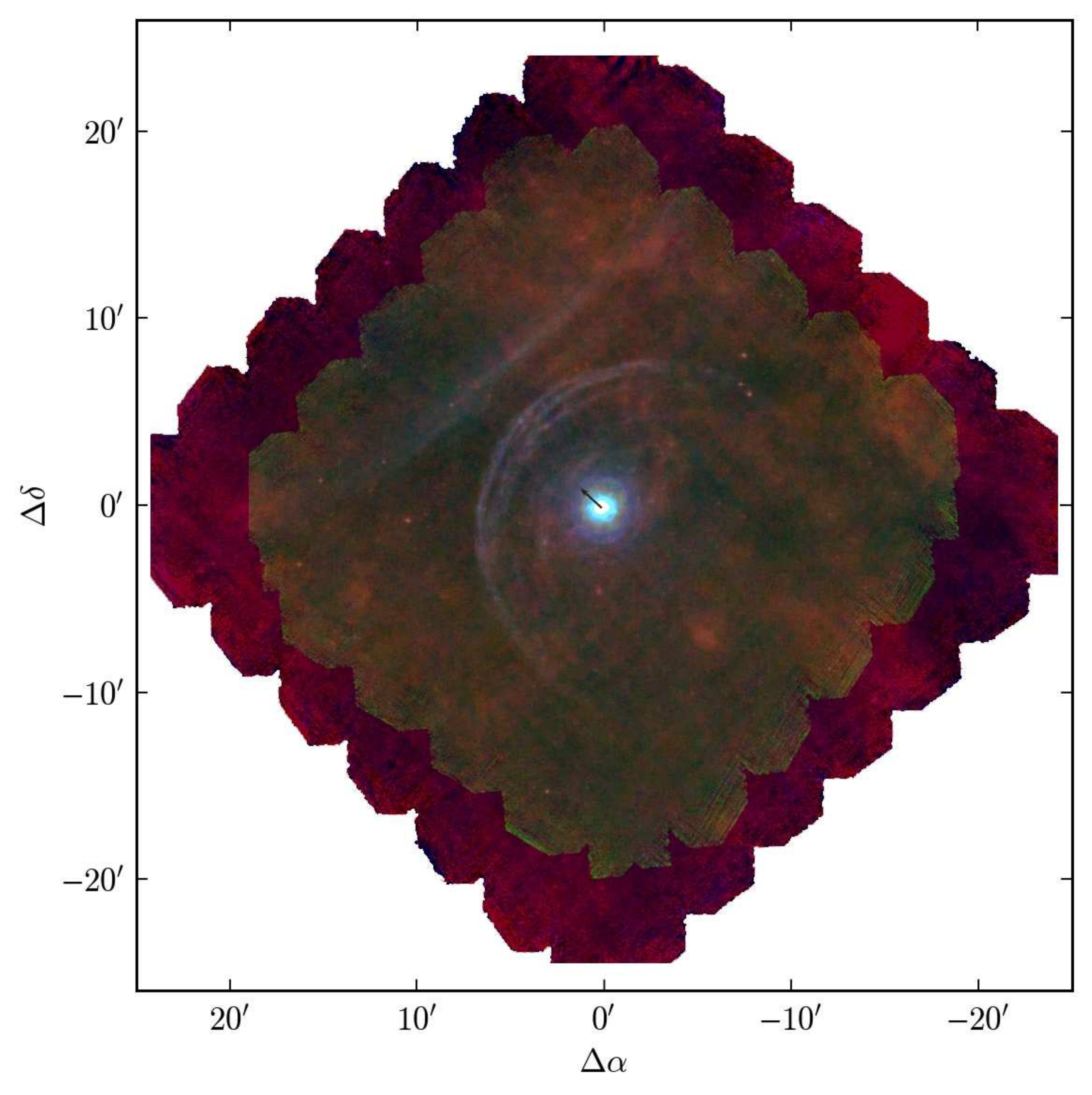}}}
 \caption{Composite colour image of the Herschel PACS  images of Betelgeuse. North is up, east to the left. Blue is the PACS 70\,$\mu$m image, green is PACS 100\,$\mu$m, and red PACS 160\,$\mu$m. 
 The black arrow indicates the direction of the space velocity of the star. The multiple arc-like structure is situated at $\sim$6-7\arcmin\ to the north-east of the central target, the linear bar at $\sim$9\arcmin. The contrast in the figure is best visible on screen.}
\label{Fig:Herschel_large}
\end{figure*}

\subsection{Betelgeuse} \label{SECT:BETELGEUSE}

Many detailed studies have already described some characteristics of the complex atmosphere, chromosphere, and dusty envelope of Betelgeuse. In this section, a  summary is given of those properties that are important for this study. 

Betelgeuse is a low-amplitude variable, whose variation was first described in 1840 by Sir John Herschel \citep{Herschel1840MNRAS...5...11H}. \citet{Schwarzschild1975ApJ...195..137S} attributed these fluctuations to a changing granulations pattern formed by a few giant convection cells covering the surface of the star. Recently, a magnetic field of about 1 Gauss has been detected using the Zeeman effect \citep{Auriere2010A&A...516L...2A}. It is very faint, but may play an important role in shaping the convection of Betelgeuse.

Using NRAO's Very Large Array (VLA), combined with Hipparcos data, \citet{Harper2008AJ....135.1430H} determined the parallax of Betelgeuse to be 5.07$\pm$1.10\,mas (or a distance of 197$\pm$45\,pc). The most likely star-formation scenario for Betelgeuse is that it is a runaway star from the Orion OB1 assocation and was originally a member of a high-mass multiple system within Ori OB1a \citep{Harper2008AJ....135.1430H}. The galactic space motion in the local standard of rest (LSR) is $U^\prime$\,=\,$-12.7$, $V^\prime$\,=\,+0.5, $W^\prime$\,=\,$+28.2$\,km/s. Betelgeuse was probably formed about 10 to 12 million years ago from the molecular clouds observed in Orion, but has evolved rapidly owing to its high mass. No consensus has yet emerged regarding the star's mass, but most studies report values between 10\,\Msun\ and 20\,\Msun\ \citep[e.g.,][]{Dolan2008APS..APR.S8006D, Neilson2011arXiv1109.4562N}.


Ultraviolet and H$\alpha$ images prove there is a chromosphere with a spatial extent of $\sim$4\,\Rstar\ \citep{Altenhoff1979A&A....73L..21A, Goldberg1981ASSL...88..301G, Hebden1987ApJ...314..690H}, where the hot plasma has a temperature of around 5000\,K \citep{Gilliland1996ApJ...463L..29G}. However, observations made with the VLA by \citet{Lim1998Natur.392..575L} suggest that much cooler ($\sim$1000--3000\,K) gas co-exists at the same radial distances from the central target, and must be much more abundant because it dominates the radio emission. In addition, narrow-slit spectroscopy of Betelgeuse at 11.15\,$\mu$m with the ISI by \citet{Verhoelst2006A&A...447..311V} reveals that silicate dust forms at distances beyond $\sim$20\,\Rstar, but that Al$_2$O$_3$ may form as close as $\sim$2\,\Rstar. This apparent dichotomy is solved when adopting a scenario in which a few inhomogeneously distributed large convective cells are responsible for lifting the cooler photospheric gas into the atmospheres and for 
producing shock waves, which could heat localized, but spatially distributed regions of the atmosphere to chromospheric temperatures \citep{Lim1998Natur.392..575L}.

Observations taken over the past few years have proven that the regions just above the stellar surface show some clear deviations from spherical symmetry. \citet{Kervella2009A&A...504..115K} obtained adaptive optics images of Betelgeuse in the 1.04--2.17\,$\mu$m range in ten narrow-band filters. These images show that the CSE of Betelgeuse has a complex and irregular structure, with in particular a bright `plume' extending in the southwestern quadrant up to a radius of at least six times the photosphere. They proposed that the `plume' might be linked to either the presence of a convective hot spot on the photosphere or to polar mass loss possibly due to rotation of the star. Using AMBER on the VLTI, \citet{Ohnaka2009A&A...503..183O} found from CO lines that the atmosphere has an inhomogeneous velocity field, which could by explained by the gas moving outward and inward in a patchy pattern.  This gives support to the idea of convective motion in the upper atmosphere or intermittent mass ejections in clumps or 
arcs. Recently, \citet{Kervella2011A&A...531A.117K} observed Betelgeuse with VLT/VISIR in the $N$ and $Q$ band. The images showed a bright, extended and complex CSE within $\sim$2.5\arcsec\ of the central star with the appearance of a partial circular shell between 0.5 and 1.0\arcsec, several knots, and filamentary structures. The ring-like structure might be related to the dust condensation zone.

Although some dust species, such as silicates and corundum, have been identified in the CSE around Betelgeuse, the envelope shows a remarkable deficiency in dust \citep{Skinner1987MNRAS.224..335S, Verhoelst2006A&A...447..311V}. In addition, the molecular content is unusually low. There is evidence that only a fraction of carbon is associated into CO \citep{Huggins1987ApJ...313..400H}. The extended, warm chromosphere is clearly unfavourable for the survival of molecules and the formation of dust species. Detailed studies by \citet{Glassgold1986ApJ...306..605G} show that the most common form of all species are neutral atoms and first ions.

A parsec-size  shell around Betelgeuse was detected with IRAS by \citet{Noriega1997AJ....114..837N}, and later on with AKARI by \citet{Ueta2008PASJ...60S.407U} (see Appendix~\ref{SEC:previous}). The shell is asymmetric with a mean radius of 6\arcmin\ and the outermost structure at $\sim$7\arcmin. \citet{Noriega1997AJ....114..837N} proposed that the shell is confined by the ram pressure of the ISM, hence represents the bow shock around Betelgeuse. To the north-east of Betelgeuse, a ridge or filament was detected, which was said to belong to the surrounding ISM. Neither IRAS nor AKARI could clearly resolve the bow shock and filament structure. The derived dust temperatures in the bow shock and filament vary between $\sim$10\,K and 40\,K \citep{Ueta2008PASJ...60S.407U}.

Recently, \citet{LeBertre2012arXiv1203.0255L} has detected a detached \ion{H}{i} gas shell of $\sim$2\arcmin\ in radius surrounding Betelgeuse, and reported the detection of atomic hydrogen associated with the far-infrared arc located 6\arcmin\ north-east of Betelgeuse.

In this paper, we focus on the morphological appearance of the inner envelope and bow shock surrounding Betelgeuse. The thermodynamical structure and chemical content of the envelope and bow shock will be discussed in \citet{Decin2012b} (from here on called Paper~II). In Sect.~\ref{SECT:observations}, the different observations with their respective calibrations are described. The observational properties of the inner envelope and bow shock structure are analysed in Sect.~\ref{SECT:results}. With the aid of hydrodynamical simulations, 
 the origin of the multiple arcs is discussed in Sect.~\ref{SECT:ORIGIN_ARCS}. The origin of the linear bar is described in Sect.~\ref{SECT:ORIGIN_BAR}. The conclusions are summarized in Sect.~\ref{SECT:conclusions}.

\section{Observations} \label{Observations} \label{SECT:observations}

In this section, an overview is given of the different observations toward Betelgeuse and its extended envelope, the data-reduction for each data-set is described, and some first results are deduced from the data.

\subsection{Herschel PACS and SPIRE images} \label{SECT:PACS_SPIRE_images}

Infrared images were obtained using the Photodetector Array Camera and Spectrometer \citep[PACS,][]{Poglitsch2010A&A...518L...2P} and the Spectral and Photometric Imaging Receiver \citep[SPIRE]{Griffin2010A&A...518L...3G} on board the Herschel satellite \citep{Pilbratt2010A&A...518L...1P}, and they are part of the MESS guaranteed-time key programme \citep{Groenewegen2011A&A...526A.162G}. For both instruments, the `scan map' observing mode 
was used with medium (20\arcsec/s) scan speed in the PACS 70, 100, and 160\,$\mu$m filter settings and 30\arcsec/s for the SPIRE 250, 350, and 500\,$\mu$m filters. To create a uniform coverage, avoid striping artefacts, and increase redundancy, two observations at orthogonal scan directions were concatenated. The PACS images were obtained on September 13, 2010 (OBSIDS 1342204435 and 1342204436) and on March 29, 2012 (OBSIDS 1342242656 and 1342242657), the SPIRE images on March 11, 2010 (OBSID 1342192099).

The data reduction was performed using the \textit{Herschel Interactive Processing Environment} (HIPE) \citep[see][for more details]{Groenewegen2011A&A...526A.162G}. To remove low-frequency noise, which causes additive brightness drifts, the {\sc{scanamorphos}} routine \citep{Roussel2011} was applied. Deconvolution of the data was done using the method as described in \citet{Ottensamer2011}. 

The PACS instrument offers a pixel size of 3.2\arcsec\ in the
70\,$\mu$m and 100\,$\mu$m bands and of 6.4\arcsec in the 160\,$\mu$m band, but the images
presented in this paper are oversampled by a factor of 3.2, resulting
in a sampling of 1\arcsec\ and 2\arcsec\ per pixel, respectively. The PACS FWHM
at 70, 100, and 160\,$\mu$m is 5.7\arcsec, 6.7\arcsec, and 11.4\arcsec, respectively.  The pixel size of the three SPIRE bands is 6\arcsec\ at 250\,$\mu$m, 10\arcsec\ at 350\,$\mu$m, and 14\arcsec\ at 500\,$\mu$m, with an FWHM of the beam size being 18.1\arcsec, 25.2\arcsec, and 36.6\arcsec, respectively.

\begin{figure}[!htp]
        \centerline{\resizebox{0.48\textwidth}{!}{\includegraphics{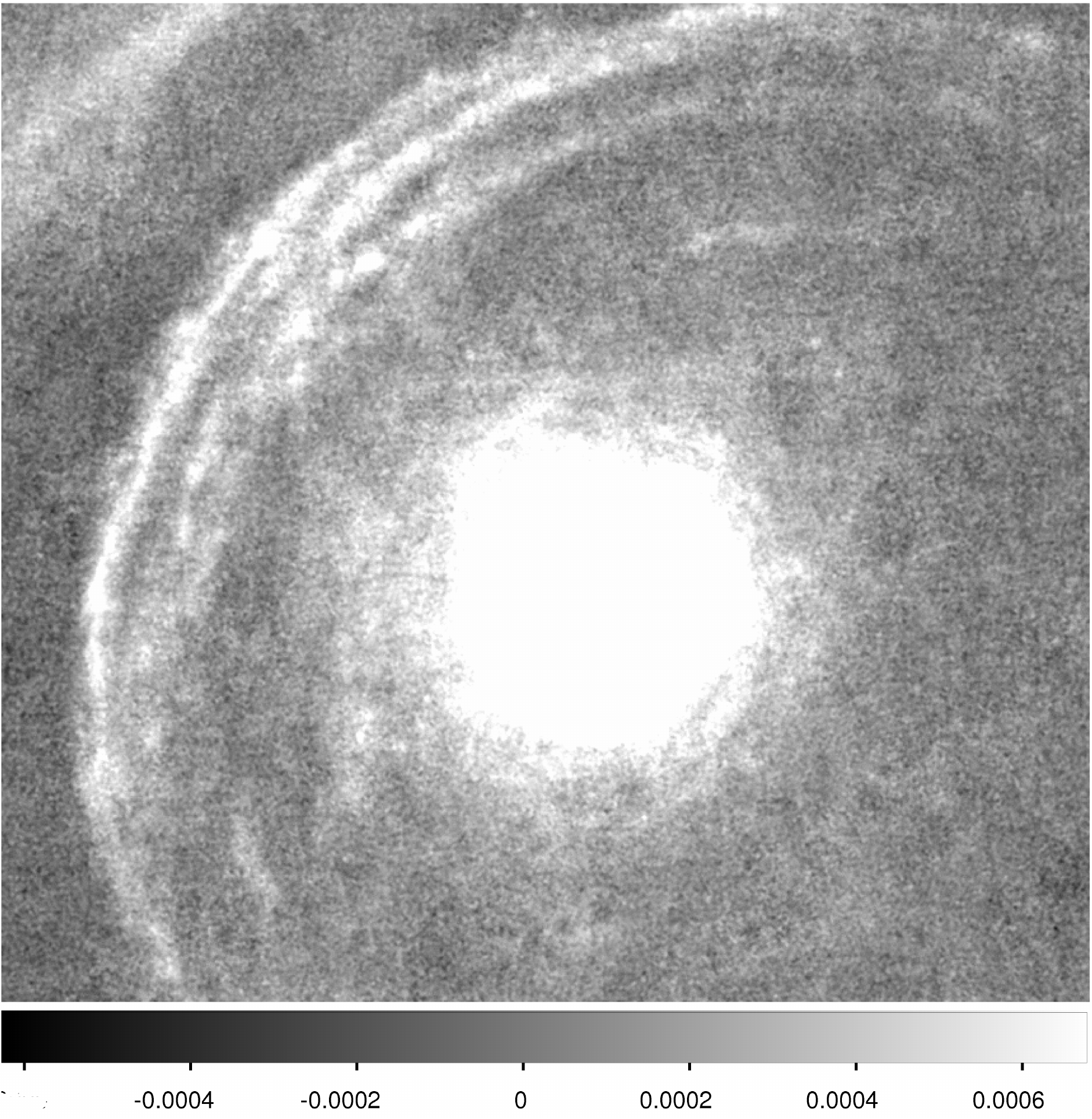}}}
\caption{Zoom into the multiple arc-like structure seen in the bow shock region of Betelgeuse as detected with the PACS 70\,$\mu$m filter. North is up, east to the left. The FOV is 855\arcsec$\times$765\arcsec. Flux units are in Jy/pixel.}
\label{Fig:Herschel}
\end{figure}

The individual scan maps are plotted in Fig.~\ref{FIG:ALL_HERSCHEL}, and a false-colour image displaying all the PACS data is shown in Fig.~\ref{Fig:Herschel_large}. Remarkably, a  multiple arc-like structure is detected at $\sim$6-7\arcmin\ away from the central star, together with an enigmatic linear bar-like structure at $\sim$9\arcmin. A closer zoom into the arc-like structure is displayed in Fig.~\ref{Fig:Herschel}. The inner radius of the closest arc is at 280\arcsec\ (as measured from the central target). The other most prominent arcs have an inner radius at 280\arcsec, 310\arcsec, 350\arcsec\ and 375\arcsec. The brightest parts of the arcs are reasonably well fitted by concentric ellipses (see Fig.~\ref{FIG:PACS70_ellipse} in the Appendix~\ref{App:additional_figures}). The width of the arcs is typically $\sim$20\arcsec. Clearly, substructures of the size of several arcseconds, probably related to density differences, are seen in the arcs. The arc-like structure is probably related to the CSM-ISM 
interaction phase, when the circumstellar material collides at a wind speed of $\sim$15\,km/s with the ISM, creating a bow shock (see Sect.~\ref{SECT:ORIGIN_ARCS} for more details).
The inner ridge of the linear bar is at 535\arcsec\ from the central star. The length is $\sim$1600\arcsec. The width is $\sim$30\arcsec\ in the northern part; moving down towards the south-east, the bar widens and splits into two parts.  The angle between the direction of space velocity (drawn in Fig.~\ref{Fig:Herschel_large}) and the linear bar is $\sim$80\deg.

In the whole sample of 78 AGB stars and red supergiants surveyed in the framework of the MESS GTKP, Betelgeuse is the most spectacular source, because it is the only one showing this multiple-arc structure and the presence of a linear bar. The linear bar and the full arc were already detected with IRAS \citep{Noriega1997AJ....114..837N} and AKARI \citep{Ueta2008PASJ...60S.407U}, but their individual structures could not be resolved (see Appendix~\ref{SEC:previous}).

\begin{figure*}[htp]
    \begin{minipage}[t]{.48\textwidth}
        \centerline{\resizebox{\textwidth}{!}{\includegraphics{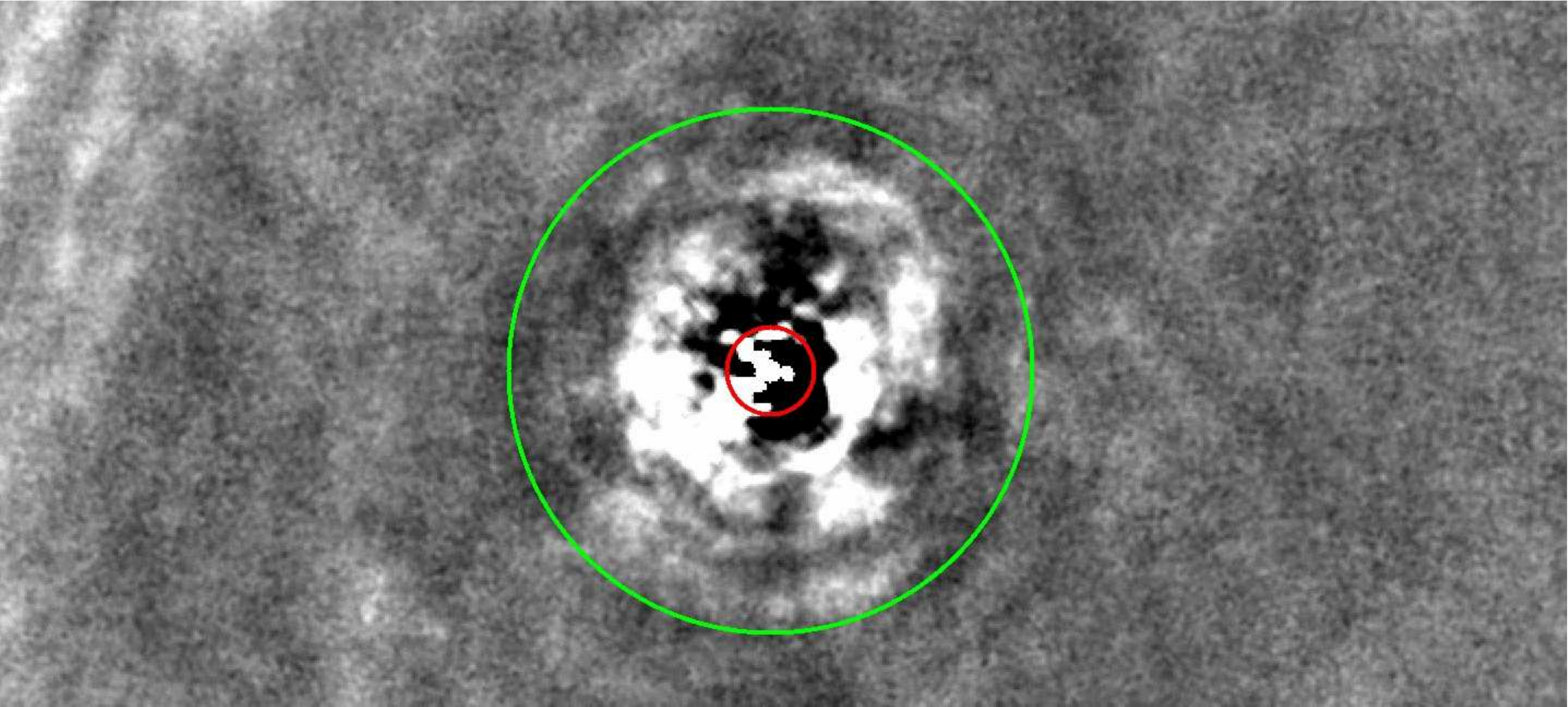}}}
    \end{minipage}
    \hfill
    \begin{minipage}[t]{.48\textwidth}
\centerline{\resizebox{\textwidth}{!}{\includegraphics{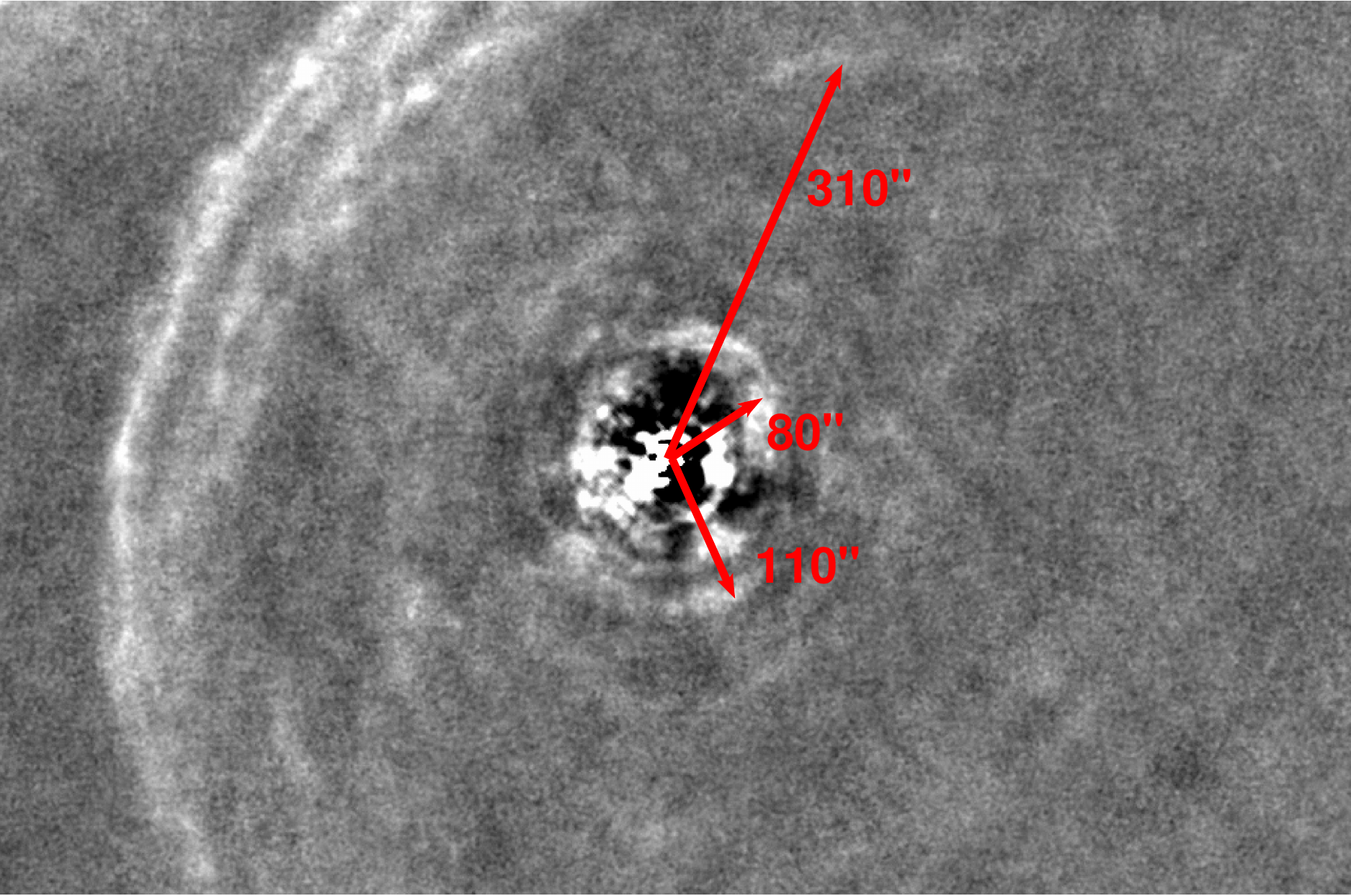}}}
    \end{minipage}
\caption{Oversampled PACS 70\,$\mu$m image of Betelgeuse after subtraction of the smooth halo of the circumstellar envelope. North is up, east to the left. The pixel size is 1\arcsec. In the left-hand panel, the power $\alpha$ has been taken as constant over the full region ($\alpha$\,=\,1.78). The field-of-view (FOV) is 720\arcsec$\times$325\arcsec. The red circle with a radius of 20\arcsec\ shows the region where some PSF artefacts are still visible; the green circle indicates a region with radius of 2\arcmin. In the right-hand panel two independent profiles were fitted before and beyond 50\arcsec\ to suppress the high flux values around 50\arcsec\ so as to better visualize the complex structure around 50--60\arcsec\ in the southeastern region. The field-of-view (FOV) is 970\arcsec$\times$640\arcsec. Distances to a few asymmetric structures are indicated in red.}
\label{Fig:AOri_inner}
\end{figure*}

To emphasize the shell morphology in the inner envelope, we removed the extended envelope halo by subtracting a smooth, azimuthally averaged profile represented by a power law $r^{-\alpha}$ for $r<$240\arcsec\ \citep[see][]{Decin2011A&A...534A...1D}. The method has been tested in detail to determine if artefacts from the point-spread function (PSF)  still affect the data (Royer et al., \textit{in prep.}). Only the region within 15-20\arcsec\ from the central target should be interpreted with care owing to the complex PSF and the bright contribution from the central target at these infrared wavelengths. The resulting image is shown in Fig.~\ref{Fig:AOri_inner}. The inner envelope structure is clearly non-homogeneous and testifies to a turbulent mass-loss history. A detailed discussion of the inner envelope structure is presented in Sect.~\ref{SECT:results_inner}.

\subsection{WISE data} \label{SECT:WISE}

\begin{figure*}[htp]
 \includegraphics[width=0.98\textwidth]{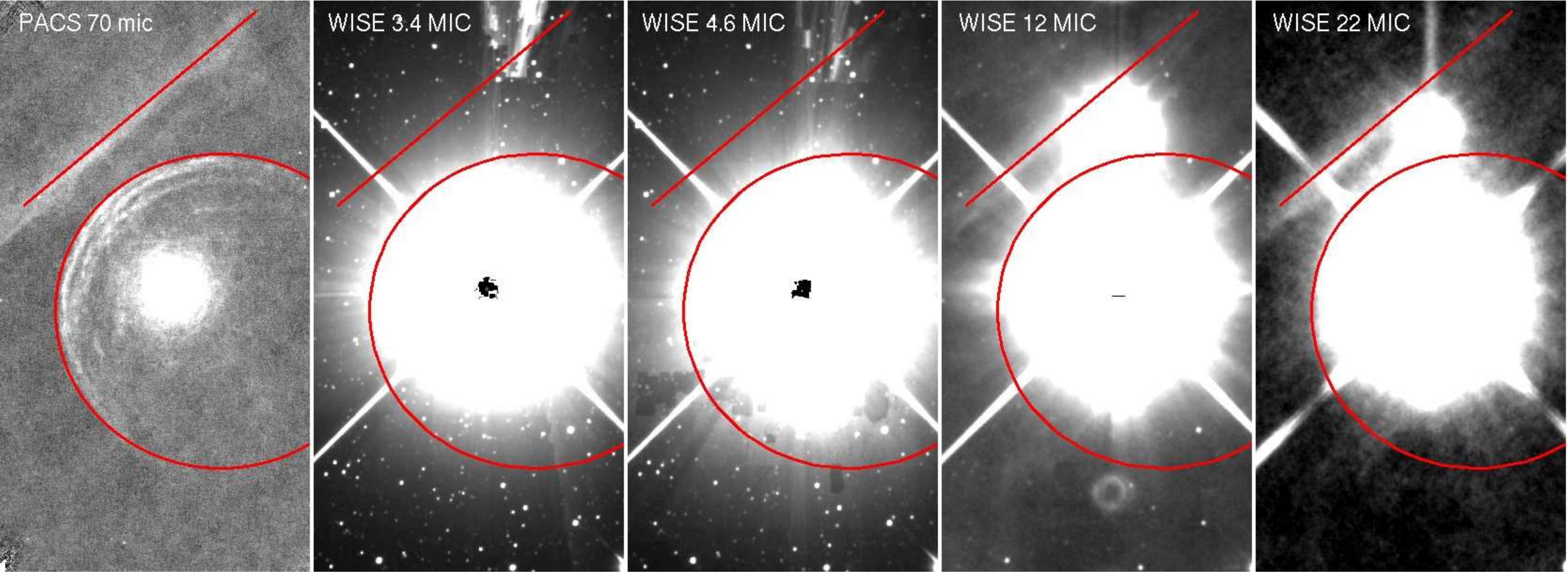}
\caption{Comparison between the WISE observations and the PACS 70\,$\mu$m image. An ellipse and line are added to guide the eye to the arc and bar-like structure. The pixel size of the WISE data is 1.375\arcsec. The bright central star causes latents and glint artefacts in the images.}
\label{FIG:wise}
\end{figure*}

The complex environment around Betelgeuse has also been imaged by the Wide-field Infrared Survey Explorer \citep[WISE,][]{Wright2010AJ....140.1868W}. WISE has four photometric bands centred at 3.4, 4.6, 12, and 22\,$\mu$m, with an angular resolution of 6.1\arcsec, 6.4\arcsec, 6.5\arcsec, and 12\arcsec, respectively. The 5$\sigma$ point source sensitivity in unconfused regions on the ecliptic is better than 0.08, 0.11, 1, and 6\,mJy, respectively. The WISE All-Sky Data were released on March 14, 2012. The WISE data images are shown in Fig.~\ref{FIG:wise}. The extreme brightness of Betelgeuse at these infrared wavelengths makes it very difficult to find any nearby associated emission structures. The linear bar is visible only at 12 and 22\,$\mu$m, with a (background-subtracted) surface brightness of $\sim$1.2\,MJy/sr and $\sim$1.9\,MJy/sr, respectively. 
and 5.2269e$-$5 for the 22\,$\mu$m band.

\subsection{GALEX observations}

\begin{figure*}[htp]
 \centering \includegraphics[width=0.98\textwidth]{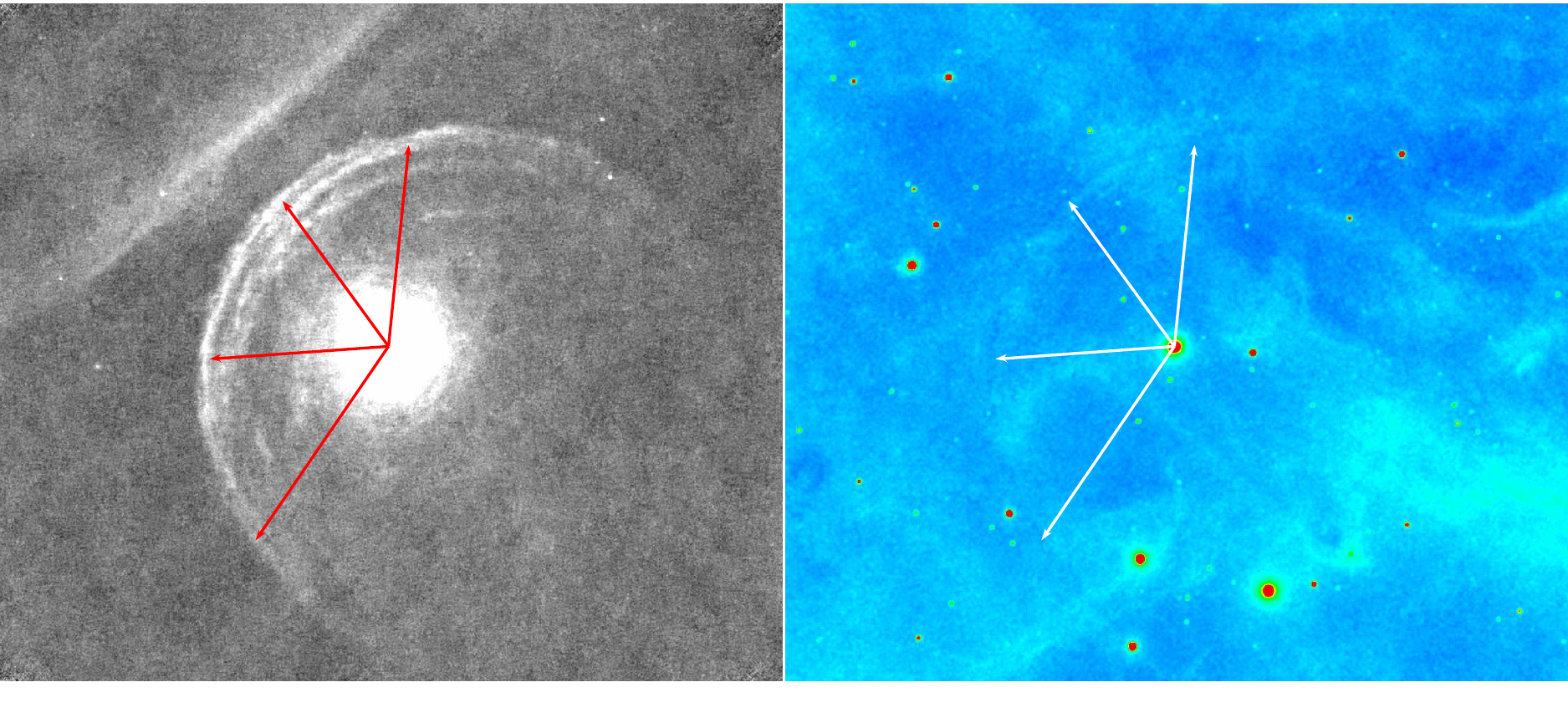}
\caption{Comparison between PACS 70\,$\mu$m (left) and GALEX FUV image (right, after applying a boxcar smoothing with a kernel radius of 2).  The white arrows indicate some faint extra extinction in the GALEX FUV image, which coincides with the outermost arc detected with Herschel. (The contrast in the figure is best viewed on screen.)}
\label{Fig:Galex}
\end{figure*}

Betelgeuse was also observed with the \emph{Galaxy Evolution Explorer} satellite (GALEX) on November 29, 2008. The pipeline-calibrated FUV and NUV images were retrieved from the \textit{GALEX} archive. The bandpass is 1344--1786\,\AA\ for an angular resolution of 4.5\arcsec\ for the FUV filter, and 1771--2831\,\AA\ for a 6\arcsec\ angular resolution for the NUV filter \citep{Morrissey2005ApJ...619L...7M}. The pixel size is 1.5\arcsec$\times$1.5\arcsec. The integration time for Betelgeuse was 48916\arcsec. Some very faint extra \textit{extinction} accompanying the outermost arc around Betelgeuse was detected in the FUV filter (see Fig.~\ref{Fig:Galex}). The FUV flux of the faint extra extinction is $\sim$7\% lower than in the adjacent pixels. The NUV filter is unfortunately satured by a ghost of the central star.

Until now, bright UV \textit{emission} in the wind-ISM interaction region has only been detected in two AGB stars: $o$~Cet \citep{Martin2007Natur.448..780M} and CW~Leo \citep{Sahai2010ApJ...711L..53S}{; $\alpha$ Ori is the first target for which extra UV extinction in the wind-ISM interaction zone is reported}.  No UV emission or extinction is seen in the region of the linear bar. 
The origin of the FUV/NUV emission of the bow shock around $o$~Cet has been attributed to H$_2$ molecules in the cold gas that are collisionally excited by hot electrons from the post-shock gas \citep{Martin2007Natur.448..780M}. The origin of the FUV/NUV emission in the interaction zone between the ISM and the stellar wind of CW~Leo has not been modelled in detail, although it was suggested by \citet{Sahai2010ApJ...711L..53S} that also here collisionally excited H$_2$ emission might be  the origin. The FUV/NUV emission around CW~Leo peaks at slightly larger distances from the central target than the  PACS/SPIRE infrared flux excess \citep[see Fig.~2 in][]{Ladjal2010A&A...518L.141L}. We note that for the planetary nebula NGC6720, dust and H$_2$ are co-spatial, and it is argued that H$_2$ has been formed on grain surfaces \citep{vanHoof2010A&A...518L.137V}. 
If the FUV extinction toward the bow shock around Betelgeuse is the FUV counterpart of the outer arcs as detected with PACS, this suggests that background UV radiation is absorbed by dust in the bow shock.


\subsection{GALFA \ion{H}{i} 21\,cm observations}

\ion{H}{i} 21\,cm observations in the vicinity of Betelgeuse were extracted from the results of the Galactic Arecibo L-Band Feed Array \ion{H}{i} (GALFA-\ion{H}{i}) survey \citep{Peek2011ApJS..194...20P}. GALFA-\ion{H}{i} is a survey of the Galactic interstellar medium with a spatial resolution of $\sim$4\arcmin, covering a large area (13\,000 deg$^2$) with a high spectral resolution (0.18\,km/s) from $-$700\,$<v_{\rm{LSR}}<$\,+700\,km/s  in the 21-cm line hyperfine transition of neutral hydrogen conducted at the Arecibo Observatory. 
{ALFA is a seven-beam feed array, with the beams arranged on the sky in a hexagon. Each beam is slightly elliptical along the zenith-angle direction, approximately 3.3\arcmin$\times$3.8\arcmin\ FWHM, with a gain of $\sim$11\,K/Jy for the central beam and $\sim$8.5\,K/Jy for the six other beams.}
Details on the data reduction are given in \citet{Peek2011ApJS..194...20P}. Typical noise levels are 80\,mK RMS in an integrated 1\,km/s channel, corresponding to an integration time of 30\,sec. 

\begin{figure*}[htp]
 \includegraphics[width=\textwidth]{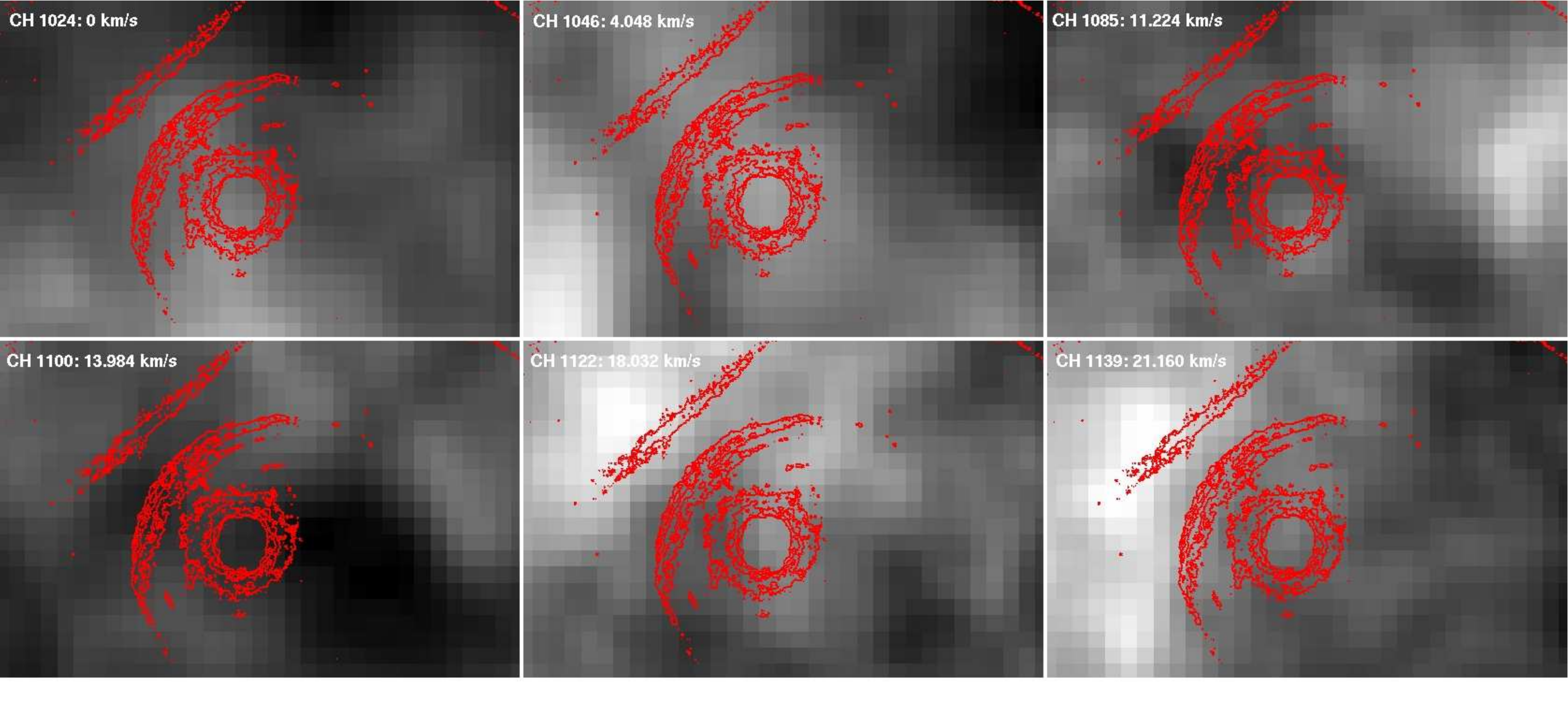}
\caption{Comparison between the GALFA-\ion{H}{i} 21 cm observations (1\arcmin\ square pixels) and the 0.3\,mJy countours of the PACS 70\,$\mu$m image (red). The GALFA-\ion{H}{i} data cube channel (CH) and corresponding velocity are indicated in the upper left corner.}
\label{Fig:GALFA}
\end{figure*}

\begin{figure}[htp]
\centering \includegraphics[height=0.35\textwidth,angle=180]{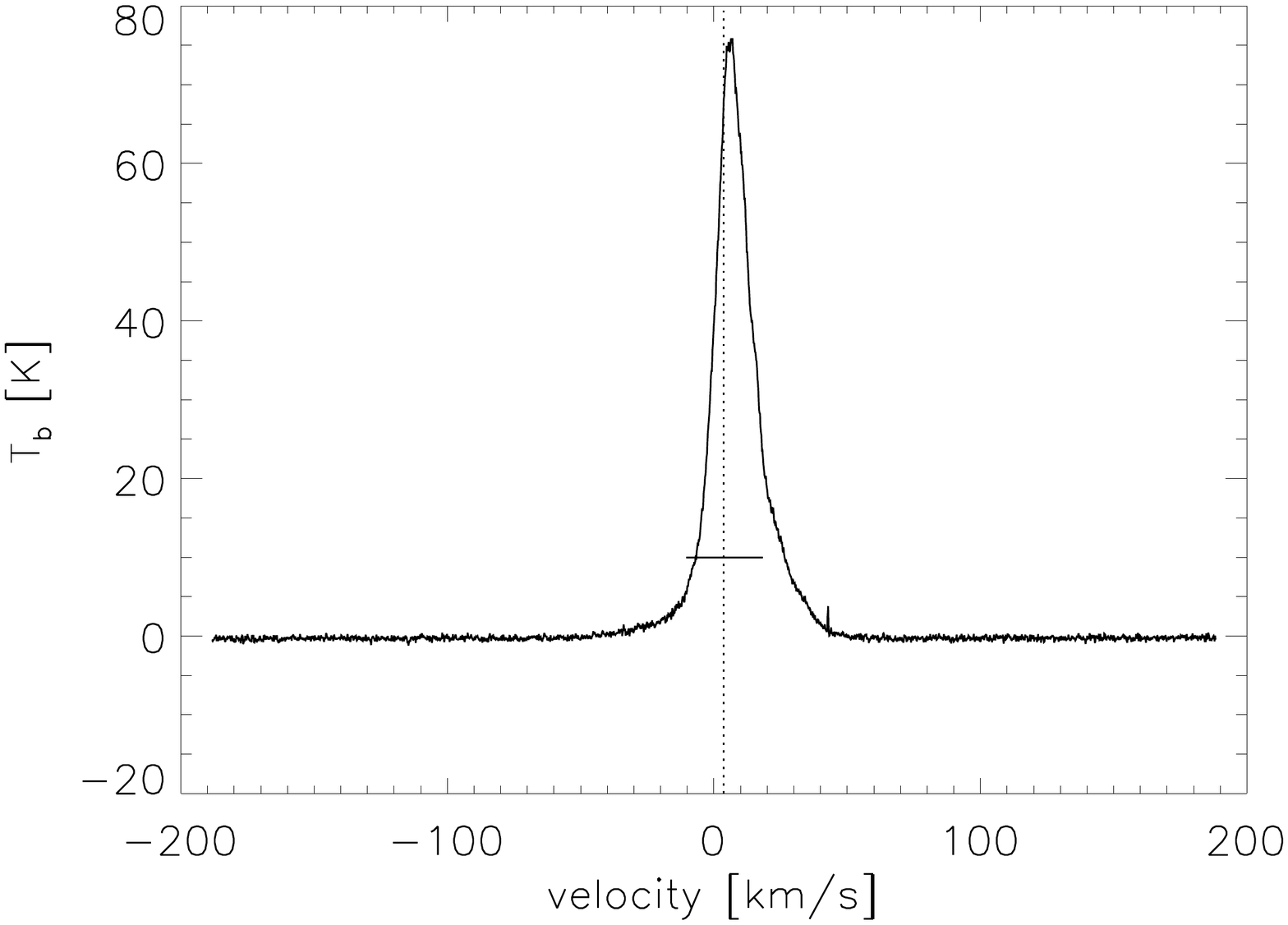}
\caption{{GALFA-\ion{H}{i} spectrum obtained on the position of Betelgeuse. The vertical line marks the CO $v_{\rm{LSR}}$ of Betelgeuse and the horizontal bar the total linewidth of the CO emission (twice the terminal velocity of the wind).}}
\label{FIG:HI_spectrum}
\end{figure}

\begin{figure*}[htp]
 \centering\includegraphics[height=0.8\textwidth, angle=180]{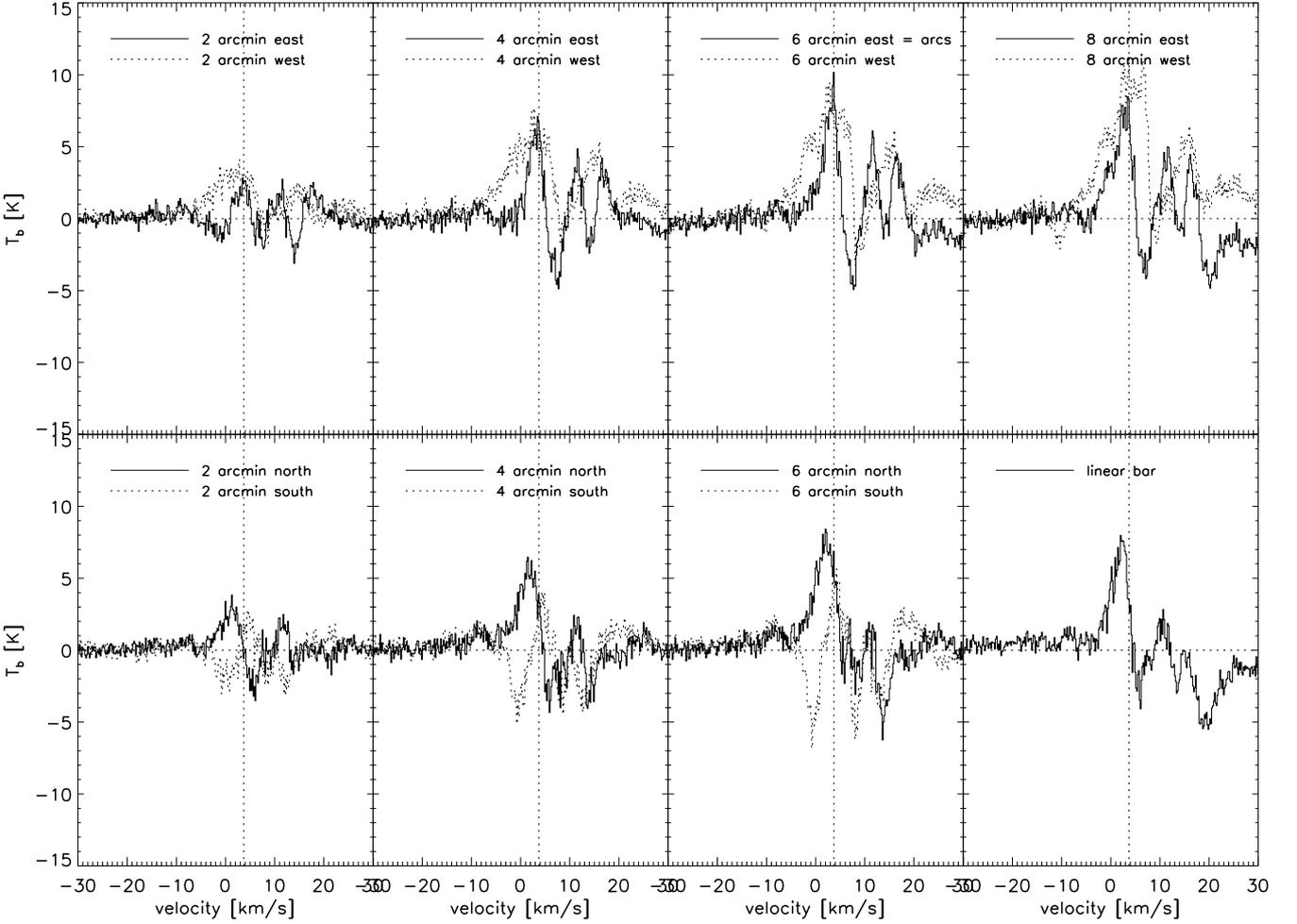}
\vspace*{3ex}
\caption{{$\alpha$ Ori spectra \ion{H}{i} spectra obtained by taken the off-positions in different directions (as indicated in each panel). The offset position at 6\arcmin\ to the east is situated in the arc detected with Herschel.  The right bottom panel shows the spectrum when the offset position is taken in the linear bar. The vertical line marks the CO $v_{\rm{LSR}}$ of Betelgeuse.} }
\label{FIG:HI_spectrum_OFFSET}
\end{figure*}

\begin{figure*}[htp]
 \centering\includegraphics[width=0.8\textwidth, angle=180]{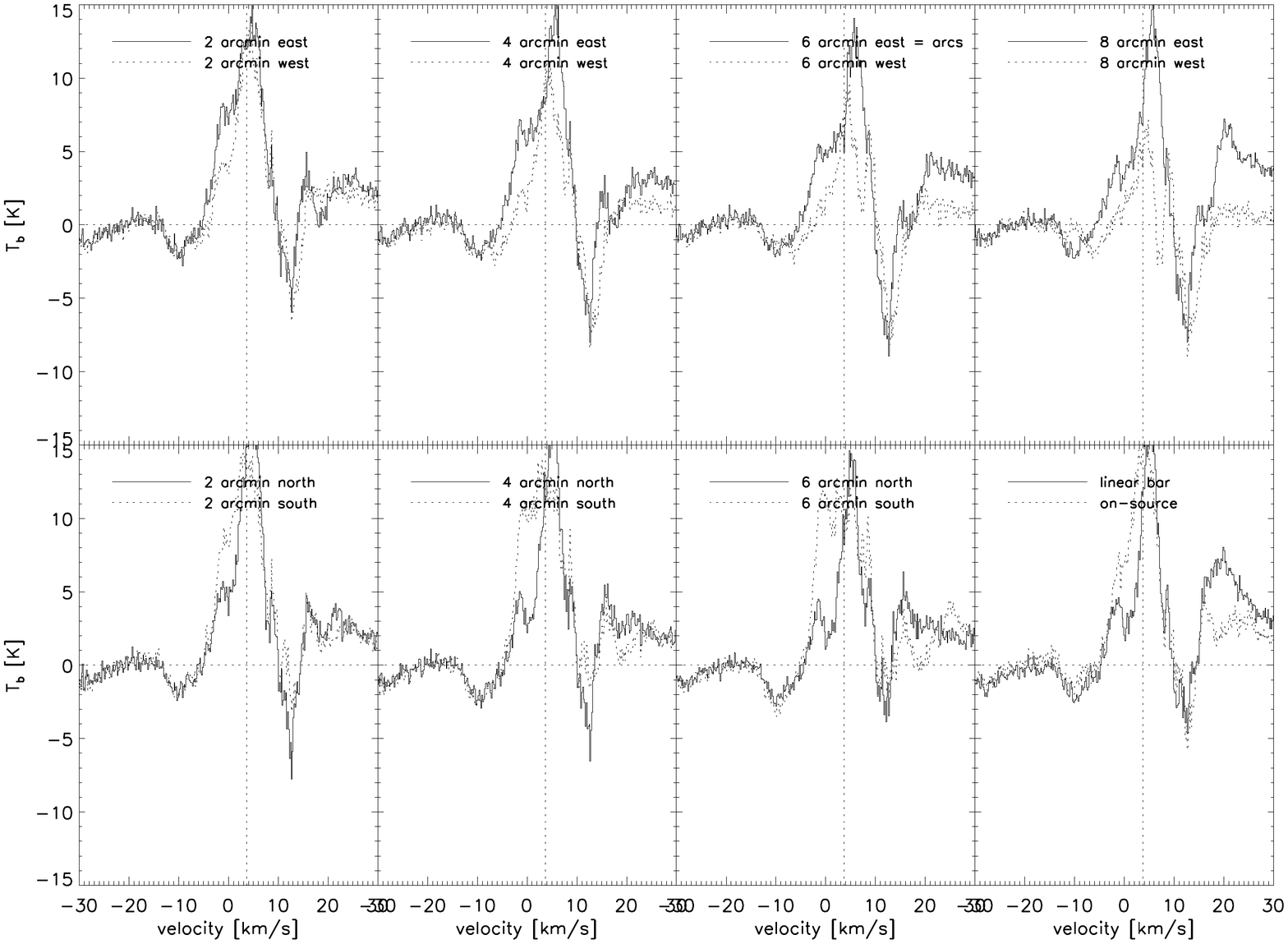}
\vspace*{3ex}
\caption{{\ion{H}{i} spectra at different positions in the envelope, arcs, and linear bar. The  spectrum of the 'off-position' taken at 11\arcmin\ to the west and 10\arcmin\ to the north of the central target has been subtracted. The position at 6\arcmin\ to the east is situated in the arc detected with Herschel.  The right bottom panel shows the on-source Betelgeuse$-$off-source \ion{H}{i} spectrum and the \ion{H}{i} spectrum in the linear bar. The vertical line marks the CO $v_{\rm{LSR}}$ of Betelgeuse.} }
\label{FIG:HI_spectrum_OFFSET2}
\end{figure*}

Several selected data cubes are displayed in Fig.~\ref{Fig:GALFA}. At the CO $v_{\rm{LSR}}$ velocity of $\sim$4\,km/s \citep{DeBeck2010A&A...523A..18D}, {\ion{H}{i} emission is clearly detected. In Fig.~\ref{FIG:HI_spectrum}, we present the \ion{H}{I} spectrum obtained on the position of Betelgeuse. The spectrum is in good agreement with the Nan\c{c}ay RadioTelescope (NRT) and Leiden/Argentina/Bonn (LAB) data presented in Fig.~3 by \citet{LeBertre2012arXiv1203.0255L}. The emission peaks at 6.992\,km/s, and no distinctive feature is seen in the velocity range of the circumstellar CO emission. It illustrates that in the direction of Betelgeuse, the \ion{H}{i} emission is dominated by galactic emission due to interstellar atomic hydrogen along the line of sight \citep{LeBertre2012arXiv1203.0255L}.
The \ion{H}{i} spectrum was also inspected at different offset positions. Taking the central star in the `on-position' and the `off-positions' at different offsets gave the  \ion{H}{i} spectrum of Betelgeuse (see Fig.~\ref{FIG:HI_spectrum_OFFSET}). Although affected by interstellar confusion, emission at the CO $v_{\rm{LSR}}$ is clearly detected in the spectra with the offsets taken in the east-west direction.   Fig.~\ref{FIG:HI_spectrum_OFFSET2} shows the  \ion{H}{i} spectrum at different offset positions for which the off-position was taken in a region outside the envelope, arcs or linear bar (i.e.\ at 11\arcmin\ to the west and 10\arcmin\ to the north of the central target). The interstellar confusion is seen for instance in the emission feature at $-$3\,km/s, and the absorption at +12\,km/s.
The Betelgeuse \ion{H}{i} line profile has an FWHM of $\sim$3.5\,km/s. The integrated line intensity is $\sim$2--7\,Jy\,km/s. Assuming a smooth outflow and using the standard relation, $M_{\rm{\ion{H}{i}}} = 2.36 \times 10^{-7} D^{2} \int{S_\nu\,dv}$, with $M_{\rm{\ion{H}{i}}}$ in \Msun, the distance $D$ in pc, and $\int{S_\nu\,dv}$ in Jy\,km/s, this  translates to an envelope mass of $\sim$0.02--0.07\,\Msun\ in atomic hydrogen at 197\,pc. As discussed already, the density in the inner  envelope is not homogeneous, for which reason this mass value should be considered as very approximate.}

\begin{figure}[htbp]
\includegraphics[width=0.48\textwidth]{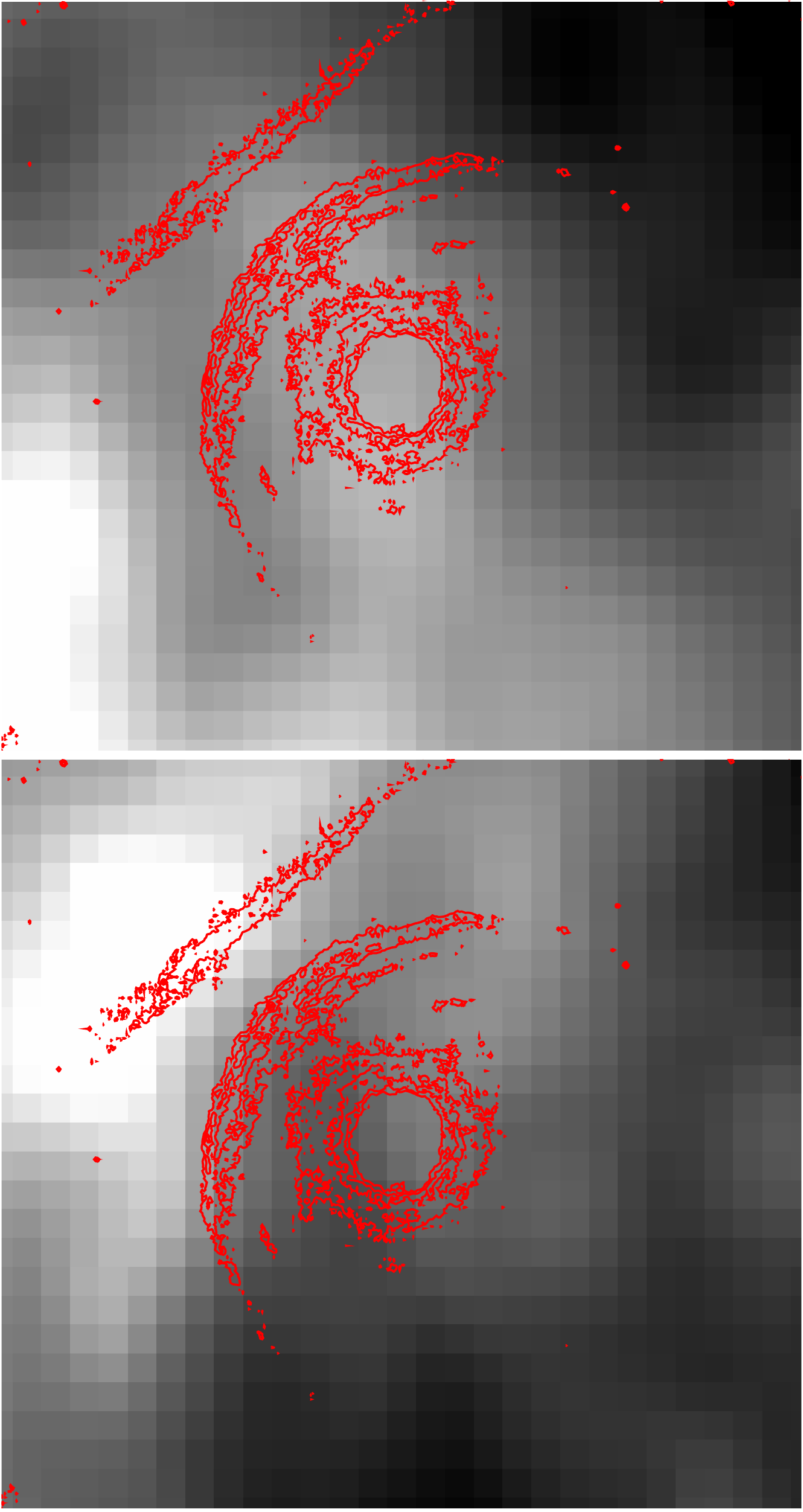}
 \caption{Comparison between the GALFA-HI emission, summed from $-1.472$ to 8.832\,km/s (top) and from 17.848 to 21.528\,km/ (bottom), and the 0.3\,mJy contours of the PACS 70\,$\mu$m image.}
\label{FIG:GALFA_PACS70}
\end{figure}

Significant emission is also seen at other velocities, of which a few examples are shown in Fig.~\ref{Fig:GALFA}. Using VLA data, \citet{LeBertre2012arXiv1203.0255L} note a seemingly spatial association between the \ion{H}{i} emission integrated over the range $-1.5$ to $8.9$\,km/s and the IRAS 60\,$\mu$m image (see their Fig.~9). This coincidence is not seen between the GALFA-\ion{H}{i} and Herschel 70\,$\mu$m data, probably owing to the low spatial resolution of the GALFA-\ion{H}{i} data (see upper panel in Fig.~\ref{FIG:GALFA_PACS70}).

The bottom panels in Fig.~\ref{Fig:GALFA}, showing the emission at a velocity of 18 and 21.17\,km/s, notably display an alignment of the \ion{H}{i} emission with the bar and arc-like structures detected in the Herschel PACS images (see also bottom panel in Fig.~\ref{FIG:GALFA_PACS70}). This \ion{H}{i} emission probably has a galactic ISM origin, and might be part of cold, low-density, atomic gas structures (see also Sect.~\ref{SECT:ORIGIN_BAR}). To gain further insight into the origin of the bar-like structure seen in the PACS observations, we  compared the galactic \ion{H}{i} emission at different velocities in a larger region covering a right ascension from 05h39m56s to 05h59m33s and a declination from 07$\deg$21\arcmin10.23\arcsec\ to 9$\deg$42\arcmin30.47\arcsec. From these data, it is clear that the ISM contains a lot of small-scale structure and that most of the emission seen in the direction of Betelgeuse is not related in any way to the mass loss of Betelgeuse. Because it is situated beyond the Local 
Bubble, Betelgeuse probably lies near the edge of a cavity of low neutral hydrogen density and close to a steep gradient in ISM column density between $l$\,=\,180\deg--200\deg\ \citep[see also][]{Harper2008AJ....135.1430H}. The ISM small-scale structure in the direction of Betelgeuse has also been detected by \citet{Knapp1988ApJ...331..974K} via the observation of discrete molecular CO clouds, probably associated with the $\lambda$ Ori molecular cloud complex.

\section{Observational properties of the infrared (extended) envelope around Betelgeuse}\label{SECT:results}

In this section, some observational properties as deduced from the infrared data for the inner envelope structure,  and the arcs and linear bar  are described in Sects.~\ref{SECT:results_inner} and~\ref{SECT:results_outer}, respectively.

\subsection{Inner envelope structure}\label{SECT:results_inner}

{Figure~\ref{Fig:AOri_inner} shows that the inner envelope structure of Betelgeuse is clearly non-homogeneous. The detection of non-spherical structures has already been reported on a sub-arcsecond-scale \citep[e.g.][]{Lim1998Natur.392..575L, Kervella2009A&A...504..115K, Kervella2011A&A...531A.117K} (see Sect.~\ref{SECT:BETELGEUSE}).}
The Herschel images show the first evidence of a high degree of non-homogeneity of the material lost by the star beyond 15\arcsec, which even persists until the material collides with the ISM. As described in Sect.~\ref{SECT:PACS_SPIRE_images} and shown in Fig.~\ref{Fig:AOri_inner}, very pronounced asymmetries are visible within 110\arcsec\ from the central star, although some weaker flux enhancements are visible until $\sim$300\arcsec\ away.  The typical angular extent for these arcs in the inner envelope ranges from $\sim$10\deg to 90\deg. {Considering that  asymmetries are seen on a subarcsecond-scale close to the stellar atmosphere and in the dust formation region, the Herschel images suggest that the 
arc-like structures seen in the free-flowing envelope of Betelgeuse (i.e.\ before the material collides with the ISM) are the relics of a non-homogeneous mass-loss process.  }

 {The radial distance  of the most pronounced asymmetries (within 2\arcmin) agrees with the size of a ring of \ion{H}{i} emission recently detected by \citet{LeBertre2012arXiv1203.0255L}, although no counterpart to the \ion{H}{i} emission plateau further away from the star in the south-western region is seen in the Herschel images. The Herschel and \ion{H}{i} observations seem to suggest that the mean gas and dust density (averaged over 360\deg) in the inner envelope drastically changed at $\sim$2\arcmin\ (or {32\,000\,yr} ago for an expansion velocity of {3.5\,km/s}). This might be due to, e.g., a change in mass-loss rate or the creation of an inner bow shock inside a fragmented, filamentary halo (see Sect.~\ref{SECT:constraints_arcs_obs}).}

The angular extent of the arc-like structures in the inner envelope of Betelgeuse is  considerably smaller than for the AGB star CW~Leo, the only other target in the MESS sample whose inner envelope has already been studied in detail \citep{Decin2011A&A...534A...1D}. In CW~Leo, almost spherical, ring-like, non-concentric dust arcs where detected until 320\arcsec, having  an angular extent between $\sim$40\deg\ and $\sim$200\deg. The shells of CW~Leo have a typical width of 5\arcsec--8\arcsec, and the shell separation varies in the range of $\sim$10\arcsec--35\arcsec, corresponding to $\sim$500\,--1\,700\,yr. This (again) suggests that the mass-loss mechanism of the AGB star CW~Leo, based on pulsations and radiation pressure on dust grains that formed non-isotropically, does not hold for Betelgeuse, which is an irregular variable with small amplitude variations. {As described in Sect.~\ref{SECT:BETELGEUSE}, the mass-loss process might by convection-induced}, which would yield locally strong variations in gas 
and dust density. 


Assuming that dust is the main contributor to the Herschel/PACS images (see Paper~II), one can derive the dust temperature in the arcs from a  modified blackbody of the form $B_\nu(T) \cdot \lambda^{-\beta}$, as expected for a grain emissivity $Q_{\rm{abs}} \sim \lambda^{-\beta}$ with the emissivity index $\beta$ ranging between one\footnote{It is generally admitted from Kramers-K\"onig relations that 1 is a lower limit for the spectral index \citep{Emerson1988felm.conf...21E}.} \citep[typical of, e.g., layered amorphous silicate grains,][]{Knapp1993ApJS...88..173K} and two \citep[typical for, e.g., crystalline silicate grains,][]{Tielens1987ASSL..134..397T, Mennella1998ApJ...496.1058M}. When $\beta$ is equal to 2.0, the dust temperature  for the clumps beyond 60\arcsec\ ranges from $\sim$25 to 65\,K. For $\beta$ equal to 1.0, the dust temperature increases to values between 35\,K and 140\,K  (see Fig.~\ref{FIG:info_Nick} in Sect.~\ref{SECT:constraints_arcs_obs}).

The Herschel images prove that clumpy structures are prevalent over the full envelope and might eventually have an impact on  the shape of the bow shock structure.

\subsection{Multiple arcs and linear bar} \label{SECT:results_outer}



\subsubsection{Orientation of the bow shock}
Using the Herschel PACS images, \citet{Cox2012A&A...537A..35C} derived a de-projected stand-off distance for the outer arcs of 4.98\arcmin\ and a position angle of 47.7\deg. Based on the distance of 131\,pc and proper motion values of \citet{vanLeeuwen2007A&A...474..653V}, a space motion vector inclination $i_{\rm LSR}$ of 8.9\deg\ for a $v_{\rm rad,LSR}$ of 3.4\,km/s was deduced. From the AKARI data,  \citet{Ueta2008PASJ...60S.407U} deduced  a de-projected stand-off distance for the outer arcs of 4.8\arcmin$\pm$0.1\arcmin, a position angle of 55\deg$\pm$2\deg\ and an inclination angle for the outer arcs of 56\deg$\pm$4\deg, where the bow shock cone is oriented to the plane of the sky. From the difference between the galactic space-velocity components of the bow shock cone and the heliocentric galactic space-velocity components of Betelgeuse, \citet{Ueta2008PASJ...60S.407U} find that the ISM around Betelgeuse flows at $\sim$11\,km/s into the position angle of $\sim$95\deg\ out of the plane of the sky (
towards us). Differences between the Herschel \citep{Cox2012A&A...537A..35C} and AKARI \citep{Ueta2008PASJ...60S.407U} results are mainly because of using different values for the radial velocity, proper motion, and distance.
 \citet{Ueta2008PASJ...60S.407U} and \citet{Cox2012A&A...537A..35C} have deduced that the peculiar velocity of the star with respect to the ISM is between 24 and 33\,km/s for an interstellar hydrogen nucleus density between 1.5 and 1.9\,cm$^{-3}$.

\subsubsection{Instabilities in the arcs}

The arc-like structure around Betelgeuse does not show clear evidence of large-scale instabilities, created by e.g.\ Kelvin-Helmholtz (KH), Rayleigh-Taylor (RT), Vishniac non-linear thin shell \citep[NTSI][]{Vishniac1994ApJ...428..186V}, or transverse acceleration \citep[TAI][]{Dgani1996ApJ...461..927D} instabilities. This is in contrast to other targets in the MESS sample as, e.g., R~Leo for which the Herschel image shows clear signatures of RT instabilities that are slightly deformed by to the action of the KH instabilities (see Fig.~\ref{Fig:RLeo_hydro} in the Appendix~\ref{Sect:RLeo}).
 The upper limit on the maximum length for instabilities in the arcs around Betelgeuse as traced from the Herschel images is $\sim$30\arcsec. Although suggested by \citet{Ueta2008PASJ...60S.407U}, vortex shedding \citep{Wareing2007ApJ...660L.129W} downstream in the tail of the bow shock is not seen in the Herschel images. 
With a wind velocity of $\sim$14.5\,km/s (as deduced from low rotational CO transitions, see Paper~II), the ratio between the peculiar velocity of the star, $v_\star$ and the wind velocity, $v_w$, is more than 1 and instabilities in the shell of the bow shock are expected \citep{Dgani1996ApJ...461..927D}.  This is also clear from the hydrodynamical simulations presented by \citet{vanMarle2011ApJ...734L..26V} for parameters typical of a red supergiant resembling Betelgeuse. The non-occurrence of large-scale instabilities can put constraints on the physical parameters determining the morphology of the bow shock (see Sect.~\ref{SECT:constraints_arcs_hydro}).



\subsubsection{Dust temperature in the arcs and linear bar}

Both the multiple arcs and the linear bar are best visible in the PACS images. 
The surface brightness in the multiple-arc like structure and linear bar has been measured for the different PACS filters using aperture photometry (see Table~\ref{Table_surface_brightness}). Uncertainties are $\sim$15\%, mainly dominated by the absolute flux uncertainty. The PACS and AKARI fluxes are consistent with each other for the arc-like region, with the exception of the AKARI 160\,$\mu$m point which is a factor $\sim$3 higher. However, the flux measurements in the bar-like structure are quite different (see Fig.~\ref{Fig:Akari}).  We note that both the arcs and the bar are extremely faint in the AKARI LW images (see  Appendix~\ref{SEC:previous}), and contamination cannot be excluded. These data points will therefore be neglected.

\begin{table}[htp]
 \caption{Background-subtracted average surface brightness values [in units of MJy/sr] as derived from the Herschel/PACS and WISE data (this paper) and from the AKARI as listed by \citet{Ueta2008PASJ...60S.407U} for the arc and bar-like structure near Betelgeuse.}
\label{Table_surface_brightness}
\vspace*{-1ex}
\centering
\begin{tabular}{lrcc}
\hline
\rule[0mm]{0mm}{3mm}& & ARC & BAR \\
\hline
\rule[0mm]{0mm}{3mm}     & 70\,$\mu$m & 17.9$\pm$2.69 & 16.3$\pm$2.45 \\
PACS & 100\,$\mu$m & 9.7$\pm$1.46 & 9.8$\pm$1.47 \\
     & 160\,$\mu$m & 4.1$\pm$0.61 & 5.3$\pm$0.80 \\
\hline
 \rule[0mm]{0mm}{3mm}& 65\,$\mu$m & 5--25 & $\sim$10 \\
 & 90\,$\mu$m & 5--20 & $\sim$10 \\
\raisebox{1.5ex}[0pt]{AKARI} &140\,$\mu$m & 5--10 & $\sim$15 \\
 & 160\,$\mu$m & 10--15 & $\sim$20 \\
\hline
\rule[0mm]{0mm}{3mm} & 12\,$\mu$m & & $\le$0.74\\
\raisebox{1.5ex}[0pt]{WISE} & 22\,$\mu$m & & 1.53$\pm$0.31\\
\hline
\end{tabular}
\end{table}

\begin{figure}[htp]
 \includegraphics[angle=180,width=0.48\textwidth]{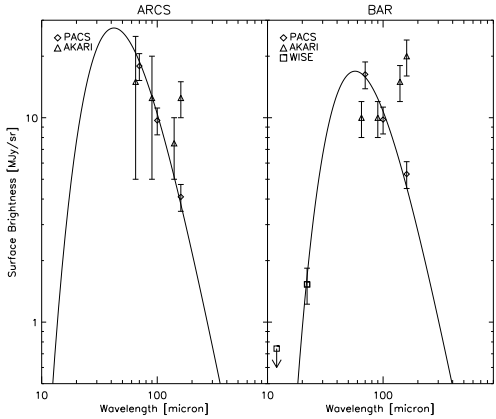}
\caption{Average surface brightness values derived from the  Herschel/PACS, AKARI, and WISE  data for the arc (left) and bar-like (right) structure around Betelgeuse. Shown in full black line is the best-fit modified blackbody to the Herschel data with a temperature of 85\,K ($\beta$\,=\,1.0) for the arcs and 63\,K ($\beta$\,=\,1.0) for the bar. The AKARI 140 and 160\,$\mu$m data for the arcs and bar are probably contaminated.
}
\label{Fig:Akari}
\end{figure}

{Neither [\ion{O}{i}] line emission at 63\,$\mu$m nor [\ion{C}{ii}] at 157\,$\mu$m is detected with Herschel in the bow shock region or linear bar (see Paper~II for more details), which is the reason we assume that dust emission is the main contributor to the Herschel/PACS images.}
Using a modified blackbody with a spectral index $\beta$ varying between one \citep[typical for, e.g., layered amorphous silicate grains or amorphous carbon grains,][]{Knapp1993ApJS...88..173K, Dupac2003A&A...404L..11D} and two \citep[typical of, e.g., crystalline silicate grains or graphitic grains,][]{Tielens1987ASSL..134..397T, Mennella1998ApJ...496.1058M, Dupac2003A&A...404L..11D}, we derived the (mean) dust temperature from the Herschel data using a Monte-Carlo uncertainty estimation with normally distributed noise.
For the arcs, a (mean) dust temperature of 87$\pm$7\,K ($\beta$\,=\,1.04$\pm$0.12) is derived and for the bar 64$\pm$2\,K ($\beta$\,=\,1.0$\pm$0.05) (see Fig.~\ref{Fig:Akari})\footnote{The Herschel/SPIRE flux values are well off the fit, presumably due to contamination by ISM dust.}. For the arcs, 91\% of the $\beta$-values have values $<$1.1, and the maximum value for $\beta$ is 1.88 with only 0.4\% having values in the interval between 1.75 and 1.90. For the bar, the maximum $\beta$-value is 1.02. Low values for the spectral index have already been found in circumstellar environments \citep[e.g.,][]{Knapp1993ApJS...88..173K}. Our result of a low spectral index (around 1) for quite high dust temperatures around 65\,K (bar) and 85\,K (arcs) is in good agreement with the inverse temperature dependence of the dust spectral index as derived by \citet{Dupac2003A&A...404L..11D}. According to \citet{Dupac2003A&A...404L..11D}, a low spectral index value might be explained by (1)~the occurrence of large grains, (2)
~the fact that warm regions could harbor aggregates of silicates, porous graphite, or amorphous carbon, having a spectral index around 1, or (3)~an intrinsic dependence of the spectral index on the temperature due to quantum processes, where in the case of high temperatures processes such as thermally-activated relaxation processes and temperature-dependent absorption could dominate. 
The chemical composition of the arcs might contain  layered amorphous silicate grains \citep[cfr.][]{Verhoelst2006A&A...447..311V} or larger grains \citep[as recently detected around the AGB star W~Hya,][]{Norris2012Natur.484..220N} explaining the low $\beta$-index. In the case where the linear bar has an interstellar origin (see Sect.~\ref{SECT:ORIGIN_BAR}), amorphous silicates --- an important component of the interstellar dust budget --- might explain the low $\beta$-index as well. However,  very small grains whose nature is thought to be carbonaceous \citep{Desert1986A&A...159..328D} might also be the origin of the low $\beta$-index. Observations in the near-IR are prerequisites to constraining  the detailed chemical content in the arcs and in the bar.

A modified blackbody with a temperature around 63\,K can explain the WISE 22\,$\mu$m data of the linear bar, but not the WISE 12\,$\mu$m data. Admittedly, the flux in the WISE 12\,$\mu$m band is quite uncertain and should be interpreted as an upper limit due to the large contribution of the central target (see Fig.~\ref{FIG:wise}). It might be that the Herschel and AKARI images show the emission by larger dust grains compared to the WISE data, which can be fitted with a modified blackbody of $\sim$145\,K.  Smaller grains indeed attain higher temperatures, hence emit at shorter wavelengths, than large grains, since the former absorb more efficiently per unit mass.

\begin{figure*}[htp]
     \begin{minipage}[t]{.48\textwidth}
        \centerline{\resizebox{\textwidth}{!}{\includegraphics{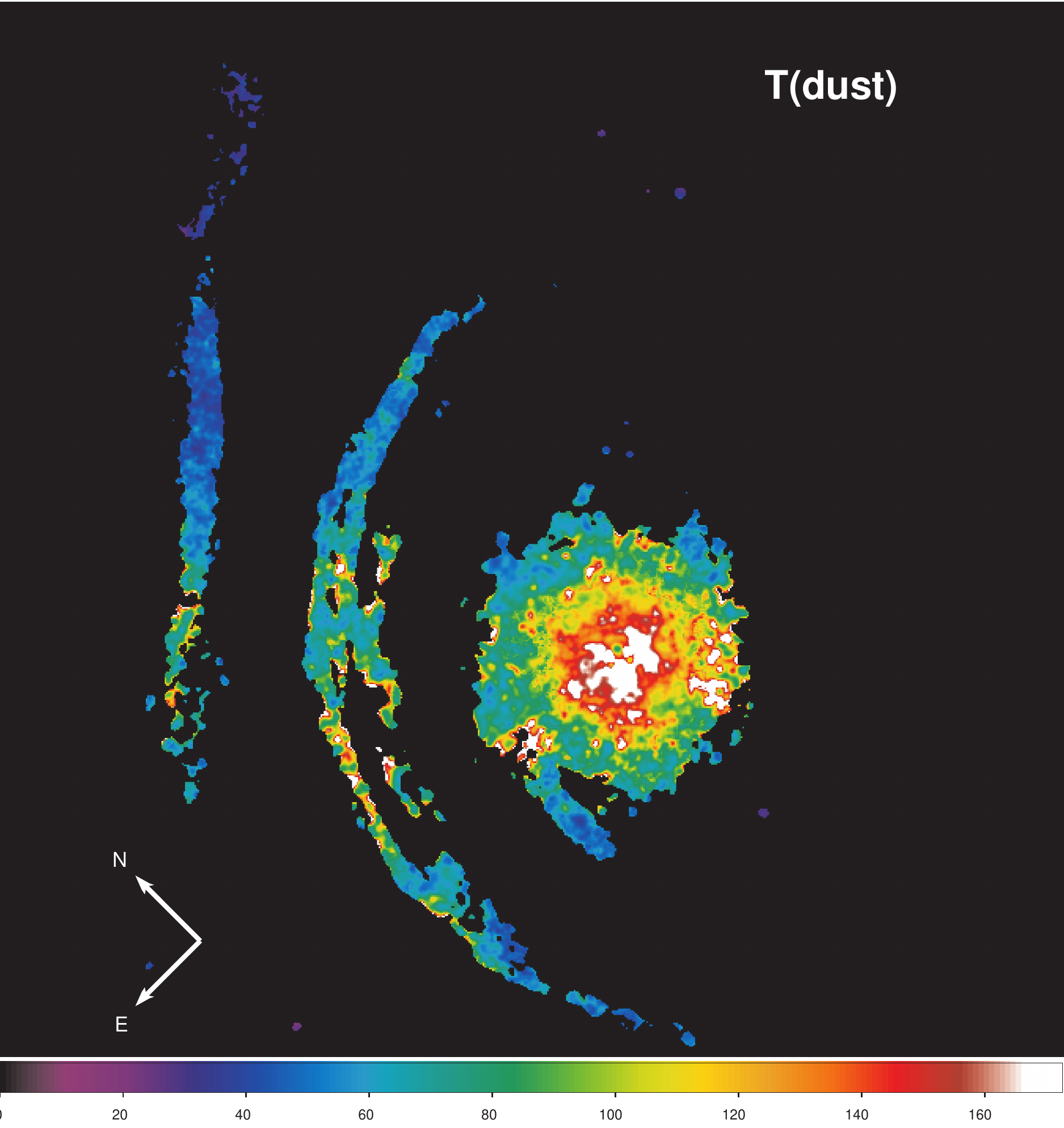}}}
    \end{minipage}
 \hfill
     \begin{minipage}[t]{.48\textwidth}
        \centerline{\resizebox{\textwidth}{!}{\includegraphics{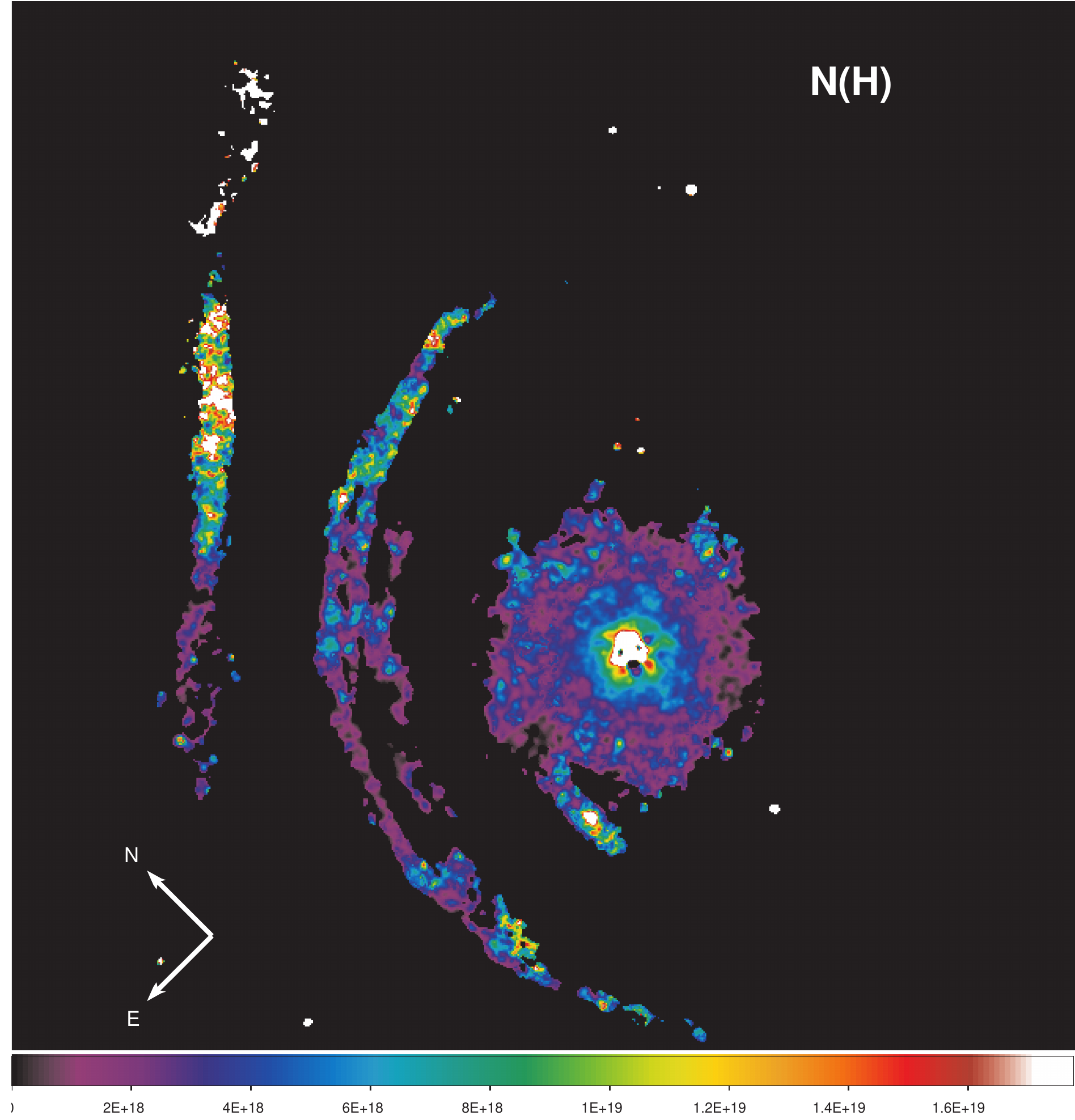}}}
    \end{minipage}
     \begin{minipage}[t]{.48\textwidth}
        \centerline{\resizebox{\textwidth}{!}{\includegraphics{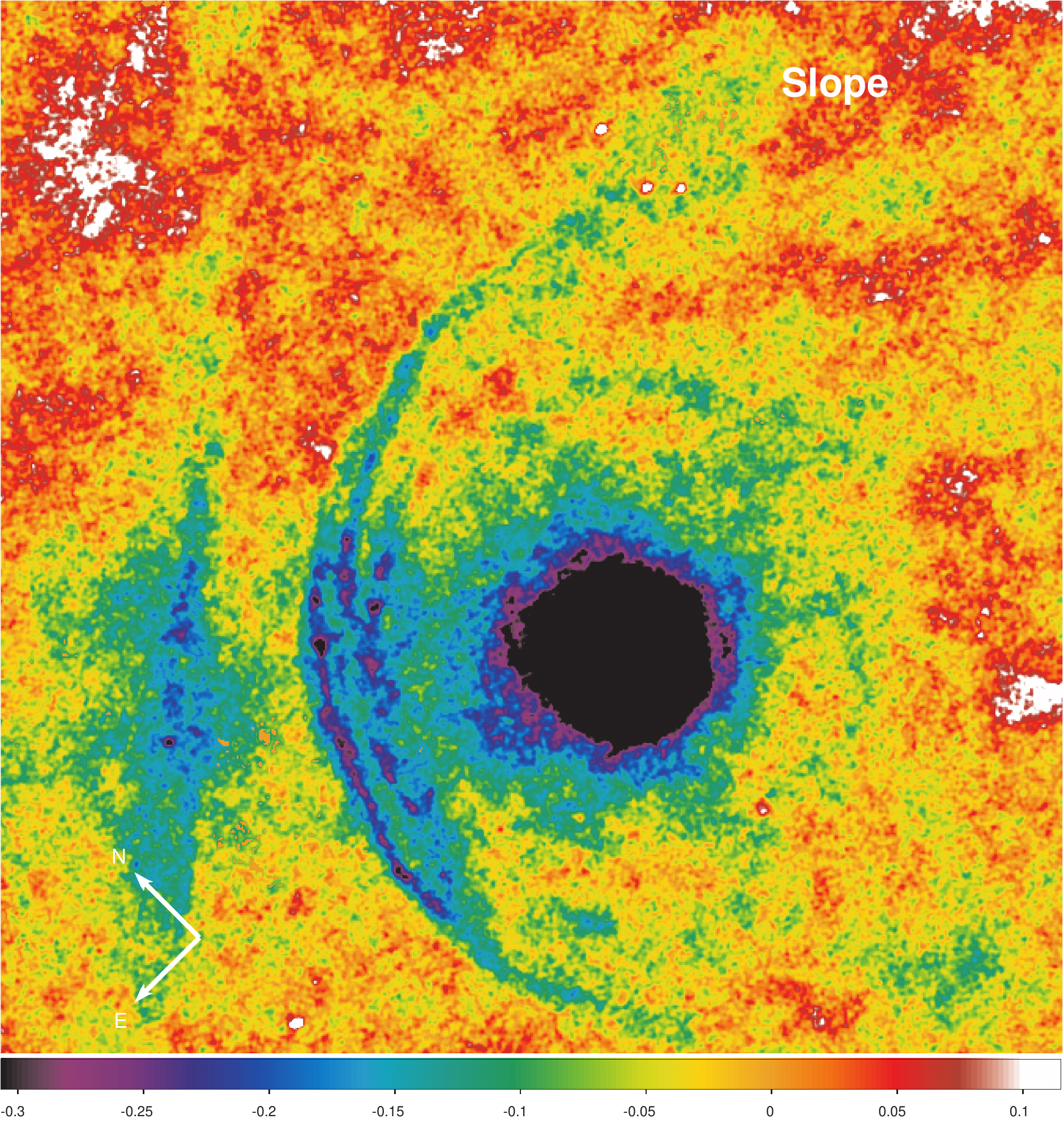}}}
    \end{minipage}
 \hfill
     \begin{minipage}[t]{.48\textwidth}
        \centerline{\resizebox{\textwidth}{!}{\includegraphics{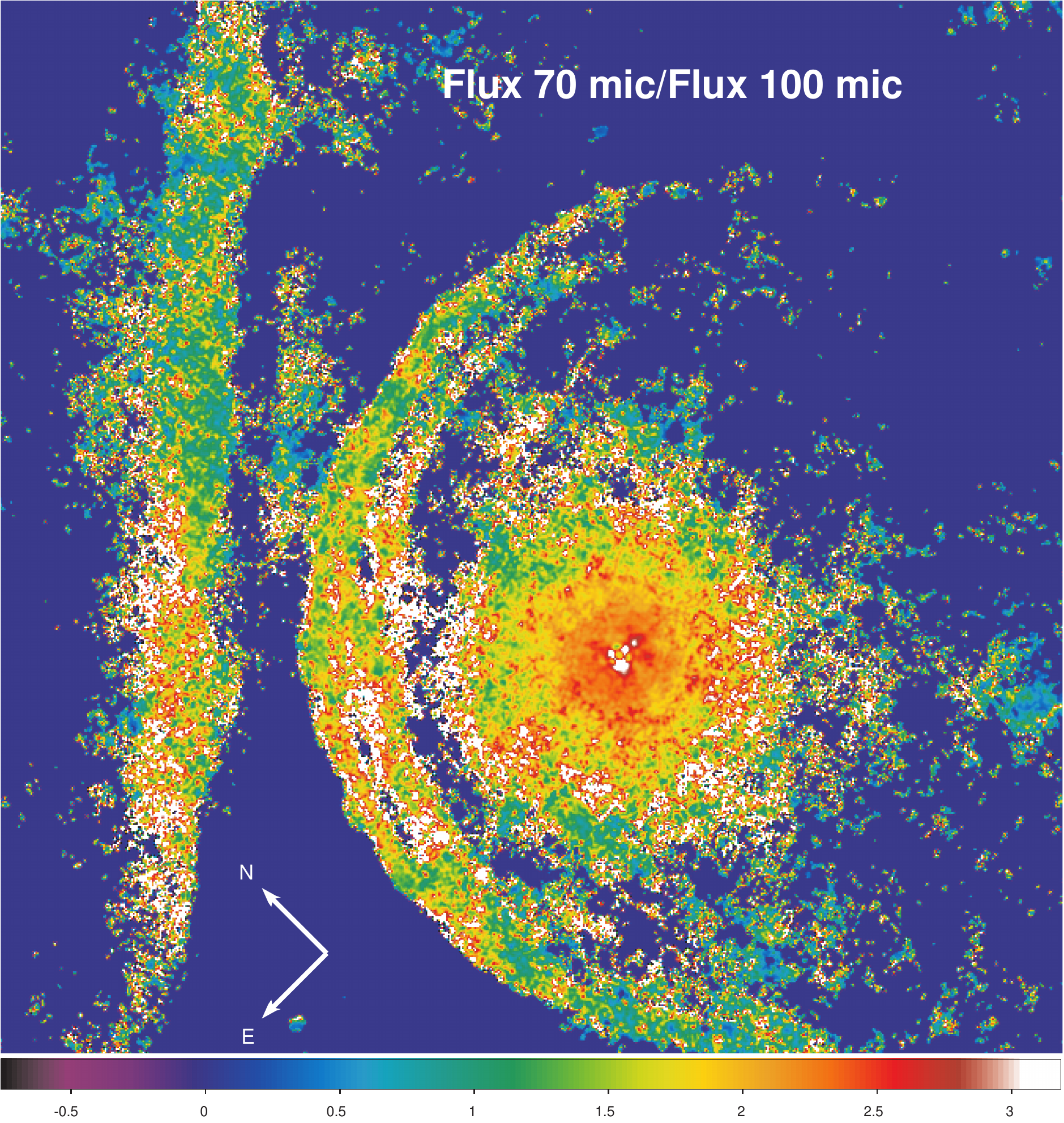}}}
    \end{minipage}
\caption{Dust temperature [K] ($\beta=1$, upper left), hydrogen column density [cm$^2$] (upper right), slope of the linear fit to the flux values at 70, 100, and 160\,$\mu$m (bottom left), and ratio between the flux values at 70 and 100\,$\mu$m (bottom right) as derived from the Herschel PACS images. See text for more details.}
\label{FIG:info_Nick}
\end{figure*}

The dust temperature map (for $\beta=1$) and hydrogen column density\footnote{for a dust opacity $\kappa_\nu=\kappa_0 (\nu/\nu_0)^\beta$, with $\nu_0$\,=\,1000\,GHz, $\kappa_0$\,=\,0.1\,cm$^2$/g, and $\beta$\,=\,1 \citep{Beckwith1990AJ.....99..924B}} in the arcs and linear bar as derived from the Herschel PACS data are shown in Fig.~\ref{FIG:info_Nick}. A lower limit for the flux values of 9, 8, and 4\,MJy/sr was used for the 70, 100, and 160\,$\mu$m data, respectively. Mainly in the eastern region, the flux values in the 160\,$\mu$m image are too low to derive the dust temperature and hydrogen column density properly. More physical information on the eastern region can be obtained from either the slope of a linear fit to the flux values at 70,100, and 160\,$\mu$m or from the ratio of the flux at 70\,$\mu$m to the flux at 100\,$\mu$m (see bottom panels in Fig.~\ref{FIG:info_Nick}).The arcs and bar have comparable dust colour temperatures.
The dust temperature is higher both in the multiple arcs and in the linear bar in the direction of space motion of Betelgeuse. A gradient in temperature is visible, with the temperature decreasing (and column density increasing) at larger distances from the central target.  

Considering energy conservation, one can show that the energy flux of the gas entering the shock front per unit area and per unit time is around five orders of magnitude lower than the stellar radiation energy flux at the distance of the arcs and bar. As a result, the additional heating of the dust grain from the hot post-shock gas is negligible, and the temperature of the dust grain is dictated by radiative equilibrium conditions when only taking the stellar radiation into account (see also {Sect.~\ref{SECT:cold_medium}). Assuming a smooth continuous outflow, the temperature of an iron or amorphous silicate grain only heated by the stellar radiation is in the order of ranges from about 45\,K to 60\,K at a distance of 280--530\arcsec\ away from the central star (see Appendix~\ref{App:Tdust}). The correspondence between the mean dust temperatures derived from the data and these theoretical calculations is satisfactory, taken the assumptions of the theoretical model into account. Specifically, one can imagine 
that the dust temperatures in a clumpy envelope might indeed be somewhat higher than the values shown in Fig.~\ref{Fig:Robin}, since stellar photons can probably reach the outer envelope regions more easily. In addition, smaller grains will attain a higher temperature since they absorb more efficiently per unit mass.
}

\subsubsection{Mass of the arcs and linear bar}

For an opacity, $\kappa_\lambda$, of 185\,cm$^2$/g at 70\,$\mu$m, 90\,cm$^2$/g at 100\,$\mu$m, and of 35\,cm$^2$/g at 160\,$\mu$m \citep{Mennella1998ApJ...496.1058M} and assuming a dust-to-gas ratio of 0.002 \citep{Verhoelst2006A&A...447..311V}, we deduce a total gas and dust mass of $\sim$$2.4\times10^{-3}$\,\Msun\ for the arcs and  $\sim$$2.1\times10^{-3}$\,\Msun\ for the linear bar. {The derived (gas+dust) mass is very sensitive to the assumed dust temperature and dust-to-gas ratio. Lowering the dust temperature in both the arcs and bar to 30\,K would yield a mass of 0.07\,\Msun\ and 0.029\,\Msun, respectively. Assuming that the linear bar has an ISM origin (see Sect~\ref{SECT:ORIGIN_BAR}) with a dust-to-gas ratio around 0.01, the mass in the bar would be around 0.01\,\Msun.}

{Assuming a smooth outflow, a first theoretical estimate for the mass in the bow shock shell can be derived from the equation \citep{Mohamed2012A&A...541A...1M}
\begin{equation}
 M_{\rm shell} = \frac{\Mdot \, R_{\rm SO}}{v_{\rm w}}\,,
\label{Eq:mass_shell}
\end{equation}
with $R_{\rm SO}$ the stand-off distance of the bow shock and $v_{\rm w}$ the wind velocity. 
This expression assumes the mass is distributed over $4 \pi$ steradian in a spherical shell and only takes the contribution from the stellar wind into account. Thus for $\theta \la 90\deg$, the expected shell mass would be half of $M_{\rm shell}$. A bow shock shell mass of $2.4\times10^{-3}$\,\Msun\ for the arcs would imply a mass-loss rate of $2 \times 10^{-7}$\,\Msun/yr (for a wind velocity of 14.5\,km/s). However, CO and \ion{H}{i} observations tracing the inner envelope regions yield a mass-loss rate that is one order of magnitude higher, $\sim$$2 \times 10^{-6}$\,\Msun/yr \citep[e.g., Paper~II;][]{Rodgers1991ApJ...382..606R, LeBertre2012arXiv1203.0255L}. This hints that the mean mass-loss rate might have varied substantially, as also suggested in Sect.~\ref{SECT:results_inner}. }

The hydrodynamical models presented in Sect.~\ref{SECT:cold_medium} yield a mass in the arcs of $\sim$0.1\,\Msun\ (including 90\% CSM and 10\% ISM material). 
According to \citet{Mohamed2012A&A...541A...1M}, the observed low mass in the arcs might indicate that the bow shock created by the RSG wind has a very young age, of the order of 20\,000\,yr, and may not yet have reached a steady state.
But, the existence of (non-homogeneous) clumps in the inner wind travelling all the way to the bow shock region might also explain the low  mass in the arcs as derived  from the Herschel images.



\section{Discussion: Origin of multiple arcs} \label{SECT:ORIGIN_ARCS}

In this section, we focus on the origin of the multiple arcs seen in the bow shock around Betelgeuse. Section~\ref{SECT:constraints_arcs_obs} gives details on some observational properties. In Sect.~\ref{SECT:constraints_arcs_hydro}, the observational data are compared with hydrodynamical simulations. In Sect.~\ref{Sect:Origin_arcs_discussion}, we discuss the origin of the multiple arcs based on the observational and theoretical constraints. 

\subsection{Constraints from the observations}
\label{SECT:constraints_arcs_obs}

From the observations described in previous sections, two concerns might be important for constraining the origin of the multiple arcs seen in the bow shock region of Betelgeuse.
(1)~The different arcs are reasonably well fitted with concentric ellipses (see Fig.~\ref{FIG:PACS70_ellipse} in the Appendix~\ref{App:additional_figures}).
(2)~The inner envelope shows clear evidence of a pronounced asymmetric clumpy structure. This leads to the hypothesis that the origin of the different arcs is a projection effect on the plane of the sky of the contact discontinuity created from the collision between different inhomogeneous mass-loss events and the ISM. As shown in Fig.~6 of \citet{Cox2012A&A...537A..35C}, different inclination angles lead to projected outlines of the contact discontinuity, where concentric ellipses with a smaller inclination angle yields a smaller projected size of the bow shock outline. This would mean that the arcs are shaped by an interaction of the inhomogeneous stellar mass loss at past and present epochs. 

Another explanation for the different arcs (and also linear bar) might be inferred from the 
 recent results of \citet{vanMarle2011ApJ...734L..26V}, who shows that small grains follow the movement of the gas mass elements, while larger CSM grains tend to  penetrate deeper into the ISM  (see also Sect.~\ref{SECT:constraints_arcs_hydro}). This analysis takes only the inertia of the grains into account. However, since the grains are charged, they will gyrate  around the magnetic field lines. Assuming a gas temperature in the shock of 10\,000\,K and a shock velocity equal to the space velocity, the typical Larmor radius for a grain with a size of 5\,nm is $\sim 1 \times 10^{14}$\,cm, while larger grains with a size of 1\,$\mu$m have a Larmor radius in the order of $4 \times 10^{18}$\,cm. With a typical width of $\sim$20\arcsec\ for the arcs, grains with a size $\la$0.1\,$\mu$m are position-coupled to the hot gas. One then could postulate that arcs further away from the central target could contain larger dust grains ($\ga$0.1\,$\mu$m), while the inner arcs contain the smallest grains coupled to the gas. 
In the case of randomized mass-loss variations (as suggested by the data tracing the inner envelope), one could consider the possibility that each mass-loss event has a favourable grain-size spectrum (i.e., the grain size distribution function reaches a maximum for some specific grain sizes). An analogous situation is obtained by \citet{Fleischer1994PhDT.......101F}, who modelled the dust formation in the case of a dynamical atmosphere and showed that far out in the wind, where the grain growth is definitely stopped, the grain-size spectrum shows distinct grain-size peaks. 
A larger grain size implies a higher drift velocity and a larger Larmor radius, which eventually might lead to a bow shock arc further away from the central target. However, it remains remarkable that only Betelgeuse in the whole MESS sample shows the appearance of multiple arcs in the bow shock region, while the same argument as given above might hold for other targets in the sample, especially the Mira-type variables with high-amplitude variations being favourable for a grain-size spectrum with strong distinct peaks. In addition, the dust temperature maps (Fig.~\ref{FIG:info_Nick}) show no significant temperature difference between the different arcs, while it is well known that smaller dust grains are generally hotter than larger grains since the former absorb more efficiently per unit mass.


The observed fragmentation of the bow shock might also be understood in terms of the effects predicted by \citet{Dgani1998ApJ...495..337D}. They postulate that RT instabilities might fragment a bow shock in the direction of motion, enabling the ISM to flow into the inner parts of the envelope and creating a bow shock  well inside the almost spherical, but very filamentary, haloes. For targets close to the Galactic plane (as is the case of Betelgeuse, with $b=-9\deg$), the ISM magnetic field can make RT instabilties very efficient. This might explain the recent detection by \citet{LeBertre2012arXiv1203.0255L} of a detached \ion{H}{i} shell with a radius of 2\arcmin. Magnetic fields would suppress certain modes but accentuate others, changing the appearance of the envelope, and potentially leading to `RT rolls' (or stripes). An example resembling the image of Betelgeuse in shown in Fig.~3a in \citet{Dgani1998ApJ...495..337D}. We should remark, however, that the calculations as performed by \citet{Dgani1998ApJ...495..337D} were for the case of a planetary nebula, which have much higher stellar wind velocities than red supergiants or AGB stars. In the sample of 78 AGB stars and supergiants analysed by \citet{Cox2012A&A...537A..35C}, Betelgeuse is the target closest to the Galactic plane ($z=-5$\,pc), and it is the only target showing this multiple arc-like structure. Four other targets in the MESS sample are at similar distances from the Galactic plane ($z<20$\,pc), one being a supergiant ($\mu$~Cep) which resembles Betelgeuse in different aspects, including a similar dust mass-loss rate to  Betelgeuse \citep{Josselin2007A&A...469..671J}. The main difference for parameters influencing the bow shock morphology is the wind velocity, which is a factor 2.4 higher in $\mu$~Cep \citep{DeBeck2010A&A...523A..18D}. While for Betelgeuse $v_\star/v_w > 1$, this ratio is $<$1 for $\mu$~Cep in which case it has been predicted by \citet{Dgani1996ApJ...461..927D} that the bow shock is stabler. The Herschel images presented by \citet{Cox2012A&A...537A..35C} show, that (possible) instabilities in the bow shock of $\mu$~Cep have comparable sizes to Betelgeuse, while different arcs are indeed not seen. 


If present, the strength of the ISM magnetic field can be predicted from the angular separation between the arcs. An angular separation of 30\arcsec\ at a distance  of 300\arcsec\ yields an Alfv\'en speed in the pre-shocked ISM of $\sim$4\,km/s \citep[Eq.~4 in][]{Dgani1998RMxAC...7..149D}. For an ISM density of 1.9\,cm$^{-3}$ \citep{Cox2012A&A...537A..35C} and assuming that mainly \ion{H}{i} is present in the bow shock, this yields a magnetic field of 3\,$\mu$G. The average strength of the total magnetic field in the Milky Way is about 6\,$\mu$G near the Sun \citep{Biermann2001pteu.conf..543B} and increases to 20--40\,$\mu$G in the Galactic centre region. Dense clouds of cold molecular gas can host fields of up to several mG strength \citep{Heiles2005LNP...664..137H}.


\subsection{Constraints from hydrodynamical modelling}
\label{SECT:constraints_arcs_hydro}


For insight into the origin of the complex CSM-ISM interface and the non-occurrence of large-scale instabilities in the arcs, we have simulated the hydrodynamical evolution of the envelope of Betelgeuse. The simulations also show if dust grains are good tracers for the gas-driven dynamics in the bow shock. The hydrodynamical computations were done using {\tt MPI-AMRVAC} code \citep{Keppens2012JCoPh.231..718K, vanMarle2011ApJ...734L..26V}.  More details on the numerical method and initial conditions are given in Appendix~\ref{App_numerics}.

\begin{table}[htp]
 \caption{Physical parameters of the envelope of Betelgeuse. }
\label{Table:input_AOri}
\begin{tabular}{lc}
 \hline \hline
Parameter & Value \\
\hline
Photospheric temperature & $T_\star$\,=\,3500\,K$^{(a)}$\\
Stellar radius & $R_\star$\,=\,7.5$\times$10$^{13}$\,cm$^{(a)}$\\
Stellar mass & $M_\star$\,=\,20\,\Msun$^{(b)}$ \\
Mass loss rate & \Mdot\,=\,$2.5\times10^{-6}$\,\Msun/yr$^{(c)}$ \\
Dust-to-gas ratio & $\psi$\,=\,0.002$^{(d)}$\\
Distance & D=197\,pc$^{(e)}$ \\
Terminal wind velociy & $v_\infty$\,=\,14.5\,km/s$^{(c)}$\\
Space velocity w.r.t.\ ISM & $v_\star$\,=\,28.3\,km/s$^{(f)}$\\
\hline
\end{tabular}\\
$^{(a)}$~\citet{Rodgers1991ApJ...382..606R}; 
$^{(b)}$~\citet{Dolan2008APS..APR.S8006D};
$^{(c)}$~paper~II;
$^{(d)}$~\citet{Verhoelst2006A&A...447..311V}; 
$^{(e)}$~\citet{Harper2008AJ....135.1430H};
$^{(f)}$~\citet{Ueta2008PASJ...60S.407U}
\end{table}

The simulation was run for five  different situations, using variations of a basic model to explore several aspects of the bow shock morphology and its dependence on the properties of the wind and the ambient medium. The physical parameters for the basic model are given in Table~\ref{Table:input_AOri}. The gas density in the ambient medium is set at $3 \times 10^{-24}$\,g cm$^{-3}$ (or $n_{\rm{H}} \approx 2$\,cm$^{-3}$) and a temperature of 100\,K (reflecting the temperature of a cold neutral medium). The five different simulations have the following specifications:
\begin{itemize}
 \item The first simulation (A) uses the basic wind parameters that are specified in Table~\ref{Table:input_AOri} and the ambient medium specified above. We assume that all dust grains have the same size, 5\,nm, and density, 3.3\,g\,cm$^{-3}$. 

 \item The second simulation (B) is the same as simulation A, except for the size of the dust grains, 
       which is increased to 100\,nm to demonstrate the behaviour of large dust grains. 

 \item The third simulation (C) has a warm ambient medium of 8\,000\, K, reflecting the temperature of a warm neutral or partially ionized medium. 
       The ambient medium is assumed to be heated by an outside radiation source. 
       Therefore, whereas the ambient medium is kept at a minimum temperature of 8\,000\,K, 
       the stellar wind, which is protected from UV-radiation by the ionized gas of the bow shock, has a minimum temperature of 100\,K. 

 \item The fourth simulation (D) has the same ambient medium temperature as simulation~C, but 
       with lower ambient medium density ($5\times10^{-25}$\,g\,cm$^{-3}$ ) and higher space velocity ($v_\star=72.5$\,km/s). This simulation was added to compare our results with the recent  hydrodynamical models of \citet{Mohamed2012A&A...541A...1M}.
 
 \item For the fifth simulation (E), as for the first and second, we set the minimum temperature to 100\,K throughout the entire domain. 
       However, instead of a smooth wind, we assume a periodic variation in the mass loss rate. 
       For 1\,000\, years the mass-loss rate is assumed to be high {(\Mdot\,=\,$2.5\times10^{-5}$\,\Msun/yr)}, then, for the next 9\,000\, years, it is low {(\Mdot\,=\,$2.5\times10^{-8}$\,\Msun/yr)}. This `picket-fence' type variation is then repeated periodically.
The two mass-loss rates are related as $\Mdot_{\rm high}\,=\,1000\,\Mdot_{\rm low}$. 
The total amount of mass lost over the 10\,000\, year period is the same as for the first two simulations. 
The period of 10\,000\, years reflects the typical timescale for AGB thermal pulses.
\end{itemize}

The effects of changing one or more input parameters in the simulations are discussed in Sects.~\ref{SECT:cold_medium}--~\ref{Sect_hydro_clumps}. A link to the online movies is provided in the online Appendix~\ref{app-online}. A reflection on the origin of the multiple arcs based on the outcome of the simulations is given in Sect.~\ref{Sect:Origin_arcs_discussion}.

\subsubsection{Cold ambient medium}\label{SECT:cold_medium}
\begin{figure*}[htp]
    \begin{minipage}[t]{.495\textwidth}
        \centerline{\resizebox{\textwidth}{!}{\includegraphics{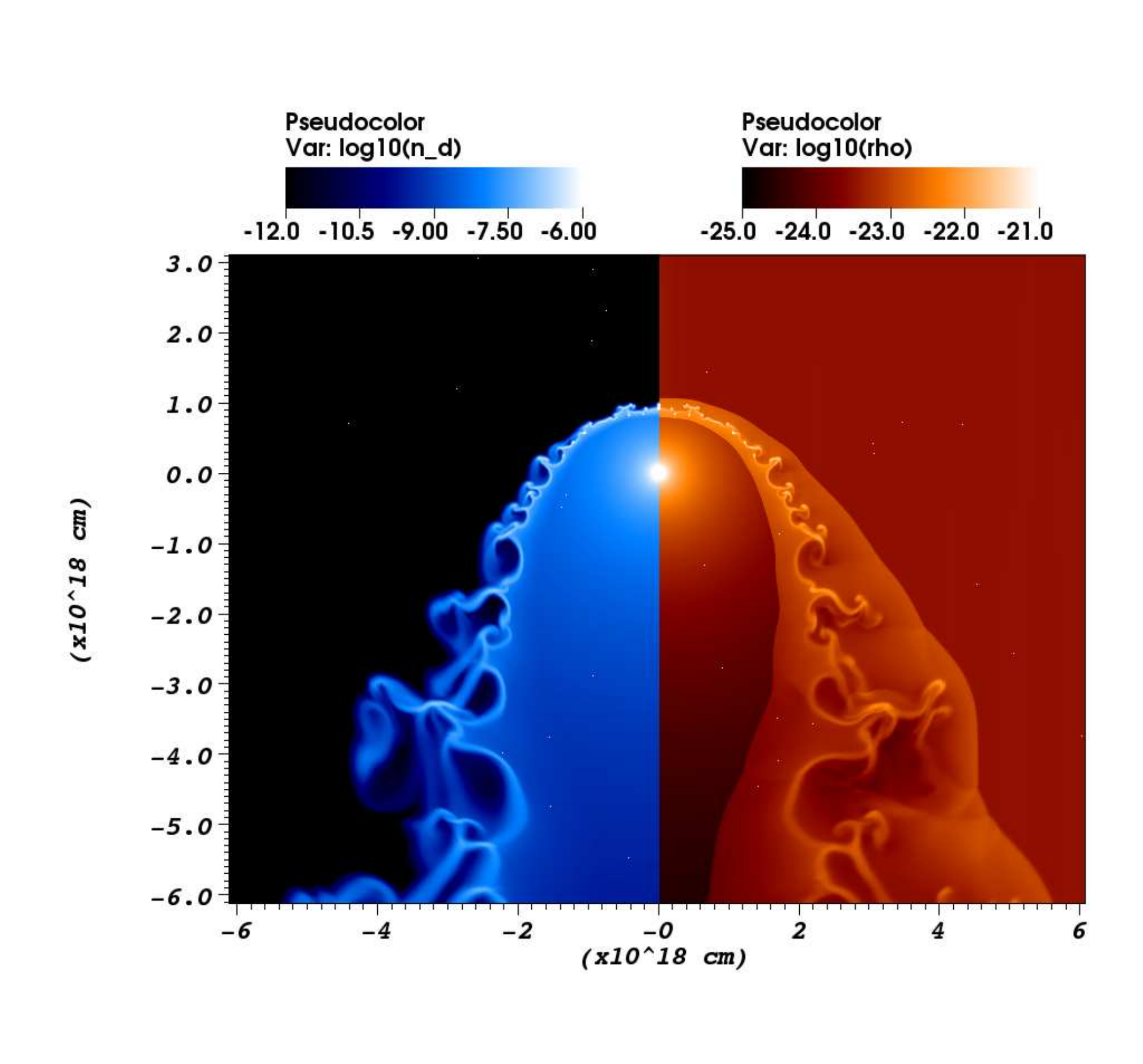}}}
    \end{minipage}
  \hfill
    \begin{minipage}[t]{.495\textwidth}
        \centerline{\resizebox{\textwidth}{!}{\includegraphics{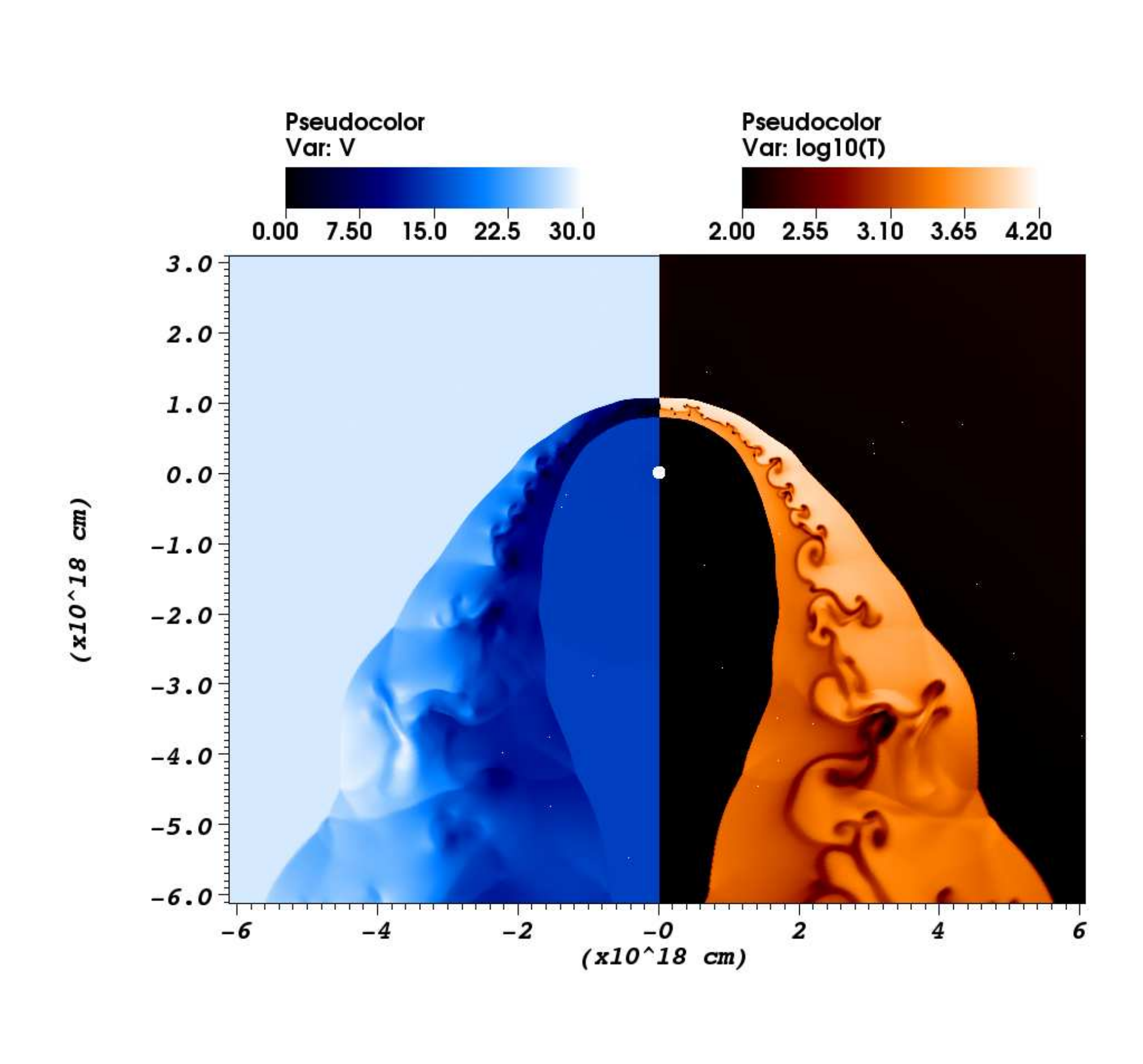}}}
    \end{minipage}
\caption{The dust particle density [cm$^3$] and gas density [g\,cm$^{-3}$] (left); and the gas velocity [km/s] and gas temperature [K] (right) 
for a simulation with an ambient medium temperature of 100\,K (simulation A). 
Although the shocks are smooth, the contact discontinuity shows extensive instabilities, which start small at the front of the shock and 
grow as they move downstream. The dust concentration follows the contact discontinuity. 
}
 \label{fig:ISM100K}
\end{figure*}

The  result for simulation~A with an ambient medium temperature of 100\,K is shown in Fig.~\ref{fig:ISM100K}. After a simulation time of 300\,000\,yr, the place of the bow shock interaction has stabilized:  the termination shock\footnote{location where the wind velocity goes from supersonic to subsonic values} occurs at 0.25\,pc (or 262\arcsec) and the bow\-shock\footnote{location where the ISM velocity goes from supersonic to subsonic values}  at 0.34\,pc (or $\sim$356\arcsec), in good agreement with the observations (see Sect.~\ref{SECT:results_outer}, neglecting the inclination angle). The turbulent  astropause/astrosheath\footnote{location where the ISM pressure equals the CSM pressure} thus has a width of $\sim$90\arcsec. 
 Although both the wind termination shock and the forward shock are smooth, the contact discontinuity, which separates the shocked wind from the shocked ambient medium, shows instabilities. These do not penetrate the shocks on either side, but cause deviations in the shape of the forward shock. 

Figure~\ref{fig:ISM100K} shows three types of instabilities in the gas.
(1)~A thin layer of shocked wind material, right on the contact discontinuity, has a higher density and lower temperature than the surrounding gas. 
This is the result of the radiative cooling instability, which causes dense gas to cool faster. 
As a result, dense gas loses thermal pressure and is compressed, which increases the density and therefore the cooling rate. 
(2)~The high density gas of the shocked wind extends ``fingers'' into the shocked ambient medium, which form in the forward shock and increase in size 
as they travel downstream. 
These are Rayleigh-Taylor instabilities, resulting from low density, but high-pressure, material (the shocked ambient medium) pushing back denser shocked wind. 
(3)~Due to the shear-force between the shocked wind and the shocked ambient medium, the Rayleigh-Taylor instabilities deform 
and show a circular motion. 
These are Kelvin-Helmholtz instabilities and are primarily visible downstream from the star. 
All three features were also found by  \citet{vanMarle2011ApJ...734L..26V}. 

The temperatures in the shocked gas are lower than the kinetic energy of the free-streaming wind and the ambient medium. For a fully adiabatic shock, the temperatures at the front of the bow shock would be about 27\,300\,K for the shocked interstellar medium and 
around 7\,200\,K for the shocked wind. Instead, the maximum temperature for the shocked interstellar medium is  14\,000\,K and the shocked wind does not reach more than 4\,000\,K. 
\citet{vanMarle2011ApJ...734L..26V} find similar temperatures. 
This is mainly due to the effective radiative cooling. Also, as the thermalized gas moves downstream, it decompresses, trading temperature for an increasing velocity, which causes the tail of the shock to be cooler than the front. 
This can be seen by comparing the absolute velocity and temperature of the gas (right panel of Fig.~\ref{fig:ISM100K}).

Comparing the gas and dust density (left panel of Fig.~\ref{fig:ISM100K}) shows that the dust particles penetrate the shocked wind layer, but are brought to a stop at the contact discontinuity, which they tend to follow. In that sense, small dust grains are good tracers of the gas-driven dynamics at the contact discontinuity. 
This confirms the earlier results found by \citet{vanMarle2011ApJ...734L..26V}. As a result, the instabilities at the contact discontinuity should also be visible in the infrared (assuming that the observations have sufficient resolution, which is the case for the Herschel instruments). Since the dust is concentrated at the contact discontinuity, other features that are typical of the shocks would most likely not show up in infrared images. 
For example, in the region behind the star, the wind termination shock curves back toward the axis of motion as its thermal pressure balances against the ram pressure of the wind. 
This effect is completely invisible in the dust. The behaviour of the circumstellar environment over the full 200\,000\, year period, 
starting from the moment when radiative cooling is introduced, is presented online in animated form (Appendix~\ref{app-online}). 
This movie shows that the number of instabilities increases with the introduction of radiative cooling. These instabilities form in the front of the bow shock and then travel downstream.

The total mass of the gas trapped inside the bow shock region, found by integrating the density between the termination shock and bow shock for $0\leq\theta\leq90\deg$, 
is approximately $0.1\, \Msun$ of which $90\%$ is shocked wind material. 
The shell mass is considerably higher than the value obtained from the Herschel data (see Sect.~\ref{SECT:results_outer}).
As already suggested, a possible explanation may be that  the envelope around Betelgeuse is very clumpy or that Betelgeuse is a very young RSG star so that the bow shock is not yet fully formed \citep{Mohamed2012A&A...541A...1M}. This could also explain the smoothness of the bow shock, since instabilities need time to form. Most of the mass is situated at the sides of the star, where the volume of the shock region is much larger; 
e.g., when integrated only over $0\leq\theta\leq45\deg$ the total mass becomes $1.8\times10^{-2}$\,\Msun, of which about one third is shocked ISM.

Assuming radiative-equilibrium balance, the heating of the dust grain due to energy transfer from the hot post-shock gas can be estimated. An approximation is given by Eq.~(5.40) from \citet{Tielens2005pcim.book.....T}, which assumes the absorption efficiency of the dust as a function of wavelength to be a simple power law:
$Q(\lambda)=Q_0(\lambda / \lambda_0)^\beta$, with $Q_0=1$, $\lambda_0=2\pi\,a$ and $\beta=1$. 
As a result, the dust temperature only depends on the mean intensity of the radiation field $J$ and the size of the dust grains $a$, 
\begin{equation}
 T_d~\propto~\biggl(\frac{J}{a}\biggr)^{0.2}\,.
\end{equation}
The total energy flux entering the shock front is
\begin{eqnarray}
 F_{\rm{total}} & = & F_{\rm wind} + F_{\rm ISM} \nonumber \\
                & = & \frac{1}{2} \Mdot \frac{v_w}{4\, \pi\, r^2} + \frac{1}{2} \rho_{\rm ISM}\, v_\star^3  \nonumber\\
                & \approx & 5.5 \times 10^{-5}\,{\rm erg\ cm^{-2}\ s^{-1}}\,.
\end{eqnarray}
Assuming that only 20\% of the energy flux is converted to radiation, the additional heating from the dust grain due to the hot post-shock gas is only $\sim$6\,K, while the heating due to the stellar radiation is one order of magnitude higher (see Fig.~\ref{Fig:Robin} in the Appendix~\ref{App:Tdust}).

\begin{figure*}[htp]
    \begin{minipage}[t]{.495\textwidth}
        \centerline{\resizebox{\textwidth}{!}{\includegraphics{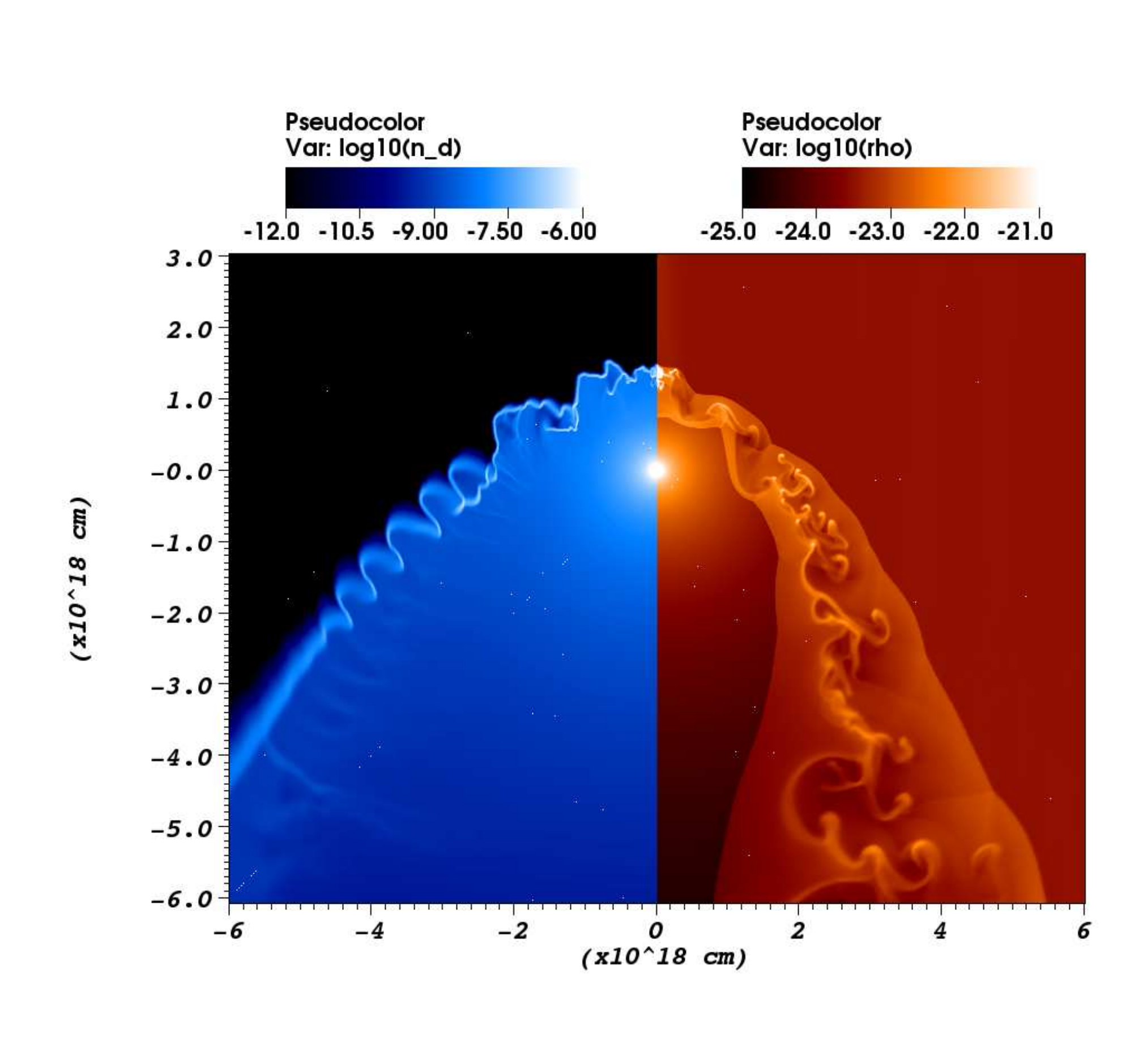}}}
    \end{minipage}
  \hfill
    \begin{minipage}[t]{.495\textwidth}
        \centerline{\resizebox{\textwidth}{!}{\includegraphics{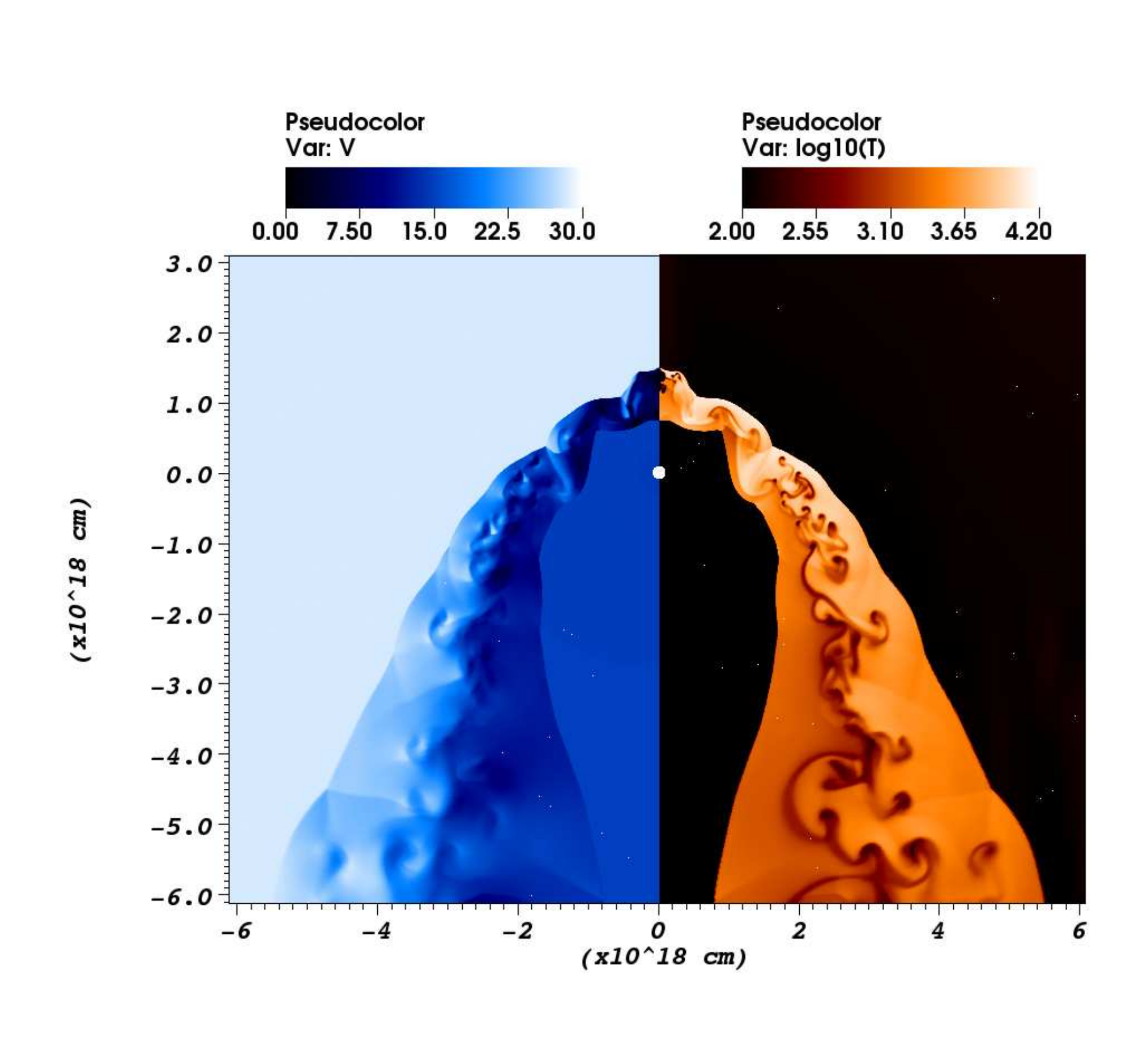}}}
    \end{minipage}
\caption{Similar to Fig~\ref{fig:ISM100K}, but for a simulation with large (100\,nm) dust grains (simulation B). The dust distribution is completely different, tracing the forward shock, rather than 
the contact discontinuity, and even penetrating the unshocked interstellar medium (see text for more details).}
 \label{fig:largegrain}
\end{figure*}

The effect of grain size on the distribution of dust in the bow shock region is demonstrated in Fig.~\ref{fig:largegrain}. This shows the result for simulation~B for a larger grain with size 100\,nm. 
Instead of following the contact discontinuity, the larger dust grains are concentrated at the forward shock and, in the tail of the bow shock, 
even penetrates the undisturbed ambient medium. 
This behaviour, which was also shown in \citet{vanMarle2011ApJ...734L..26V}, is the result of the larger momentum of large grains, compared to the drag force. 
The large grains are less tightly coupled to the gas and can therefore penetrate deeper into the bow shock region and the ambient medium beyond. However, since larger grains are less numerous than smaller ones, the infrared morphology of the bow shock will mainly be determined by the smaller dust species.

\subsubsection{Warm ambient medium}
\begin{figure*}[htp]
    \begin{minipage}[t]{.495\textwidth}
        \centerline{\resizebox{\textwidth}{!}{\includegraphics{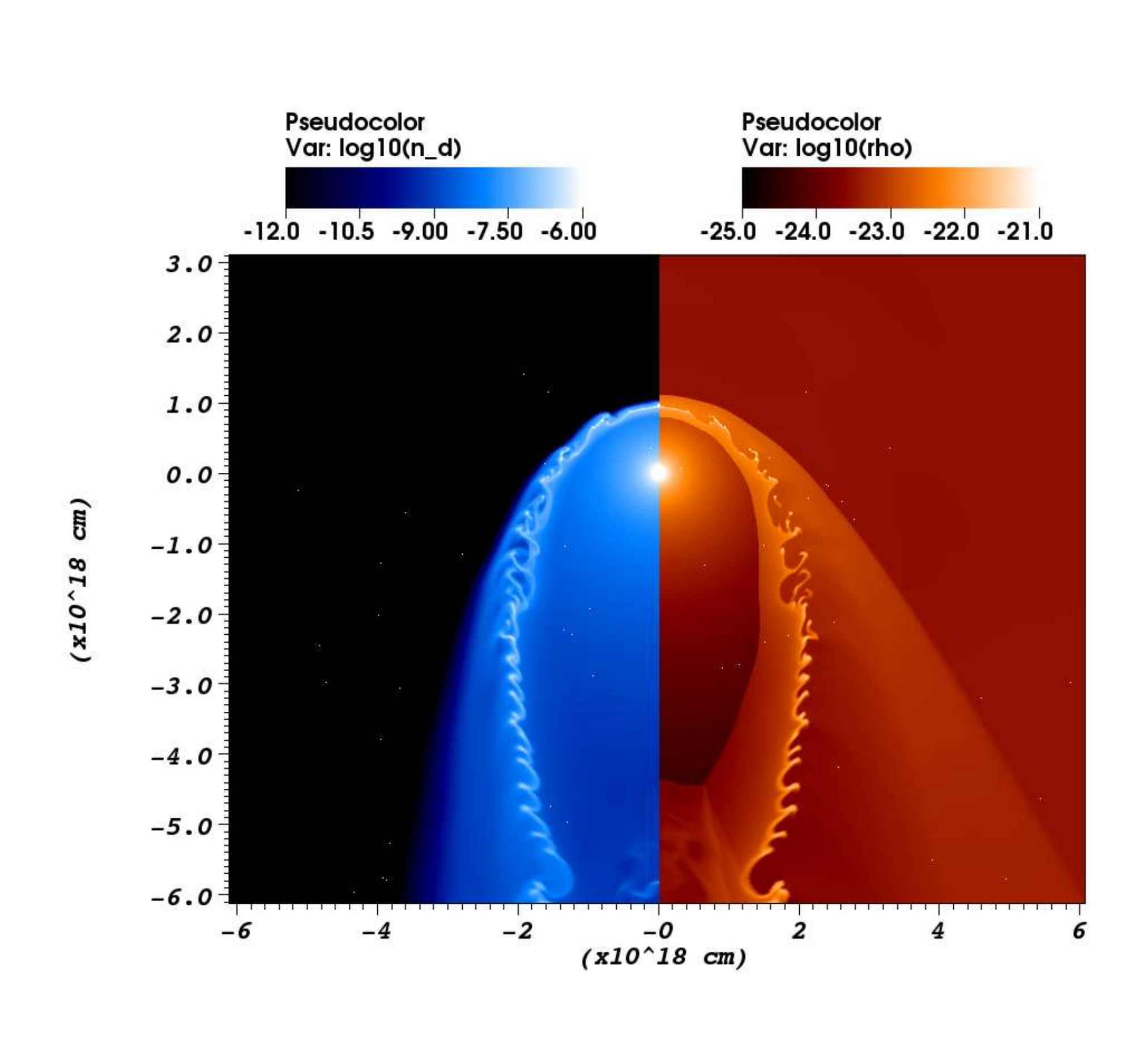}}}
    \end{minipage}
  \hfill
    \begin{minipage}[t]{.495\textwidth}
        \centerline{\resizebox{\textwidth}{!}{\includegraphics{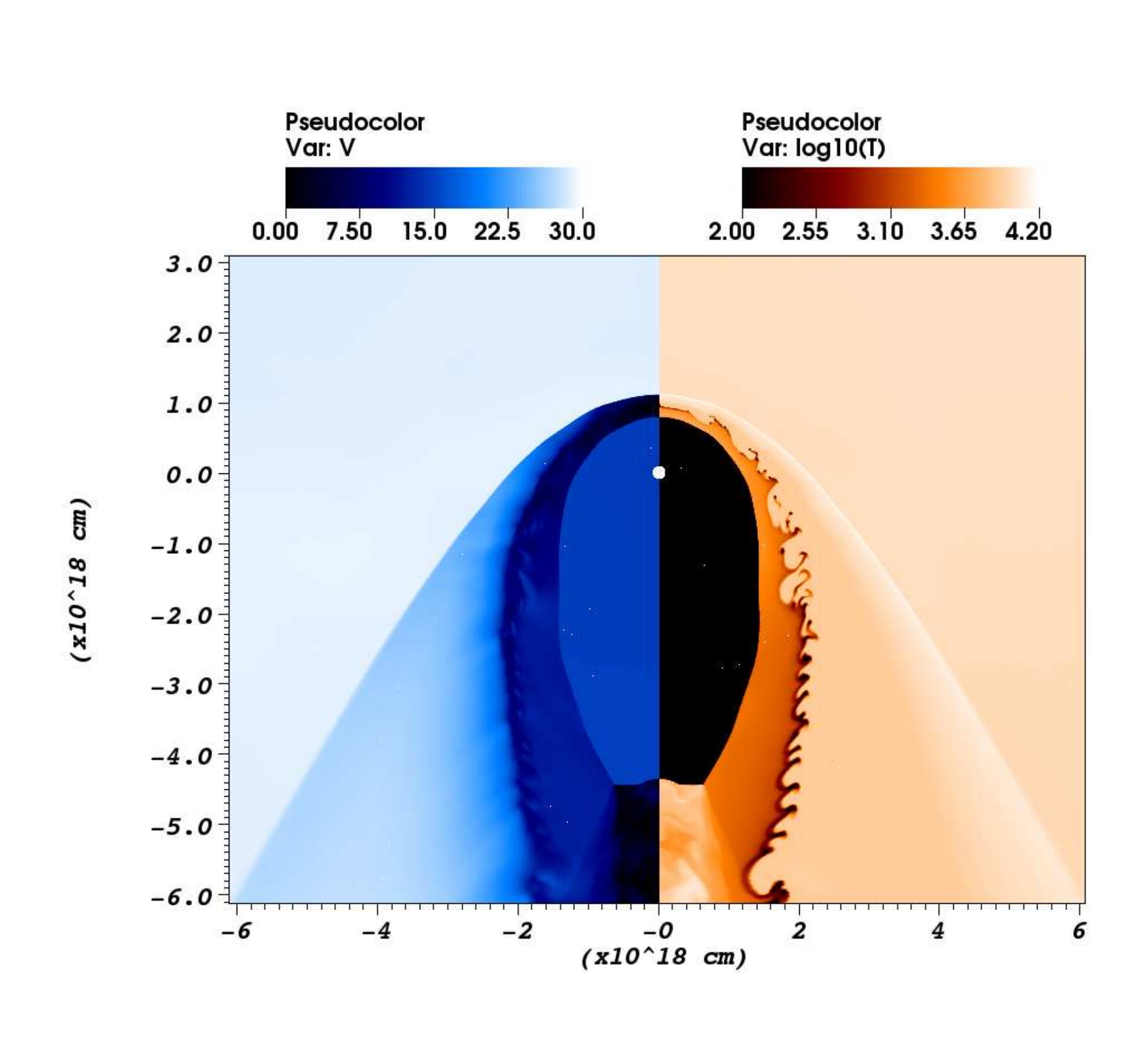}}}
    \end{minipage}
\caption{Similar to Fig~\ref{fig:ISM100K}, but with an ambient medium temperature of 8\,000\,K (simulation C). 
The instabilities are much smaller. Since the shocked interstellar matter does not cool below 8\,000\,K, the shocked ambient medium layer extends further outward than in Fig.~\ref{fig:ISM100K}. On the other hand, the shocked wind region is more restricted.}
 \label{fig:ISM8000K}
\end{figure*}

When the ambient medium is kept at a minimum temperature of 8\,000\,K (simulation~C), the results look different, as shown in Fig.~\ref{fig:ISM8000K}. The Rayleigh-Taylor instabilities are much smaller. This is the result of radiative cooling. The instabilities, which are made up of stellar wind material, are allowed to cool to 100\,K, whereas the surrounding shocked ambient medium has a minimum temperature  of 8\,000\,K. Therefore, the gas in the Rayleigh-Taylor fingers has a lower thermal pressure than the surrounding gas and is compressed, inhibiting their growth. This is different from the results shown in \citet{Cox2012A&A...537A..35C}, which also showed large instabilities for the warm ambient medium. However, these may have been partially due to numerical effects of modelling a spherical wind expansion on a cylindrical grid, which lead to a large instability close to the symmetry axis. 

The layer of shocked interstellar matter extends further from the axis of motion and has an almost triangular shape. 
The combination of the smaller instabilities and the larger distance between contact discontinuity and forward shock also reduces the effect 
of the instabilities on the shape of the shock, which is completely unaffected. 

Because of the high thermal pressure of the ambient medium, the wind region is more constrained; in fact, the free-streaming wind has not yet reached the lower boundary, but is still constrained by the shocked wind. This was also seen in the results of \citet{Cox2012A&A...537A..35C}. 
As in the first simulation, the small dust grains follow the contact discontinuity. 

\begin{figure*}[htp]
    \begin{minipage}[t]{.495\textwidth}
        \centerline{\resizebox{\textwidth}{!}{\includegraphics{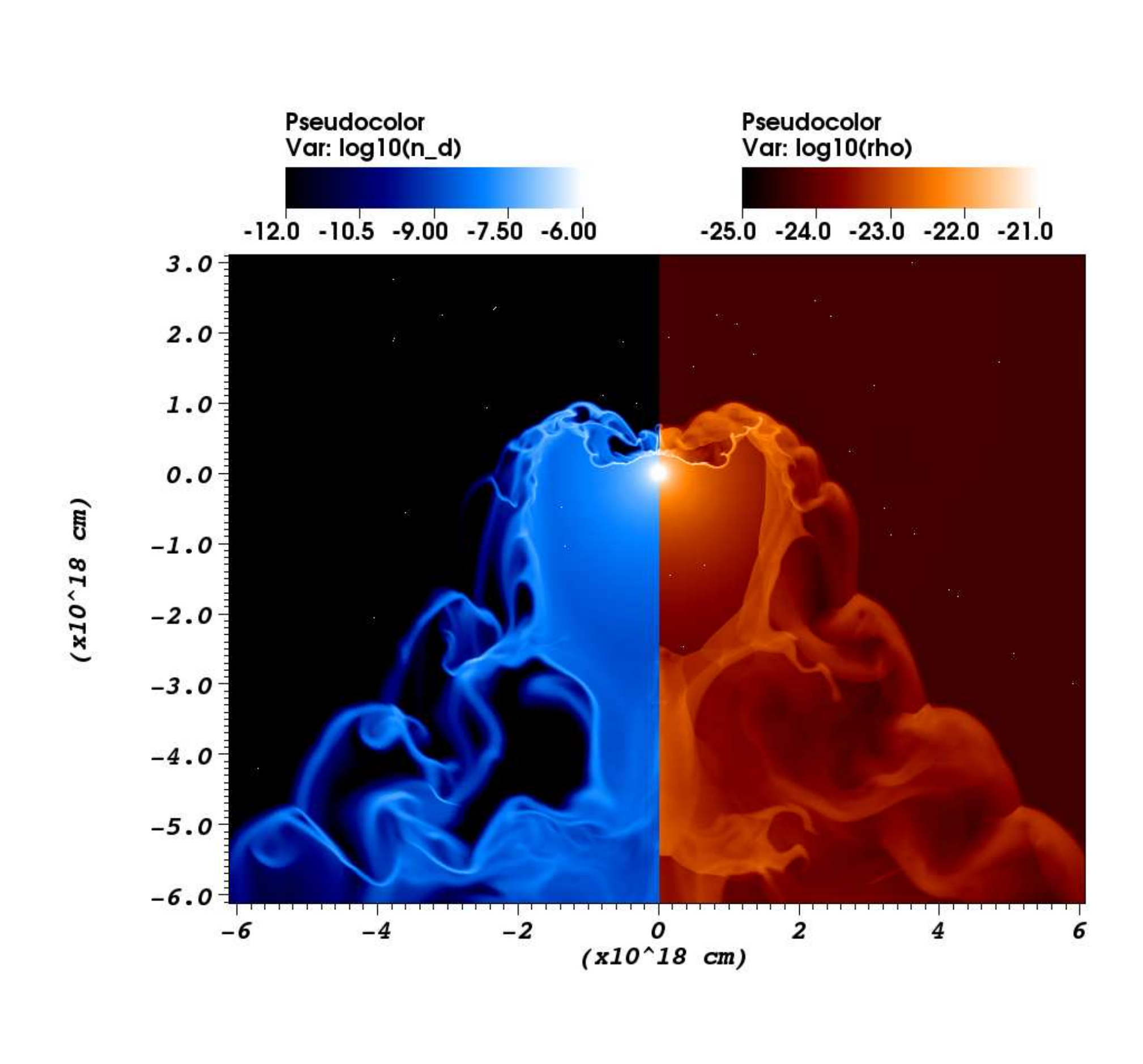}}}
    \end{minipage}
  \hfill
    \begin{minipage}[t]{.495\textwidth}
        \centerline{\resizebox{\textwidth}{!}{\includegraphics{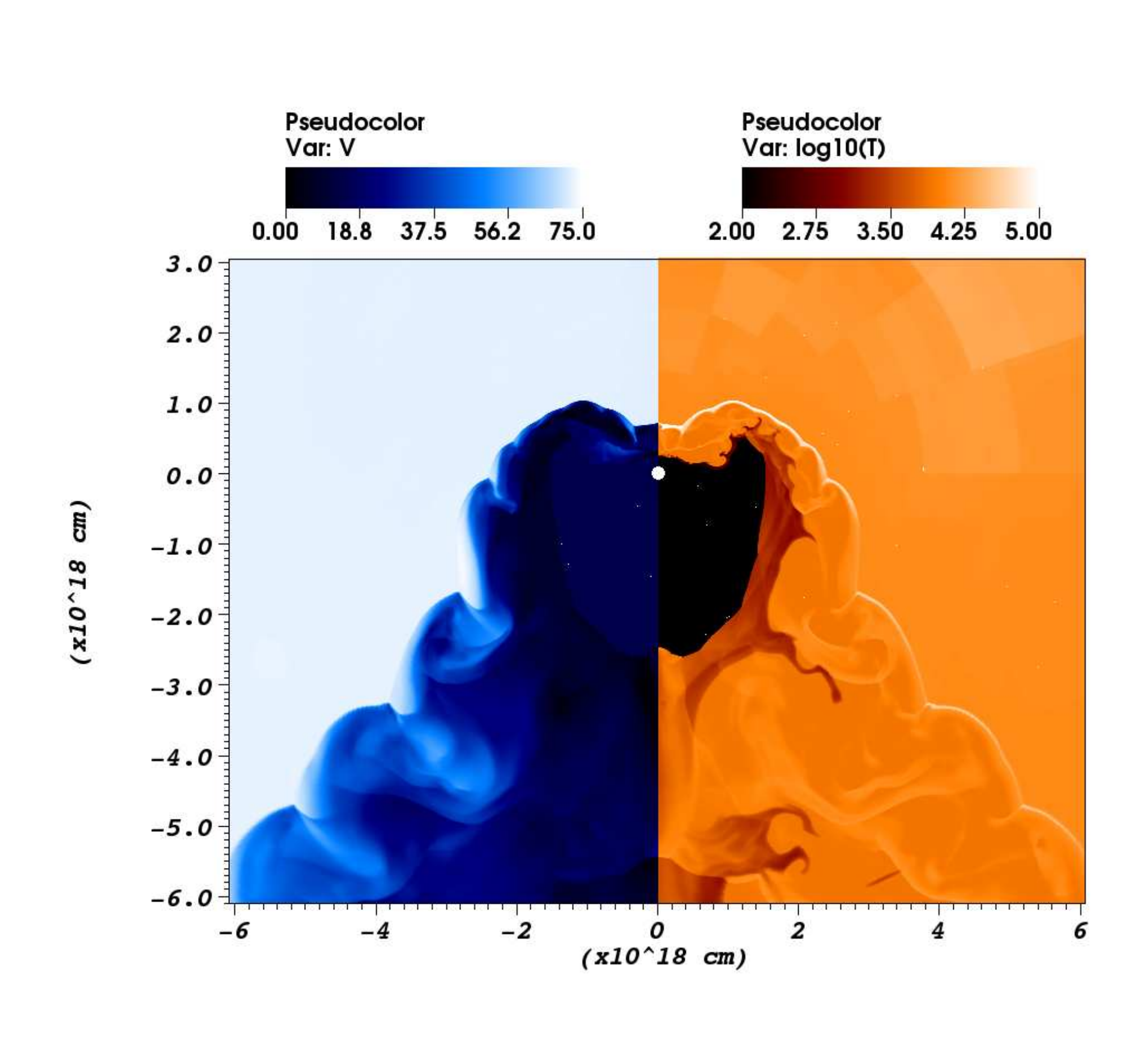}}}
    \end{minipage}
\caption{Similar to Figs.~\ref{fig:ISM100K} and \ref{fig:largegrain}, but with an ambient medium temperature of 8\,000\,K, 
an ambient medium density of $10^{-25}$\,g\,cm$^{-3}$, and a stellar velocity of 72.5\,km/s  (simulation~D). 
The bow shock is both locally and globally unstable. the shock temperature of the bow shock is high, because of the higher collision speed.}
 \label{fig:fast}
\end{figure*}

The result of simulation~D with the same high ambient medium temperature, but lower ambient medium density and higher space velocity, is shown in Fig.~\ref{fig:fast}. 
The high stellar velocity creates a stronger shear force between the shocked wind and the shocked ambient medium. 
As a result, the Kelvin-Helmholtz instabilities are much greater, leading to a large-scale instability that changes the shape of the entire bow shock. 
This is contrary to the results of \citet{Mohamed2012A&A...541A...1M}, who obtained a smooth bow shock for similar input parameters. 
This difference may be the result of a different numerical treatment (grid versus SPH), but may also result from the total timescale of the simulation, 
because Kelvin-Helmholtz instabilities need time to grow (see accompanying movie in Appendix~\ref{app-online}).
Owing to the larger collision speed at the bow shock, the shock temperatures are much higher ($\sim\,10^5$\, K) than for the other simulations. 
This largely negates the effect of the high ambient medium temperature, since its thermal energy is much lower than the kinetic energy. While the general structure of the bow shock is unstable, the local dust shell directly in front of the star is quite smooth. 
Also, the instability of the bow shock leads to a structure that curves in the opposite direction from the wind termination shock. 
Similar features have been found by \citet{Wareing2007MNRAS.382.1233W} and \citet{Cox2012A&A...537A..35C}.

\subsubsection{Clumpy mass-loss episodes} \label{Sect_hydro_clumps}

\begin{figure*}[htp]
    \begin{minipage}[t]{.495\textwidth}
        \centerline{\resizebox{\textwidth}{!}{\includegraphics{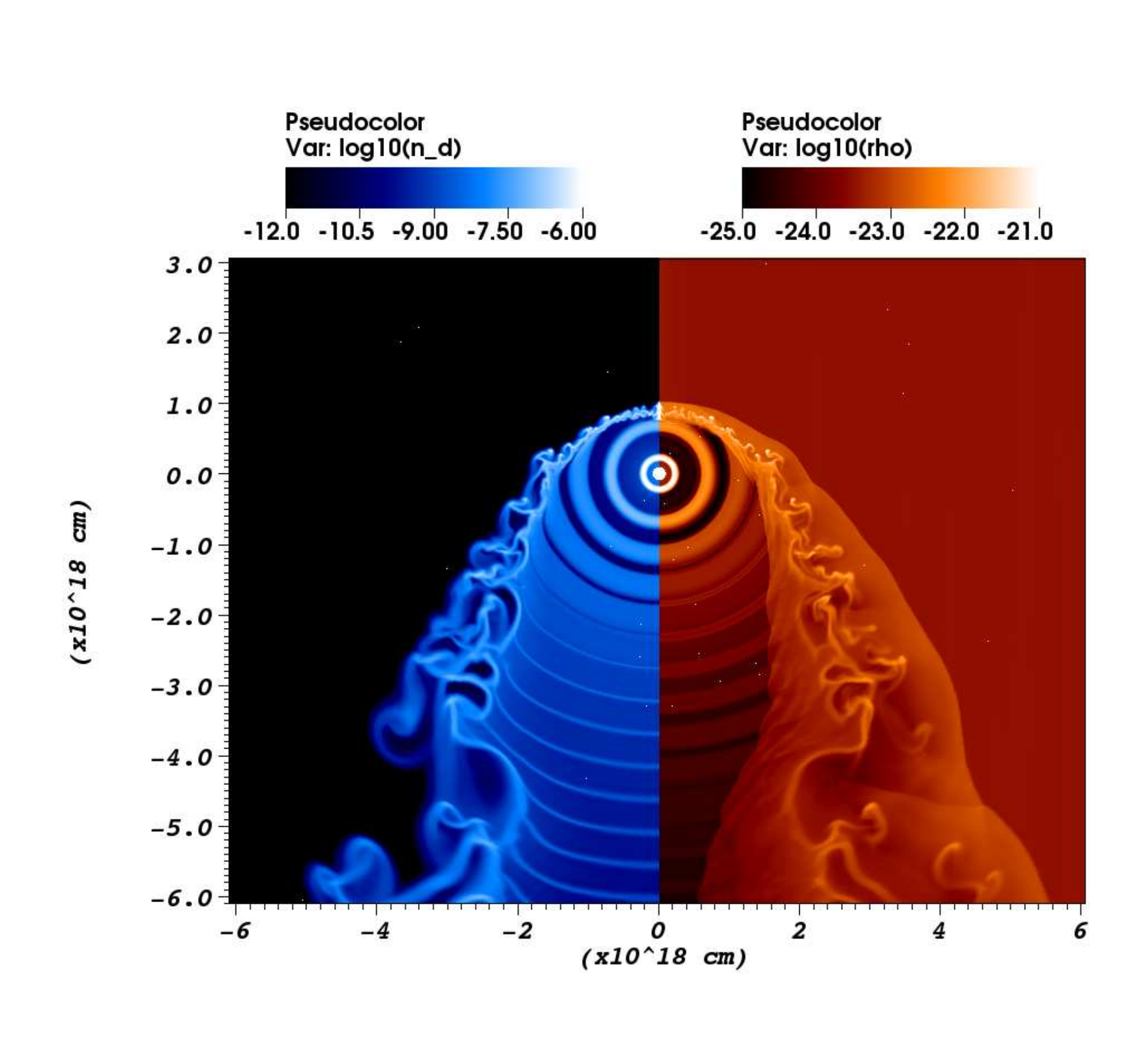}}}
    \end{minipage}
  \hfill
    \begin{minipage}[t]{.495\textwidth}
        \centerline{\resizebox{\textwidth}{!}{\includegraphics{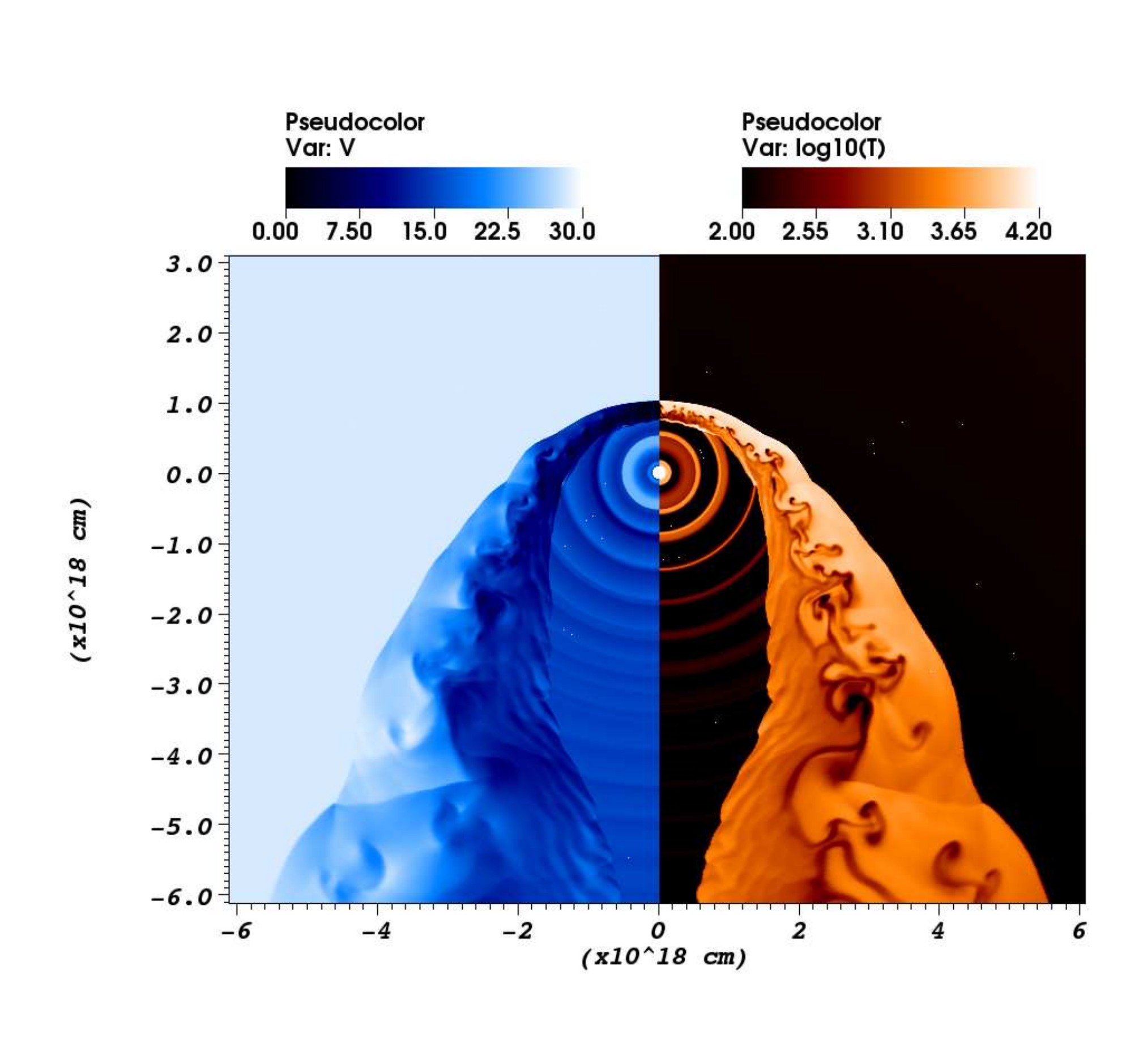}}}
    \end{minipage}
\caption{Similar to Figs.~\ref{fig:ISM100K}, \ref{fig:largegrain}, and \ref{fig:ISM8000K}, but with a variable mass-loss rate (simulation~E) . 
The instabilities are more numerous and have more small-scale structure than for the same simulation with a smooth wind. }
 \label{fig:vardM1}
\end{figure*}

\begin{figure*}[htp]
    \begin{minipage}[t]{.495\textwidth}
        \centerline{\resizebox{\textwidth}{!}{\includegraphics{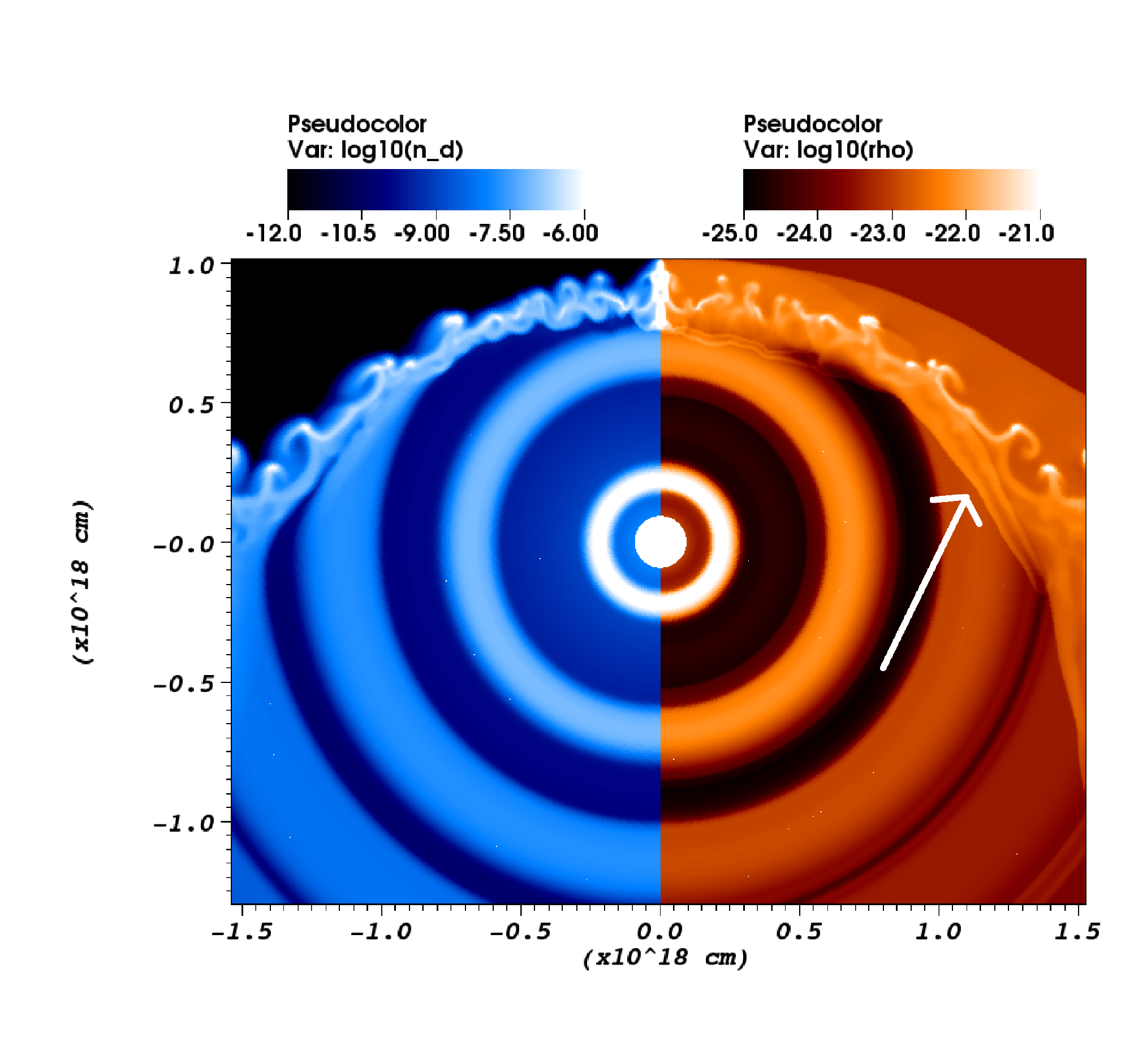}}}
    \end{minipage}
  \hfill
    \begin{minipage}[t]{.495\textwidth}
        \centerline{\resizebox{\textwidth}{!}{\includegraphics{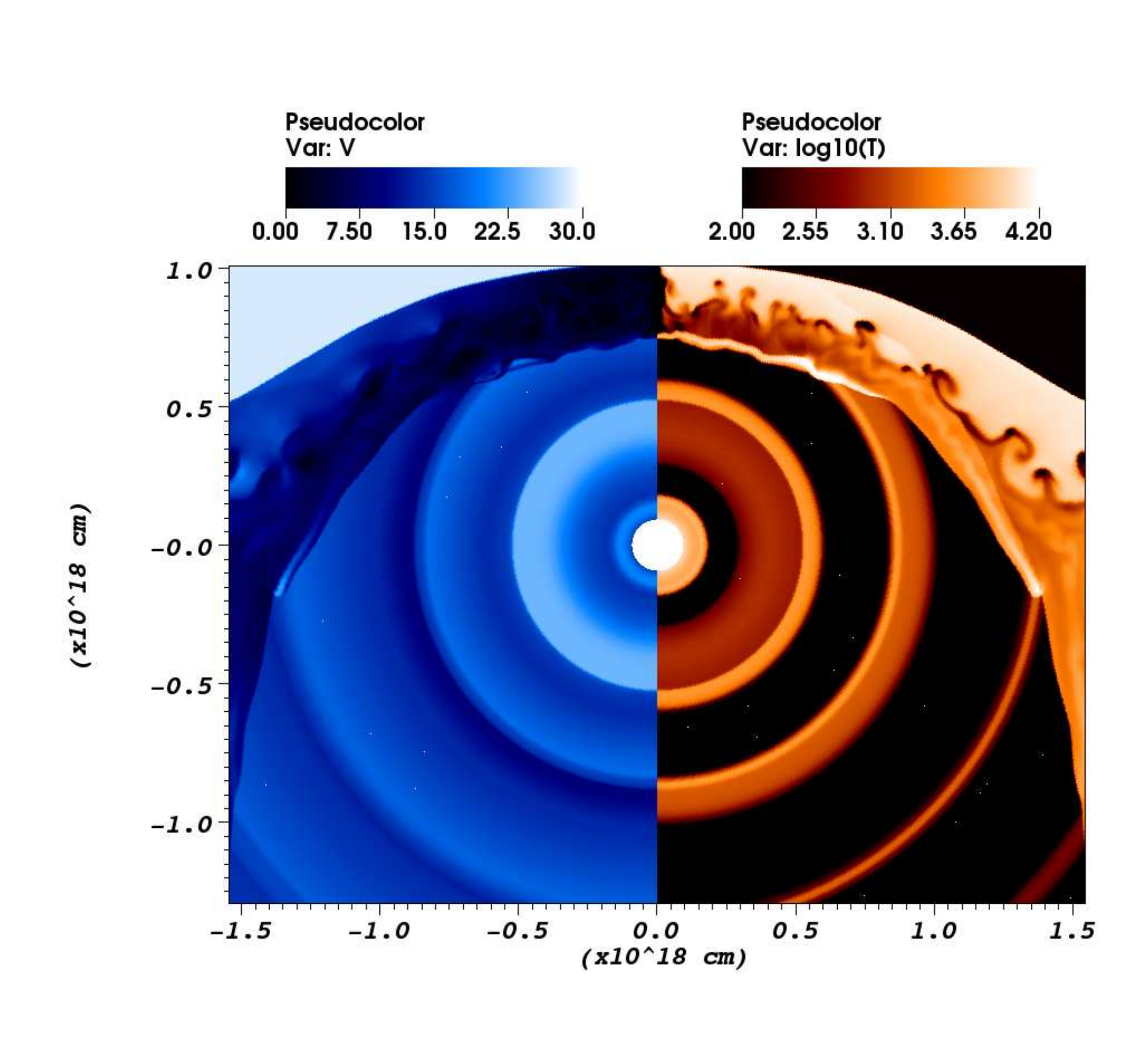}}}
    \end{minipage}
\caption{Zoom of Fig.~\ref{fig:vardM1}. The white arrow indicates the possibility of multiple arcs in the front of the bow shock region. }
 \label{fig:vardM2}
\end{figure*}

The hypothesis that the different arcs might be the result of the collision of different  clumpy mass-loss episodes having filling factors $<$1 with the surrounding ISM (see Sect.~\ref{SECT:constraints_arcs_obs}) can be checked using the hydrodynamical simulations. 
In the first instance, this can be done in a simplified 2D scheme, where the collision between the ISM and a CSE with a variable mass-loss rate has been modelled. A full 3D model is beyond the scope of this paper. As outlined above, we simulated this effect by varying the mass-loss rate by a factor 1000 over a period of 10\,000\,yr. Since the change in dust density at the edge of a clump is very steep, the mass-loss rate variations were implemented in a `picket-fence' way. 

Figure~\ref{fig:vardM1} shows the result of simulation~E with variable mass loss. 
The mass leaves the star concentrated in shells that are separated by voids with a density that is three orders of magnitude lower. 
As they travel outward, the shells expand owing to their own internal pressure. 
In the wake of the star the shells eventually merge. 
To the front of the star this does not occur because the shells hit the wind termination shock before they can expand far enough. 

The variation in mass-loss rate (and therefore in ram pressure) causes a widespread disturbance in the morphology of the bow shock region. The wind termination shock, where the isotropic thermal pressure of the shocked gas is equal to the ram pressure of the wind, 
shows a wave-like pattern in the downwind region as it conforms to the variable ram pressure of the wind. At the front of the bow shock the variation in the free-streaming wind increases the instability at the contact discontinuity. The resulting Rayleigh-Taylor and Kelvin-Helmholtz instabilities are more numerous than in the case of a smooth wind (Fig.~\ref{fig:ISM100K}) and show more small-scale structure. 

Finally, the impact of subsequent shell collisions means that multiple arc-like structures appear between the termination shock and the contact discontinuity (for a detailed view see Fig.~\ref{fig:vardM2}). 
This may be an indication of the origin of the multiple arcs observed in the bow shock of Betelgeuse. 

The animation of the simulation (Appendix~\ref{app-online}) shows the influence of the variable mass-loss rate, in terms of both the formation and growth of local features, as well 
as of the large-scale morphology of the bow shock. 
Despite the variation in ram pressure, the location of the bow shock mostly remains unchanged. 
This is the result of the period of the variation relative to the dynamical timescale of the bow shock region. The bow shock location can only change if the dynamical timescale is significantly shorter than the period of the variation. This is not the case in this simulation. 
The bow shock region has a cross-section of the bow shock of approximately 0.1\,pc at the front and up to a parsec to the sides of the star. 
With a sound speed of about 10\,km/s (for a temperature of 10\,000\,K), the timescale varies between 10\,000 and 100\,000 years. 
Since the period of the variations is only about 10\,000 years, the bow shock does not have enough time to adjust to the variations.

\subsection{Discussion} \label{Sect:Origin_arcs_discussion}

\paragraph{A. Multiple arcs:}
The hydrodynamical simulations show some  good agreement between the observations and the theoretical predictions, which might point us toward the origin of the arcs; e.g., the observed ratio between the projected --- minimal --- distance between the star and the bow shock outline in the direction of relative motion and the distance perpendicular to that, i.e.\ at $\theta = 90\deg$, is $\sim$0.66 for the different arcs. The same ratio is obtained in simulations~A, C, and E, while the ratio is significantly lower for simulations~B and D.

The hydrodynamical simulations show that the formation of dust clumps in the inner wind regions, simulated in all its generality by implying  spherically symmetric mass-loss variations, can have an impact on the observed morphology of the bow shock region, and potentially might lead to the formation of multiple arcs in the bow shock. Dust clumps with higher grain size have a larger drift velocity, large Larmor radius, and longer stopping length when  passing through the bow shock, and eventually might create a bow shock arc further away from the central target. However, as said above, we would then expect that other targets in the MESS sample would show multiple arcs in the bow shock region as well. 

An effect not included in the {\tt MPI-AMRVAC}-code is photo-ionization. In the case of the CO waves (or ripples) detected on the surface of the Orion molecular cloud, \citet{Berne2010Natur.466..947B} argues that ultraviolet radiation has created a (small) insulating photo-ablative layer, allowing the development of a KH instability with a wavelength at least one order of magnitude longer than the insulating layer. A speculative idea is that the multiple arcs found around Betelgeuse are undulations created by photo-chemical effects stimulating the growth of some instability with wavelength around the observed width of the arcs, which are then traced by the smaller dust grains that tend to follow the gas instabilities. However, also for this effect, it remains surprising that Betelgeuse is the only target in the MESS sample showing multiple arcs.

In our opinion, the most probably reason for the origin of the multiple arcs is the combined effect of (1)~a turbulent non-homogeneous mass loss, which might enable the ISM to flow into the envelope and create multiple bow shocks, together with (2)~the effect of the interstellar magnetic field, which can make some instabilities very efficient while suppressing other modes (see Sect.~\ref{SECT:constraints_arcs_obs}). The angular separation between the arcs points toward a magnetic field of $\sim$3\,$\mu$G.

\paragraph{B. Suppression of large-scale instabilities:}
Remarkably, the size of (local) density clumps tracing potential instabilities inside the arcs as seen in the Herschel images (with an upper limit to the maximum length of $\sim$30\arcsec, see Sect.~\ref{SECT:results_outer}) is often smaller than predicted by the hydrodynamical simulations.  Most of the instabilities seen in Figs.~\ref{fig:ISM100K} and~\ref{fig:vardM2} are Rayleigh-Taylor (RT) instabilities, which are slightly deformed by the action of Kelvin-Helmholtz (KH) instabilities. Thin-shell instabilities such as the NTSI \citep{Vishniac1994ApJ...428..186V} or TAI \citep{Dgani1996ApJ...461..927D} do not occur, since the shell between both shocks is too thick: with a typical thickness of the shell in the order of 0.1\,pc, the crossing time is in the order of $10^4$\,yr (for a velocity of $\sim$10\,km/s) by which time an individual mass element is already displaced by a few times $10^{17}$\,cm, preventing the growth of an instability. Thermal instabilities are also ruled out, since radiative shocks 
with a power-law cooling function ($\Lambda \propto T^{\gamma}$) are stable against thermal instabilities if $\gamma \ga 1$ \citep{Gaetz1988ApJ...329..927G}.

The only example in our simulations showing only small-scale instabilities is the case of a warm ambient medium around Betelgeuse  (with a temperature of $\sim$8000\,K) for which a strong growth of the instabilities is prevented by compressing the Rayleigh-Taylor fingers. As shown in Fig.~\ref{fig:ISM8000K}, instabilities in the direction of the space motion of the target then have typical lengths $\la 1 \times 10^{17}$\,cm (or 34\arcsec).

 {We should also realize that the size and growth rate of the instabilities are directly related to the density of the gas \citep{Chandrasekhar1961hhs..book.....C}. In our simulations, compression due to radiative cooling increases the density of the shocked wind. However, it is possible that we over-estimate the cooling. The energy loss due to radiative cooling is proportional to the ion density times the electron density.  If the gas is only partially ionized (as in the 3\,000-8\,000\,K regime), the cooling may be reduced, which in turn would reduce the compression and therefore the density contrast at the contact discontinuity, damping the  instabilities.} 

Not considered in our simulations is the effect of a magnetic field. Depending on the direction of the magnetic field with respect to the shear flow, a magnetic field might (de)stabilize KH instabilities \citep{Miura1982JGR....87.7431M, Keppens1999JPlPh..61....1K}. \citet{Miura1982JGR....87.7431M} found that in a sheared magnetohydrodynamical flow in a compressible plasma, modes with $k \Delta < 2$ are unstable, with $\Delta$ the scale length of the shear layer and $k$ the wavenumber. Compressibility and a magnetic field component parallel to the flow ($\mathbf{B}_0 \parallel \mathbf{v}_0$) are found to be stabilizing effects. For the transverse case ($\mathbf{B}_0 \perp \mathbf{v}_0$), only the fast magnetosonic mode is destabilized, but if $\mathbf{k \cdot B}_0 \ne 0$, the instability contains Alfv\'en-mode and slow-mode components as well. \citet{Keppens1999JPlPh..61....1K} studied the case of an initial magnetic field aligned with the shear flow. In the case the initial magnetic field is unidirectional 
everywhere (uniform case), the growth of KH instabilities is compressed, while a reversed field (i.e.\ a field that changes sign at the interphase) acts to destabilize the linear phase of the KH instability. As a result, the non-occurrence of large-scale instabilities might 
potentially be explained by the presence of a magnetic field. As suggested in Sect.~\ref{SECT:constraints_arcs_obs}, the detection of multiple arcs in the bow shock might also point towards the presence of an ISM magnetic field. With a strength of $\sim$3\,$\mu$G, the magnetic pressure is $\sim 3.6 \times 10^{-13}$\,dyn cm$^{-2}$, being two orders of magnitude lower than the ram pressure at the contact discontinuity. Without dedicated simulations, it is not possible to predict how the presence of a weak ISM magnetic field might influence the growth rate of the instabilities.

One might also argue that the bow shock shell is still very young so that instabilities might not have had enough time to grow. Using the prescriptions of \citet{Brighenti1995MNRAS.277...53B} the typical growth time for RT instabilities to reach a size of 30\arcsec\ at a distance of $\sim$200\,pc is only $\sim$7000\,yr, while the KH growth timescale is at least an order of magnitude smaller\footnote{However, one should realize that for KH instabilities the difference in velocity $\Delta v > 2 v_{\rm sound}$, with $\Delta v$ being at maximum $\sim$43\,km/s for these simulations.}. However, according to our simulations, the shell would not have had time yet to reach the observed stand-off distance. Changing the mass-loss rate by a factor of $\sim$10 only reduces the current stabilization time from $\sim$100\,000\,yr to $\sim$30\,000\,yr.

The clumpy structures found in the inner envelope region might also have an imprint on the appearance of the bow shock. With shock velocities below 50\,km/s, the percentage of dust mass destroyed (i.e.\ returned to the gas phase) is negligible \citep{Tielens2005pcim.book.....T}.
But, one should also take into account that dust clumps tend to dissipate. One possible depletion mechanism is heat conduction, which is thought to be unimportant in the case of cool supergiant (and AGB) winds owing to the small difference in temperature between the clump and its environment. According to \citet{Hartquist1986MNRAS.221..715H}, clouds are, however, ablated when they are in relative motion with respect to the surrounding medium. Following \citet{Smith1988MNRAS.234..625S}, we estimate 
that all dust clumps with a size smaller than $\sim 10^{15}$\,cm (or 0.3\arcsec\ at a distance of 200\,pc) will be ablated by the time the clump reaches the bow shock region; larger dust clumps might still be visible in the bow shock region.
In the case where most of the dust grains formed in the inner wind had already been ablated by the time they arrived at the bow shock region, the clumps seen in the outer arcs might also reflect ISM inhomogeneities progressively engulfed by the expanding shell.

\section{Discussion: Origin of the linear bar} \label{SECT:ORIGIN_BAR}

It is striking that the arcs and the linear bar show the same far-infrared colour temperature and density (as deduced from the PACS data). This might suggest that they are illuminated by the same source, probably Betelgeuse itself, and belong to the same physical phenomenon (i.e.\ the bow shock). 
As already argued in previous section, larger grains  are capable of leaving the shocked wind layer, and can penetrate the shocked or even unshocked ISM.  The large grains almost
completely ignore the morphology of the shocked gas shell and penetrate the unshocked ISM a considerable
distance \citep[$\sim$0.1--0.2\,pc,][]{vanMarle2011ApJ...734L..26V}. Since they are very low in number, they
will not influence the gas morphology. However, even for very large grains, the shape of the bow shock still shows some curvature \citep[see Fig.~4 in][]{vanMarle2011ApJ...734L..26V}, which is not seen in the Herschel images.
{Recently, \citet{Mackey2012arXiv1204.3925M} have suggested that the
linear bar is the relic of a blue supergiant (BSG) wind interaction with the ISM, where Betelgeuse has recently entered the red supergiant (RSG) phase.  The BSG reverse shock collapses in on the RSG wind producing an inner shocked shell (the arcs in the Herschel image). The $r^{-2}$ decrease in dust emission makes the contact discontinuity look a little more like a bar (compared to the projected density in a 2D simulation) because the strongly curved parts at larger radii are fainter. If so, \citet{Mackey2012arXiv1204.3925M} predict that the bar should eventually begin to curve back into a bow shock shape, with a radius of curvature that can be larger than the current stand-off distance of the bar suggests. The large FOV Herschel images cannot give extra support to this proposed scenario since no curvature is seen.}

We thus should consider the possibility that the linear bar is not related to Betelgeuse, but is an interstellar structure, which is by accident co-spaced with Betelgeuse. If so, the motion of the star and angular separation between the star and bar imply a collision between the two in about 17500\,yr, while the arcs and linear bar would collide in about 5000\,yr. 

The linear bar might be the edge of an interstellar cloud. The results obtained by \citet{Dickey1989ApJ...341..201D} prove that it is not unreasonable that Betelgeuse is near an interstellar cloud.
\citet{Dickey1989ApJ...341..201D} showed that a few hundred clouds are present within 1\,kpc having a column density around $10^{19}$\,cm$^{-2}$ (cfr.\ Fig.~\ref{FIG:info_Nick}).  With an estimated mass for the linear bar around $\sim$0.01\,\Msun, the molecular cloud mass spectrum  has a density of $\sim$10\,pc$^{-3}$ \citep[see Fig.~2 in][]{Dickey1989ApJ...341..201D}. Using \ion{H}{i} 21\,cm observations, \citet{Matthews2011AJ....141...60M} has recently detected an extended circumstellar wake associated with the CSE around the AGB star X~Her. Mapping shows that the star lies (in projection) near the periphery of a much larger \ion{H}{i} cloud that also exhibits signatures of interaction with the ISM. While the radial velocity of the cloud near X~Her overlaps that of the CSE, this is not the case for $\alpha$~Ori (see Figs.~\ref{Fig:GALFA}). However, the probability of a mere chance superposition in position, velocity, and direction of space motion between a cloud and target is extremely low. Betelgeuse 
might thus be close to an interstellar cloud, but have a different radial velocity.

The bar might also be a linear filament in the interstellar cirrus, as also seen in the Galactic centre, whose possible origin is linked to the Galactic magnetic field \citep{Ysef1984Natur.310..557Y, Yusef1987ApJ...322..721Y}. Linear filaments might  be caused by particle acceleration by shocks related to preexisting filamentary magnetic structure in the ISM \citep{Heyvaerts1988ApJ...330..718H} or by particle energization at reconnection sites at the ionized surfaces of molecular clouds and subsequent particle loading of ISM magnetic field lines \citep{Serabyn1994ApJ...424L..91S}. Filamentation can also be the result of synchrotron thermal instabilites \citep{Bodo1990MNRAS.244..530B}: both the thermal condensation mode and slow magnetosonic wave can create structures that are aligned with the background magnetic field.
 \citet{Rosner1996ApJ...470L..49R} show that a particle acceleration process, akin to the acceleration of the anomalous cosmic-ray component associated with the solar wind termination shock, assisted by radiative instabilities driven by synchrotron emission can provide an explanation for the filamentation process. Some fraction of accelerated particles will be line-tied to the external (ISM) magnetic field.
The local field magnetic field is parallel to the Galactic plane and follows the local spiral
arms \citep{Andreasyan1989Ap.....31..560A}, having a pitch angle of  about $-8\deg$ \citep{Han2008NuPhS.175...62H}. Optical polarization measurements in the direction of Betelgeuse on a scale of several degrees show that the position angle of the local ISM galactic field  \citep{Bingham1967MNRAS.136..347B, Heiles2000APS..APR.B7003H, Heiles2005LNP...664..137H} is  consistent with the position of the linear bar.  However, typical temperatures of interstellar cirrus clouds are around 15\,K \citep[e.g.][]{Miville2010A&A...518L.104M}, much lower than the temperatures we deduced from the Herschel maps. Only when the filamentary cirrus is by chance close to Betelgeuse can the radiation of the target illuminate and heat the filamentary structure.

\section{Conclusions}
\label{SECT:conclusions}

This paper presented new Herschel PACS and SPIRE images of the red supergiant Betelgeuse. The Herschel data show that the collision between the wind of the red supergiant and the interstellar medium has created a very complex structure. Multiple arcs at a distance of 6--7\arcmin\ are detected with typical dust temperatures in the range of 40--140\,K. At a distance of $\sim$9\arcmin\ from the central target, an intriguing linear bar is detected, with the same colour temperature as the arcs. Betelgeuse is the only target in the MESS sample showing these two phenomena.


The inner envelope structure is very inhomogeneous and testifies to a turbulent mass-loss history. The Herschel images show the first evidence of a high degree of non-homogeneous distribution for the material beyond 15\arcsec, which even persists until the material collides with the ISM. This inhomogeneity is probably linked to the giant convection cells recently detected in the outer atmosphere of Betelgeuse, which might enable localized dust creation and ejection.

The Herschel data have been complemented with data from the UV (from GALEX), near-IR (from WISE), and radio (from the GALFA-HI survey) to disentangle the different components that might result in this complex bow-shock structure and in the linear bar. The average temperature in the arcs is $\sim$85\,K and in the bar is $\sim$63\,K. A low dust emissivity index, $\beta$, around 1 is deduced from fitting the spectral energy distribution. This might point toward (1)~the occurrence of large grains, (2)~a grain composition resembling layered amorphous silicates grains, porous graphite, or carbonaceous grains, or (3)~high temperature processes influencing the quantum effects. The dust temperature in the arcs and bar is regulated by the  stellar radiation field of Betelgeuse, while heating due to the hot post-shock gas is negligible.

{The Herschel PACS images of the inner envelope and the \ion{H}{i} results by \citet{LeBertre2012arXiv1203.0255L} suggest that the mean dust and gas density might have changed considerably {$\sim$32\,000\,yr} ago (radius of $\sim$2\arcmin). The idea of a strong variation in the mean density is also supported by the low bow shock shell mass deduced from the Herschel data. This change in mean density might be from a change in mean mass-loss rate or the formation of an inner bow shock well inside a filamentary halo, in which case the ISM is able to flow into the inner parts of the envelope creating a bow shock.}

Detailed hydrodynamical simulations were performed to understand the morphology of the bow shock around Betelgeuse.
 The stand-off distance and the ratio between the project distance between the star and the bow shock outline in the direction of relative motion and the distance perpendicular to that are predicted well by the hydrodynamical simulations, except in the case of large dust grains (around 100\,nm; simulation~B) or when a low ambient medium density and high space velocity are considered (simulation~D). It is clear that small dust grains are good tracers for the gas-driven dynamics at the contact discontinuity. Most of the mass in the bow shock is situated to the sides of the star, and only $\sim$20\% is situated in a cone with angular extent $<$45\deg. The dust temperature in the bow shock is mainly determined by heating by the stellar radiation, while energy transfer from the hot post-shock gas can be neglected. Increasing the temperature in either the ambient interstellar medium or in the shocked wind prevents the growth of strong instabilities. Allowing for strong variations in the mass-loss rate in the inner 
envelope has an impact on the appearance of the bow shock, and eventually multiple arc-like structures might appear between the termination shock and the contact discontinuity.

Based on the observations and on hydrodynamical simulations, different hypotheses are formulated to explain the origin of the multiple arcs and to understand why no large-scale instabilities are seen in the bow shock region. In our opinion, the two main ingredients to explain both features are (1)~a clumpy mass-loss process and (2)~the influence of the Galactic magnetic field.  The occurrence of clumps might also explain the low total (gas and dust) mass in the arcs as deduced from the data ($\sim2.5 \times 10^{-3}$\,\Msun). The Galactic magnetic field might serve as an extra trigger to fragment the bow shock region. The angular separation between the arcs seen in the Herschel data is compatible with a  magnetic field of $\sim$3\,$\mu$G, in agreement with the average strength of the magnetic field in the Milky Way. 

The linear bar might be the edge of an interstellar cloud illuminated by Betelgeuse or a linear filament whose a possible origin is linked to the Galactic magnetic field. Since no curvature is present in the bar, we believe that the bar is not directly linked to a previous blue supergiant wind, as recently discussed by \citet{Mackey2012arXiv1204.3925M}.

\begin{acknowledgements}
 This research has made use of NASA's Astrophysics Data System Bibliographic Service and the SIMBAD database, operated at the CDS, Strasbourg, France.
PACS was developed by a consortium of institutes led by the MPE (Germany) and including UVIE (Austria); KUL, CSL, IMEC (Belgium); CEA, OAMP (France); MPIA (Germany); IFSI, OAP/AOT, OAA/CAISMI, LENS, SISSA (Italy); IAC (Spain). The development has been supported by the funding agencies BMVIT (Austria), ESA-PRODEX (Belgium), CEA/CNES (France), DLR (Germany), ASI (Italy), and CICT/MCT (Spain). SPIRE was developed by a consortium of institutes led
by Cardiff Univ. (UK) and include: Univ. Lethbridge (Canada);
NAOC (China); CEA, LAM (France); IFSI, Univ. Padua (Italy);
IAC (Spain); Stockholm Observatory (Sweden); Imperial College
London, RAL, UCL-MSSL, UKATC, Univ. Sussex (UK); and Caltech,
JPL, NHSC, Univ. Colorado (USA). The development has been
supported by national funding agencies: CSA (Canada); NAOC
(China); CEA, CNES, CNRS (France); ASI (Italy); MCINN (Spain);
SNSB (Sweden); STFC, UKSA (UK); and NASA (USA). The development has been supported
by national funding agencies: CSA (Canada); NAOC (China); CEA,
CNES, CNRS (France); ASI (Italy); MCINN (Spain); SNSB (Sweden);
STFC (UK); and NASA (USA). NLJC, PR, BV, JB, and MG acknowledge support from the Belgian Federal Science Policy Office via the PRODEX Programme of ESA. FK acknowledges funding by the Austrian Science Fund FWF under project number P23586-N16, RO under project number I163-N16.
The Galactic Arecibo L-Band Feed Array \ion{H}{i} (GALFA-\ion{H}{i}) Survey data set
was obtained with the Arecibo L-band Feed Array (ALFA) on the Arecibo
305m telescope.  Arecibo Observatory is part of the National Astronomy
and Ionosphere Center, which is operated by Cornell University under 
Cooperative Agreement with the National Science Foundation of the
 United States of America.

\end{acknowledgements}
\vspace*{-1ex}
\vspace*{-.5cm}
\bibliographystyle{aa}
\bibliography{19792}

\begin{thebibliography}{101}
\expandafter\ifx\csname natexlab\endcsname\relax\def\natexlab#1{#1}\fi

\bibitem[{{Altenhoff} {et~al.}(1979){Altenhoff}, {Oster}, \&
  {Wendker}}]{Altenhoff1979A&A....73L..21A}
{Altenhoff}, W.~J., {Oster}, L., \& {Wendker}, H.~J. 1979, \aap, 73, L21

\bibitem[{{Andreasyan} \& {Makarov}(1989)}]{Andreasyan1989Ap.....31..560A}
{Andreasyan}, R.~R. \& {Makarov}, A.~N. 1989, Astrophysics, 31, 560

\bibitem[{{Auri{\`e}re} {et~al.}(2010){Auri{\`e}re}, {Donati},
  {Konstantinova-Antova}, {Perrin}, {Petit}, \&
  {Roudier}}]{Auriere2010A&A...516L...2A}
{Auri{\`e}re}, M., {Donati}, J.-F., {Konstantinova-Antova}, R., {et~al.} 2010,
  \aap, 516, L2

\bibitem[{{Beckwith} {et~al.}(1990){Beckwith}, {Sargent}, {Chini}, \&
  {Guesten}}]{Beckwith1990AJ.....99..924B}
{Beckwith}, S.~V.~W., {Sargent}, A.~I., {Chini}, R.~S., \& {Guesten}, R. 1990,
  \aj, 99, 924

\bibitem[{{Bern{\'e}} {et~al.}(2010){Bern{\'e}}, {Marcelino}, \&
  {Cernicharo}}]{Berne2010Natur.466..947B}
{Bern{\'e}}, O., {Marcelino}, N., \& {Cernicharo}, J. 2010, \nat, 466, 947

\bibitem[{{Biermann}(2001)}]{Biermann2001pteu.conf..543B}
{Biermann}, P.~L. 2001, in Phase Transitions in the Early Universe: Theory and
  Observations, ed. H.~J. {de Vega}, I.~M. {Khalatnikov}, \& N.~G. {Sanchez},
  543

\bibitem[{{Bingham} \& {Shakeshaft}(1967)}]{Bingham1967MNRAS.136..347B}
{Bingham}, R.~G. \& {Shakeshaft}, J.~R. 1967, \mnras, 136, 347

\bibitem[{{Bodo} {et~al.}(1990){Bodo}, {Ferrari}, {Massaglia}, \&
  {Trussoni}}]{Bodo1990MNRAS.244..530B}
{Bodo}, G., {Ferrari}, A., {Massaglia}, S., \& {Trussoni}, E. 1990, \mnras,
  244, 530

\bibitem[{{Bohren} \& {Huffman}(1983)}]{Bohren1983asls.book.....B}
{Bohren}, C.~F. \& {Huffman}, D.~R. 1983, {Absorption and scattering of light
  by small particles}

\bibitem[{{Brighenti} \& {D'Ercole}(1995)}]{Brighenti1995MNRAS.277...53B}
{Brighenti}, F. \& {D'Ercole}, A. 1995, \mnras, 277, 53

\bibitem[{{Chandrasekhar}(1961)}]{Chandrasekhar1961hhs..book.....C}
{Chandrasekhar}, S. 1961, {Hydrodynamic and hydromagnetic stability}

\bibitem[{{Cox} {et~al.}(2012){Cox}, {Kerschbaum}, {van Marle}, {Decin},
  {Ladjal}, {Mayer}, {Groenewegen}, {van Eck}, {Royer}, {Ottensamer}, {Ueta},
  {Jorissen}, {Mecina}, {Meliani}, {Luntzer}, {Blommaert}, {Posch},
  {Vandenbussche}, \& {Waelkens}}]{Cox2012A&A...537A..35C}
{Cox}, N.~L.~J., {Kerschbaum}, F., {van Marle}, A.-J., {et~al.} 2012, \aap,
  537, A35

\bibitem[{{De Beck} {et~al.}(2010){De Beck}, {Decin}, {de Koter}, {Justtanont},
  {Verhoelst}, {Kemper}, \& {Menten}}]{DeBeck2010A&A...523A..18D}
{De Beck}, E., {Decin}, L., {de Koter}, A., {et~al.} 2010, \aap, 523, A18

\bibitem[{{de Vries} {et~al.}(2010){de Vries}, {Min}, {Waters}, {Blommaert}, \&
  {Kemper}}]{deVries2010A&A...516A..86D}
{de Vries}, B.~L., {Min}, M., {Waters}, L.~B.~F.~M., {Blommaert}, J.~A.~D.~L.,
  \& {Kemper}, F. 2010, \aap, 516, A86

\bibitem[{{Decin} {et~al.}(2006){Decin}, {Hony}, {de Koter}, {Justtanont},
  {Tielens}, \& {Waters}}]{Decin2006A&A...456..549D}
{Decin}, L., {Hony}, S., {de Koter}, A., {et~al.} 2006, \aap, 456, 549

\bibitem[{{Decin} {et~al.}(2012){Decin}, {Royer}, {Cox}, {Van Marle},
  {Cernicharo}, {Teyssier}, {Vandenbussche}, {Van Winckel}, {Raskin},
  {Ottensamer}, {Barlow}, {Blommaert}, {Groenewegen}, {Lim}, {Kerschbaum},
  {Posch}, \& {Waelkens}}]{Decin2012b}
{Decin}, L., {Royer}, P., {Cox}, N., {et~al.} 2012, \aap, {in prep.}

\bibitem[{{Decin} {et~al.}(2011){Decin}, {Royer}, {Cox}, {Vandenbussche},
  {Ottensamer}, {Blommaert}, {Groenewegen}, {Barlow}, {Lim}, {Kerschbaum},
  {Posch}, \& {Waelkens}}]{Decin2011A&A...534A...1D}
{Decin}, L., {Royer}, P., {Cox}, N.~L.~J., {et~al.} 2011, \aap, 534, A1

\bibitem[{{Desert} {et~al.}(1986){Desert}, {Boulanger}, {Leger}, {Puget}, \&
  {Sellgren}}]{Desert1986A&A...159..328D}
{Desert}, F.~X., {Boulanger}, F., {Leger}, A., {Puget}, J.~L., \& {Sellgren},
  K. 1986, \aap, 159, 328

\bibitem[{{Dgani}(1998)}]{Dgani1998RMxAC...7..149D}
{Dgani}, R. 1998, in Revista Mexicana de Astronomia y Astrofisica, vol. 27,
  Vol.~7, Revista Mexicana de Astronomia y Astrofisica Conference Series, ed.
  {R.~J.~Dufour \& S.~Torres-Peimbert}, 149

\bibitem[{{Dgani} \& {Soker}(1998)}]{Dgani1998ApJ...495..337D}
{Dgani}, R. \& {Soker}, N. 1998, \apj, 495, 337

\bibitem[{{Dgani} {et~al.}(1996){Dgani}, {van Buren}, \&
  {Noriega-Crespo}}]{Dgani1996ApJ...461..927D}
{Dgani}, R., {van Buren}, D., \& {Noriega-Crespo}, A. 1996, \apj, 461, 927

\bibitem[{{Dickey} \& {Garwood}(1989)}]{Dickey1989ApJ...341..201D}
{Dickey}, J.~M. \& {Garwood}, R.~W. 1989, \apj, 341, 201

\bibitem[{{Dolan} {et~al.}(2008){Dolan}, {Mathews}, \&
  {Dearborn}}]{Dolan2008APS..APR.S8006D}
{Dolan}, M., {Mathews}, G., \& {Dearborn}, D. 2008, in APS April Meeting
  Abstracts, 8006

\bibitem[{{Dupac} {et~al.}(2003){Dupac}, {Bernard}, {Boudet}, {Giard},
  {Lamarre}, {M{\'e}ny}, {Pajot}, {Ristorcelli}, {Serra}, {Stepnik}, \&
  {Torre}}]{Dupac2003A&A...404L..11D}
{Dupac}, X., {Bernard}, J.-P., {Boudet}, N., {et~al.} 2003, \aap, 404, L11

\bibitem[{{Emerson}(1988)}]{Emerson1988felm.conf...21E}
{Emerson}, J.~P. 1988, in NATO ASIC Proc. 241: Formation and Evolution of Low
  Mass Stars, ed. A.~K. {Dupree} \& M.~T.~V.~T. {Lago}, 21

\bibitem[{{Fleischer}(1994)}]{Fleischer1994PhDT.......101F}
{Fleischer}, A.~J. 1994, PhD thesis, , Technische Universit{\"a}t, Berlin, FRG,
  (1994)

\bibitem[{{Gaetz} {et~al.}(1988){Gaetz}, {Edgar}, \&
  {Chevalier}}]{Gaetz1988ApJ...329..927G}
{Gaetz}, T.~J., {Edgar}, R.~J., \& {Chevalier}, R.~A. 1988, \apj, 329, 927

\bibitem[{{Gilliland} \& {Dupree}(1996)}]{Gilliland1996ApJ...463L..29G}
{Gilliland}, R.~L. \& {Dupree}, A.~K. 1996, \apjl, 463, L29

\bibitem[{{Glassgold} \& {Huggins}(1986)}]{Glassgold1986ApJ...306..605G}
{Glassgold}, A.~E. \& {Huggins}, P.~J. 1986, \apj, 306, 605

\bibitem[{{Goldberg}(1981)}]{Goldberg1981ASSL...88..301G}
{Goldberg}, L. 1981, in Astrophysics and Space Science Library, Vol.~88,
  Physical Processes in Red Giants, ed. {I.~Iben Jr.~\& A.~Renzini}, 301--304

\bibitem[{{Griffin} {et~al.}(2010){Griffin}, {Abergel}, {Abreu}, {Ade},
  {Andr{\'e}}, {Augueres}, {Babbedge}, {Bae}, {Baillie}, {Baluteau}, {Barlow},
  {Bendo}, {Benielli}, {Bock}, {Bonhomme}, {Brisbin}, {Brockley-Blatt},
  {Caldwell}, {Cara}, {Castro-Rodriguez}, {Cerulli}, {Chanial}, {Chen},
  {Clark}, {Clements}, {Clerc}, {Coker}, {Communal}, {Conversi}, {Cox},
  {Crumb}, {Cunningham}, {Daly}, {Davis}, {de Antoni}, {Delderfield}, {Devin},
  {di Giorgio}, {Didschuns}, {Dohlen}, {Donati}, {Dowell}, {Dowell}, {Duband},
  {Dumaye}, {Emery}, {Ferlet}, {Ferrand}, {Fontignie}, {Fox}, {Franceschini},
  {Frerking}, {Fulton}, {Garcia}, {Gastaud}, {Gear}, {Glenn}, {Goizel},
  {Griffin}, {Grundy}, {Guest}, {Guillemet}, {Hargrave}, {Harwit}, {Hastings},
  {Hatziminaoglou}, {Herman}, {Hinde}, {Hristov}, {Huang}, {Imhof}, {Isaak},
  {Israelsson}, {Ivison}, {Jennings}, {Kiernan}, {King}, {Lange}, {Latter},
  {Laurent}, {Laurent}, {Leeks}, {Lellouch}, {Levenson}, {Li}, {Li},
  {Lilienthal}, {Lim}, {Liu}, {Lu}, {Madden}, {Mainetti}, {Marliani}, {McKay},
  {Mercier}, {Molinari}, {Morris}, {Moseley}, {Mulder}, {Mur}, {Naylor},
  {Nguyen}, {O'Halloran}, {Oliver}, {Olofsson}, {Olofsson}, {Orfei}, {Page},
  {Pain}, {Panuzzo}, {Papageorgiou}, {Parks}, {Parr-Burman}, {Pearce},
  {Pearson}, {P{\'e}rez-Fournon}, {Pinsard}, {Pisano}, {Podosek}, {Pohlen},
  {Polehampton}, {Pouliquen}, {Rigopoulou}, {Rizzo}, {Roseboom}, {Roussel},
  {Rowan-Robinson}, {Rownd}, {Saraceno}, {Sauvage}, {Savage}, {Savini},
  {Sawyer}, {Scharmberg}, {Schmitt}, {Schneider}, {Schulz}, {Schwartz},
  {Shafer}, {Shupe}, {Sibthorpe}, {Sidher}, {Smith}, {Smith}, {Smith},
  {Spencer}, {Stobie}, {Sudiwala}, {Sukhatme}, {Surace}, {Stevens}, {Swinyard},
  {Trichas}, {Tourette}, {Triou}, {Tseng}, {Tucker}, {Turner}, {Vaccari},
  {Valtchanov}, {Vigroux}, {Virique}, {Voellmer}, {Walker}, {Ward}, {Waskett},
  {Weilert}, {Wesson}, {White}, {Whitehouse}, {Wilson}, {Winter}, {Woodcraft},
  {Wright}, {Xu}, {Zavagno}, {Zemcov}, {Zhang}, \&
  {Zonca}}]{Griffin2010A&A...518L...3G}
{Griffin}, M.~J., {Abergel}, A., {Abreu}, A., {et~al.} 2010, \aap, 518, L3

\bibitem[{{Groenewegen} {et~al.}(2011){Groenewegen}, {Waelkens}, {Barlow},
  {Kerschbaum}, {Garcia-Lario}, {Cernicharo}, {Blommaert}, {Bouwman}, {Cohen},
  {Cox}, {Decin}, {Exter}, {Gear}, {Gomez}, {Hargrave}, {Henning},
  {Hutsem{\'e}kers}, {Ivison}, {Jorissen}, {Krause}, {Ladjal}, {Leeks}, {Lim},
  {Matsuura}, {Naz{\'e}}, {Olofsson}, {Ottensamer}, {Polehampton}, {Posch},
  {Rauw}, {Royer}, {Sibthorpe}, {Swinyard}, {Ueta}, {Vamvatira-Nakou},
  {Vandenbussche}, {van de Steene}, {van Eck}, {van Hoof}, {van Winckel},
  {Verdugo}, \& {Wesson}}]{Groenewegen2011A&A...526A.162G}
{Groenewegen}, M.~A.~T., {Waelkens}, C., {Barlow}, M.~J., {et~al.} 2011, \aap,
  526, A162

\bibitem[{{Han}(2008)}]{Han2008NuPhS.175...62H}
{Han}, J.~L. 2008, Nuclear Physics B Proceedings Supplements, 175, 62

\bibitem[{{Harper} {et~al.}(2008){Harper}, {Brown}, \&
  {Guinan}}]{Harper2008AJ....135.1430H}
{Harper}, G.~M., {Brown}, A., \& {Guinan}, E.~F. 2008, \aj, 135, 1430

\bibitem[{{Hartmann} \& {Avrett}(1984)}]{Hartmann1984ApJ...284..238H}
{Hartmann}, L. \& {Avrett}, E.~H. 1984, \apj, 284, 238

\bibitem[{{Hartquist} {et~al.}(1986){Hartquist}, {Dyson}, {Pettini}, \&
  {Smith}}]{Hartquist1986MNRAS.221..715H}
{Hartquist}, T.~W., {Dyson}, J.~E., {Pettini}, M., \& {Smith}, L.~J. 1986,
  \mnras, 221, 715

\bibitem[{{Hebden} {et~al.}(1987){Hebden}, {Eckart}, \&
  {Hege}}]{Hebden1987ApJ...314..690H}
{Hebden}, J.~C., {Eckart}, A., \& {Hege}, E.~K. 1987, \apj, 314, 690

\bibitem[{{Heiles}(2000)}]{Heiles2000APS..APR.B7003H}
{Heiles}, C. 2000, in APS April Meeting Abstracts, B7003

\bibitem[{{Heiles} \& {Crutcher}(2005)}]{Heiles2005LNP...664..137H}
{Heiles}, C. \& {Crutcher}, R. 2005, in Lecture Notes in Physics, Berlin
  Springer Verlag, Vol. 664, Cosmic Magnetic Fields, ed. {R.~Wielebinski \&
  R.~Beck}, 137

\bibitem[{{Herschel}(1840)}]{Herschel1840MNRAS...5...11H}
{Herschel}, Sir, J.~F.~W. 1840, \mnras, 5, 11

\bibitem[{{Heyvaerts} {et~al.}(1988){Heyvaerts}, {Norman}, \&
  {Pudritz}}]{Heyvaerts1988ApJ...330..718H}
{Heyvaerts}, J., {Norman}, C., \& {Pudritz}, R.~E. 1988, \apj, 330, 718

\bibitem[{{H{\"o}fner}(2008)}]{Hoefner2008A&A...491L...1H}
{H{\"o}fner}, S. 2008, \aap, 491, L1

\bibitem[{{H{\"o}fner} \& {Andersen}(2007)}]{Hoefner2007A&A...465L..39H}
{H{\"o}fner}, S. \& {Andersen}, A.~C. 2007, \aap, 465, L39

\bibitem[{{Huggins}(1987)}]{Huggins1987ApJ...313..400H}
{Huggins}, P.~J. 1987, \apj, 313, 400

\bibitem[{{Josselin} \& {Plez}(2007)}]{Josselin2007A&A...469..671J}
{Josselin}, E. \& {Plez}, B. 2007, \aap, 469, 671

\bibitem[{{Keppens} {et~al.}(2012){Keppens}, {Meliani}, {van Marle}, {Delmont},
  {Vlasis}, \& {van der Holst}}]{Keppens2012JCoPh.231..718K}
{Keppens}, R., {Meliani}, Z., {van Marle}, A.~J., {et~al.} 2012, Journal of
  Computational Physics, 231, 718

\bibitem[{{Keppens} {et~al.}(1999){Keppens}, {T{\'o}th}, {Westermann}, \&
  {Goedbloed}}]{Keppens1999JPlPh..61....1K}
{Keppens}, R., {T{\'o}th}, G., {Westermann}, R.~H.~J., \& {Goedbloed}, J.~P.
  1999, Journal of Plasma Physics, 61, 1

\bibitem[{{Kervella} {et~al.}(2011){Kervella}, {Perrin}, {Chiavassa},
  {Ridgway}, {Cami}, {Haubois}, \& {Verhoelst}}]{Kervella2011A&A...531A.117K}
{Kervella}, P., {Perrin}, G., {Chiavassa}, A., {et~al.} 2011, \aap, 531, A117

\bibitem[{{Kervella} {et~al.}(2009){Kervella}, {Verhoelst}, {Ridgway},
  {Perrin}, {Lacour}, {Cami}, \& {Haubois}}]{Kervella2009A&A...504..115K}
{Kervella}, P., {Verhoelst}, T., {Ridgway}, S.~T., {et~al.} 2009, \aap, 504,
  115

\bibitem[{{Knapp} \& {Bowers}(1988)}]{Knapp1988ApJ...331..974K}
{Knapp}, G.~R. \& {Bowers}, P.~F. 1988, \apj, 331, 974

\bibitem[{{Knapp} {et~al.}(1993){Knapp}, {Sandell}, \&
  {Robson}}]{Knapp1993ApJS...88..173K}
{Knapp}, G.~R., {Sandell}, G., \& {Robson}, E.~I. 1993, \apjs, 88, 173

\bibitem[{{Kwok}(1975)}]{Kwok1975ApJ...198..583K}
{Kwok}, S. 1975, \apj, 198, 583

\bibitem[{{Ladjal} {et~al.}(2010){Ladjal}, {Barlow}, {Groenewegen}, {Ueta},
  {Blommaert}, {Cohen}, {Decin}, {De Meester}, {Exter}, {Gear}, {Gomez},
  {Hargrave}, {Huygen}, {Ivison}, {Jean}, {Kerschbaum}, {Leeks}, {Lim},
  {Olofsson}, {Polehampton}, {Posch}, {Regibo}, {Royer}, {Sibthorpe},
  {Swinyard}, {Vandenbussche}, {Waelkens}, \&
  {Wesson}}]{Ladjal2010A&A...518L.141L}
{Ladjal}, D., {Barlow}, M.~J., {Groenewegen}, M.~A.~T., {et~al.} 2010, \aap,
  518, L141

\bibitem[{{Le Bertre} {et~al.}(2012){Le Bertre}, {Matthews}, {G{\'e}rard}, \&
  {Libert}}]{LeBertre2012arXiv1203.0255L}
{Le Bertre}, T., {Matthews}, L.~D., {G{\'e}rard}, E., \& {Libert}, Y. 2012,
  \mnras, 422, 3433

\bibitem[{{Lim} {et~al.}(1998){Lim}, {Carilli}, {White}, {Beasley}, \&
  {Marson}}]{Lim1998Natur.392..575L}
{Lim}, J., {Carilli}, C.~L., {White}, S.~M., {Beasley}, A.~J., \& {Marson},
  R.~G. 1998, \nat, 392, 575

\bibitem[{{Mackey} {et~al.}(2012){Mackey}, {Mohamed}, {Neilson}, {Langer}, \&
  {Meyer}}]{Mackey2012arXiv1204.3925M}
{Mackey}, J., {Mohamed}, S., {Neilson}, H.~R., {Langer}, N., \& {Meyer},
  D.~M.-A. 2012, \apjl, 751, L10

\bibitem[{{Maeder}(1992)}]{Maeder1992A&A...264..105M}
{Maeder}, A. 1992, \aap, 264, 105

\bibitem[{{Martin} {et~al.}(2007){Martin}, {Seibert}, {Neill}, {Schiminovich},
  {Forster}, {Rich}, {Welsh}, {Madore}, {Wheatley}, {Morrissey}, \&
  {Barlow}}]{Martin2007Natur.448..780M}
{Martin}, D.~C., {Seibert}, M., {Neill}, J.~D., {et~al.} 2007, \nat, 448, 780

\bibitem[{{Matthews} {et~al.}(2011){Matthews}, {Libert}, {G{\'e}rard}, {Le
  Bertre}, {Johnson}, \& {Dame}}]{Matthews2011AJ....141...60M}
{Matthews}, L.~D., {Libert}, Y., {G{\'e}rard}, E., {et~al.} 2011, \aj, 141, 60

\bibitem[{{Mennella} {et~al.}(1998){Mennella}, {Brucato}, {Colangeli},
  {Palumbo}, {Rotundi}, \& {Bussoletti}}]{Mennella1998ApJ...496.1058M}
{Mennella}, V., {Brucato}, J.~R., {Colangeli}, L., {et~al.} 1998, \apj, 496,
  1058

\bibitem[{{Min} {et~al.}(2009){Min}, {Dullemond}, {Dominik}, {de Koter}, \&
  {Hovenier}}]{Min2009A&A...497..155M}
{Min}, M., {Dullemond}, C.~P., {Dominik}, C., {de Koter}, A., \& {Hovenier},
  J.~W. 2009, \aap, 497, 155

\bibitem[{{Miura} \& {Pritchett}(1982)}]{Miura1982JGR....87.7431M}
{Miura}, A. \& {Pritchett}, P.~L. 1982, \jgr, 87, 7431

\bibitem[{{Miville-Desch{\^e}nes} {et~al.}(2010){Miville-Desch{\^e}nes},
  {Martin}, {Abergel}, {Bernard}, {Boulanger}, {Lagache}, {Anderson},
  {Andr{\'e}}, {Arab}, {Baluteau}, {Blagrave}, {Bontemps}, {Cohen},
  {Compiegne}, {Cox}, {Dartois}, {Davis}, {Emery}, {Fulton}, {Gry}, {Habart},
  {Huang}, {Joblin}, {Jones}, {Kirk}, {Lim}, {Madden}, {Makiwa}, {Menshchikov},
  {Molinari}, {Moseley}, {Motte}, {Naylor}, {Okumura}, {Pinheiro Gon{\c
  c}alves}, {Polehampton}, {Rod{\'o}n}, {Russeil}, {Saraceno}, {Schneider},
  {Sidher}, {Spencer}, {Swinyard}, {Ward-Thompson}, {White}, \&
  {Zavagno}}]{Miville2010A&A...518L.104M}
{Miville-Desch{\^e}nes}, M.-A., {Martin}, P.~G., {Abergel}, A., {et~al.} 2010,
  \aap, 518, L104

\bibitem[{{Mohamed} {et~al.}(2012){Mohamed}, {Mackey}, \&
  {Langer}}]{Mohamed2012A&A...541A...1M}
{Mohamed}, S., {Mackey}, J., \& {Langer}, N. 2012, \aap, 541, A1

\bibitem[{{Morrissey} {et~al.}(2005){Morrissey}, {Schiminovich}, {Barlow},
  {Martin}, {Blakkolb}, {Conrow}, {Cooke}, {Erickson}, {Fanson}, {Friedman},
  {Grange}, {Jelinsky}, {Lee}, {Liu}, {Mazer}, {McLean}, {Milliard}, {Randall},
  {Schmitigal}, {Sen}, {Siegmund}, {Surber}, {Vaughan}, {Viton}, {Welsh},
  {Bianchi}, {Byun}, {Donas}, {Forster}, {Heckman}, {Lee}, {Madore}, {Malina},
  {Neff}, {Rich}, {Small}, {Szalay}, \& {Wyder}}]{Morrissey2005ApJ...619L...7M}
{Morrissey}, P., {Schiminovich}, D., {Barlow}, T.~A., {et~al.} 2005, \apjl,
  619, L7

\bibitem[{{Neilson} {et~al.}(2011){Neilson}, {Lester}, \&
  {Haubois}}]{Neilson2011arXiv1109.4562N}
{Neilson}, H., {Lester}, J.~B., \& {Haubois}, X. 2011, ArXiv e-prints

\bibitem[{{Noriega-Crespo} {et~al.}(1997){Noriega-Crespo}, {van Buren}, {Cao},
  \& {Dgani}}]{Noriega1997AJ....114..837N}
{Noriega-Crespo}, A., {van Buren}, D., {Cao}, Y., \& {Dgani}, R. 1997, \aj,
  114, 837

\bibitem[{{Norris} {et~al.}(2012){Norris}, {Tuthill}, {Ireland}, {Lacour},
  {Zijlstra}, {Lykou}, {Evans}, {Stewart}, \&
  {Bedding}}]{Norris2012Natur.484..220N}
{Norris}, B.~R.~M., {Tuthill}, P.~G., {Ireland}, M.~J., {et~al.} 2012, \nat,
  484, 220

\bibitem[{{Ohnaka} {et~al.}(2009){Ohnaka}, {Hofmann}, {Benisty}, {Chelli},
  {Driebe}, {Millour}, {Petrov}, {Schertl}, {Stee}, {Vakili}, \&
  {Weigelt}}]{Ohnaka2009A&A...503..183O}
{Ohnaka}, K., {Hofmann}, K.-H., {Benisty}, M., {et~al.} 2009, \aap, 503, 183

\bibitem[{{Ottensamer} {et~al.}(2011){Ottensamer}, {Luntzer}, {Mecina},
  {Kerschbaum}, {Blommaert}, {Decin}, {Groenewegen}, {Posch}, {Vandenbussche},
  \& {Waelkens}}]{Ottensamer2011}
{Ottensamer}, R., {Luntzer}, A., {Mecina}, M., {et~al.} 2011, in Astronomical
  Society of the Pacific Conference Series, Vol. 445, Why Galaxies Care about
  AGB Stars II: Shining Examples and Common Inhabitants, ed. F.~{Kerschbaum},
  T.~{Lebzelter}, \& R.~F. {Wing}, 625

\bibitem[{{Paardekooper} \& {Mellema}(2006)}]{Paardekoper2006A&A...453.1129P}
{Paardekooper}, S.-J. \& {Mellema}, G. 2006, \aap, 453, 1129

\bibitem[{{Peek} {et~al.}(2011){Peek}, {Heiles}, {Douglas}, {Lee}, {Grcevich},
  {Stanimirovi{\'c}}, {Putman}, {Korpela}, {Gibson}, {Begum}, {Saul},
  {Robishaw}, \& {Kr{\v c}o}}]{Peek2011ApJS..194...20P}
{Peek}, J.~E.~G., {Heiles}, C., {Douglas}, K.~A., {et~al.} 2011, \apjs, 194, 20

\bibitem[{{Pilbratt} {et~al.}(2010){Pilbratt}, {Riedinger}, {Passvogel},
  {Crone}, {Doyle}, {Gageur}, {Heras}, {Jewell}, {Metcalfe}, {Ott}, \&
  {Schmidt}}]{Pilbratt2010A&A...518L...1P}
{Pilbratt}, G.~L., {Riedinger}, J.~R., {Passvogel}, T., {et~al.} 2010, \aap,
  518, L1

\bibitem[{{Poglitsch} {et~al.}(2010){Poglitsch}, {Waelkens}, {Geis},
  {Feuchtgruber}, {Vandenbussche}, {Rodriguez}, {Krause}, {Renotte}, {van
  Hoof}, {Saraceno}, {Cepa}, {Kerschbaum}, {Agn{\`e}se}, {Ali}, {Altieri},
  {Andreani}, {Augueres}, {Balog}, {Barl}, {Bauer}, {Belbachir}, {Benedettini},
  {Billot}, {Boulade}, {Bischof}, {Blommaert}, {Callut}, {Cara}, {Cerulli},
  {Cesarsky}, {Contursi}, {Creten}, {De Meester}, {Doublier}, {Doumayrou},
  {Duband}, {Exter}, {Genzel}, {Gillis}, {Gr{\"o}zinger}, {Henning},
  {Herreros}, {Huygen}, {Inguscio}, {Jakob}, {Jamar}, {Jean}, {de Jong},
  {Katterloher}, {Kiss}, {Klaas}, {Lemke}, {Lutz}, {Madden}, {Marquet},
  {Martignac}, {Mazy}, {Merken}, {Montfort}, {Morbidelli}, {M{\"u}ller},
  {Nielbock}, {Okumura}, {Orfei}, {Ottensamer}, {Pezzuto}, {Popesso},
  {Putzeys}, {Regibo}, {Reveret}, {Royer}, {Sauvage}, {Schreiber}, {Stegmaier},
  {Schmitt}, {Schubert}, {Sturm}, {Thiel}, {Tofani}, {Vavrek}, {Wetzstein},
  {Wieprecht}, \& {Wiezorrek}}]{Poglitsch2010A&A...518L...2P}
{Poglitsch}, A., {Waelkens}, C., {Geis}, N., {et~al.} 2010, \aap, 518, L2

\bibitem[{{Rodgers} \& {Glassgold}(1991)}]{Rodgers1991ApJ...382..606R}
{Rodgers}, B. \& {Glassgold}, A.~E. 1991, \apj, 382, 606

\bibitem[{{Rosner} \& {Bodo}(1996)}]{Rosner1996ApJ...470L..49R}
{Rosner}, R. \& {Bodo}, G. 1996, \apjl, 470, L49

\bibitem[{{Roussel}(2012)}]{Roussel2011}
{Roussel}, H. 2012, ArXiv e-prints

\bibitem[{{Sahai} \& {Chronopoulos}(2010)}]{Sahai2010ApJ...711L..53S}
{Sahai}, R. \& {Chronopoulos}, C.~K. 2010, \apjl, 711, L53

\bibitem[{{Schwarzschild}(1975)}]{Schwarzschild1975ApJ...195..137S}
{Schwarzschild}, M. 1975, \apj, 195, 137

\bibitem[{{Serabyn} \& {Morris}(1994)}]{Serabyn1994ApJ...424L..91S}
{Serabyn}, E. \& {Morris}, M. 1994, \apjl, 424, L91

\bibitem[{{Skinner} \& {Whitmore}(1987)}]{Skinner1987MNRAS.224..335S}
{Skinner}, C.~J. \& {Whitmore}, B. 1987, \mnras, 224, 335

\bibitem[{{Smith} {et~al.}(1988){Smith}, {Pettini}, {Dyson}, \&
  {Hartquist}}]{Smith1988MNRAS.234..625S}
{Smith}, L.~J., {Pettini}, M., {Dyson}, J.~E., \& {Hartquist}, T.~W. 1988,
  \mnras, 234, 625

\bibitem[{{Tielens}(2005)}]{Tielens2005pcim.book.....T}
{Tielens}, A.~G.~G.~M. 2005, {The Physics and Chemistry of the Interstellar
  Medium}, ed. {Tielens, A.~G.~G.~M.}

\bibitem[{{Tielens} \& {Allamandola}(1987)}]{Tielens1987ASSL..134..397T}
{Tielens}, A.~G.~G.~M. \& {Allamandola}, L.~J. 1987, in Astrophysics and Space
  Science Library, Vol. 134, Interstellar Processes, ed. {D.~J.~Hollenbach \&
  H.~A.~Thronson Jr.}, 397--469

\bibitem[{{Townsend}(2009)}]{Townsend:2009}
{Townsend}, R.~H.~D. 2009, \apjs, 181, 391

\bibitem[{{Ueta} {et~al.}(2008){Ueta}, {Izumiura}, {Yamamura}, {Nakada},
  {Matsuura}, {Ita}, {Tanab{\'e}}, {Fukushi}, {Matsunaga}, \&
  {Mito}}]{Ueta2008PASJ...60S.407U}
{Ueta}, T., {Izumiura}, H., {Yamamura}, I., {et~al.} 2008, \pasj, 60, 407

\bibitem[{{Ueta} {et~al.}(2006){Ueta}, {Speck}, {Stencel}, {Herwig}, {Gehrz},
  {Szczerba}, {Izumiura}, {Zijlstra}, {Latter}, {Matsuura}, {Meixner},
  {Steffen}, \& {Elitzur}}]{Ueta2006ApJ...648L..39U}
{Ueta}, T., {Speck}, A.~K., {Stencel}, R.~E., {et~al.} 2006, \apjl, 648, L39

\bibitem[{{van Hoof} {et~al.}(2010){van Hoof}, {van de Steene}, {Barlow},
  {Exter}, {Sibthorpe}, {Ueta}, {Peris}, {Groenewegen}, {Blommaert}, {Cohen},
  {De Meester}, {Ferland}, {Gear}, {Gomez}, {Hargrave}, {Huygen}, {Ivison},
  {Jean}, {Leeks}, {Lim}, {Olofsson}, {Polehampton}, {Regibo}, {Royer},
  {Swinyard}, {Vandenbussche}, {van Winckel}, {Waelkens}, {Walker}, \&
  {Wesson}}]{vanHoof2010A&A...518L.137V}
{van Hoof}, P.~A.~M., {van de Steene}, G.~C., {Barlow}, M.~J., {et~al.} 2010,
  \aap, 518, L137

\bibitem[{{van Leeuwen}(2007)}]{vanLeeuwen2007A&A...474..653V}
{van Leeuwen}, F. 2007, \aap, 474, 653

\bibitem[{{van Marle} {et~al.}(2011){van Marle}, {Meliani}, {Keppens}, \&
  {Decin}}]{vanMarle2011ApJ...734L..26V}
{van Marle}, A.~J., {Meliani}, Z., {Keppens}, R., \& {Decin}, L. 2011, \apjl,
  734, L26

\bibitem[{{Verhoelst} {et~al.}(2006){Verhoelst}, {Decin}, {van Malderen},
  {Hony}, {Cami}, {Eriksson}, {Perrin}, {Deroo}, {Vandenbussche}, \&
  {Waters}}]{Verhoelst2006A&A...447..311V}
{Verhoelst}, T., {Decin}, L., {van Malderen}, R., {et~al.} 2006, \aap, 447, 311

\bibitem[{{Villaver} {et~al.}(2012){Villaver}, {Manchado}, \&
  {Garc{\'{\i}}a-Segura}}]{Villaveretal:2012}
{Villaver}, E., {Manchado}, A., \& {Garc{\'{\i}}a-Segura}, G. 2012, \apj, 748,
  94

\bibitem[{{Vishniac}(1994)}]{Vishniac1994ApJ...428..186V}
{Vishniac}, E.~T. 1994, \apj, 428, 186

\bibitem[{{Wareing} {et~al.}(2007{\natexlab{a}}){Wareing}, {Zijlstra}, \&
  {O'Brien}}]{Wareing2007MNRAS.382.1233W}
{Wareing}, C.~J., {Zijlstra}, A.~A., \& {O'Brien}, T.~J. 2007{\natexlab{a}},
  \mnras, 382, 1233

\bibitem[{{Wareing} {et~al.}(2007{\natexlab{b}}){Wareing}, {Zijlstra}, \&
  {O'Brien}}]{Wareing2007ApJ...660L.129W}
{Wareing}, C.~J., {Zijlstra}, A.~A., \& {O'Brien}, T.~J. 2007{\natexlab{b}},
  \apjl, 660, L129

\bibitem[{{Weigelt} {et~al.}(2002){Weigelt}, {Balega}, {Bl{\"o}cker},
  {Hofmann}, {Men'shchikov}, \& {Winters}}]{Weigelt2002A&A...392..131W}
{Weigelt}, G., {Balega}, Y.~Y., {Bl{\"o}cker}, T., {et~al.} 2002, \aap, 392,
  131

\bibitem[{{Woitke}(2006{\natexlab{a}})}]{Woitke2006A&A...452..537W}
{Woitke}, P. 2006{\natexlab{a}}, \aap, 452, 537

\bibitem[{{Woitke}(2006{\natexlab{b}})}]{Woitke2006A&A...460L...9W}
{Woitke}, P. 2006{\natexlab{b}}, \aap, 460, L9

\bibitem[{{Wright} {et~al.}(2010){Wright}, {Eisenhardt}, {Mainzer}, {Ressler},
  {Cutri}, {Jarrett}, {Kirkpatrick}, {Padgett}, {McMillan}, {Skrutskie},
  {Stanford}, {Cohen}, {Walker}, {Mather}, {Leisawitz}, {Gautier}, {McLean},
  {Benford}, {Lonsdale}, {Blain}, {Mendez}, {Irace}, {Duval}, {Liu}, {Royer},
  {Heinrichsen}, {Howard}, {Shannon}, {Kendall}, {Walsh}, {Larsen}, {Cardon},
  {Schick}, {Schwalm}, {Abid}, {Fabinsky}, {Naes}, \&
  {Tsai}}]{Wright2010AJ....140.1868W}
{Wright}, E.~L., {Eisenhardt}, P.~R.~M., {Mainzer}, A.~K., {et~al.} 2010, \aj,
  140, 1868

\bibitem[{{Yusef-Zadeh} \& {Morris}(1987)}]{Yusef1987ApJ...322..721Y}
{Yusef-Zadeh}, F. \& {Morris}, M. 1987, \apj, 322, 721

\bibitem[{{Yusef-Zadeh} {et~al.}(1984){Yusef-Zadeh}, {Morris}, \&
  {Chance}}]{Ysef1984Natur.310..557Y}
{Yusef-Zadeh}, F., {Morris}, M., \& {Chance}, D. 1984, \nat, 310, 557

\end{thebibliography}

\afterpage{\clearpage}
\newpage
\begin{appendix}

\section{Previous images of the bow shock around Betelgeuse} \label{SEC:previous}
\begin{figure}[htp]
 \includegraphics[width=0.48\textwidth]{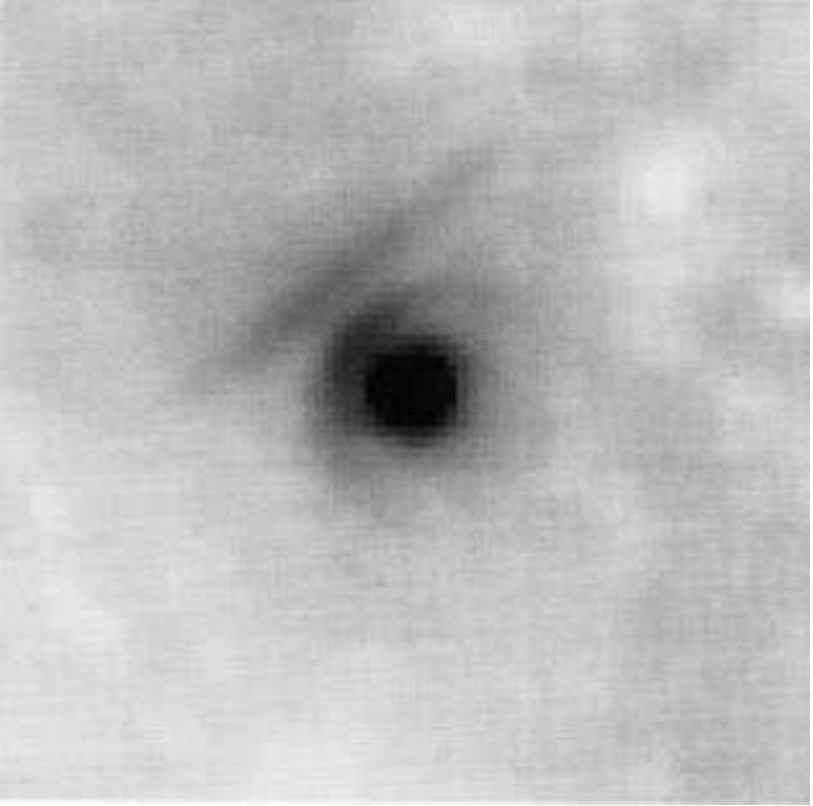}
\caption{IRAS 60\,$\mu$m image of the bow shock around Betelgeuse as detected by \citet{Noriega1997AJ....114..837N}.}
\label{Fig:IRAS60}
\end{figure}

\begin{figure}[htp]
 \includegraphics[width=0.48\textwidth]{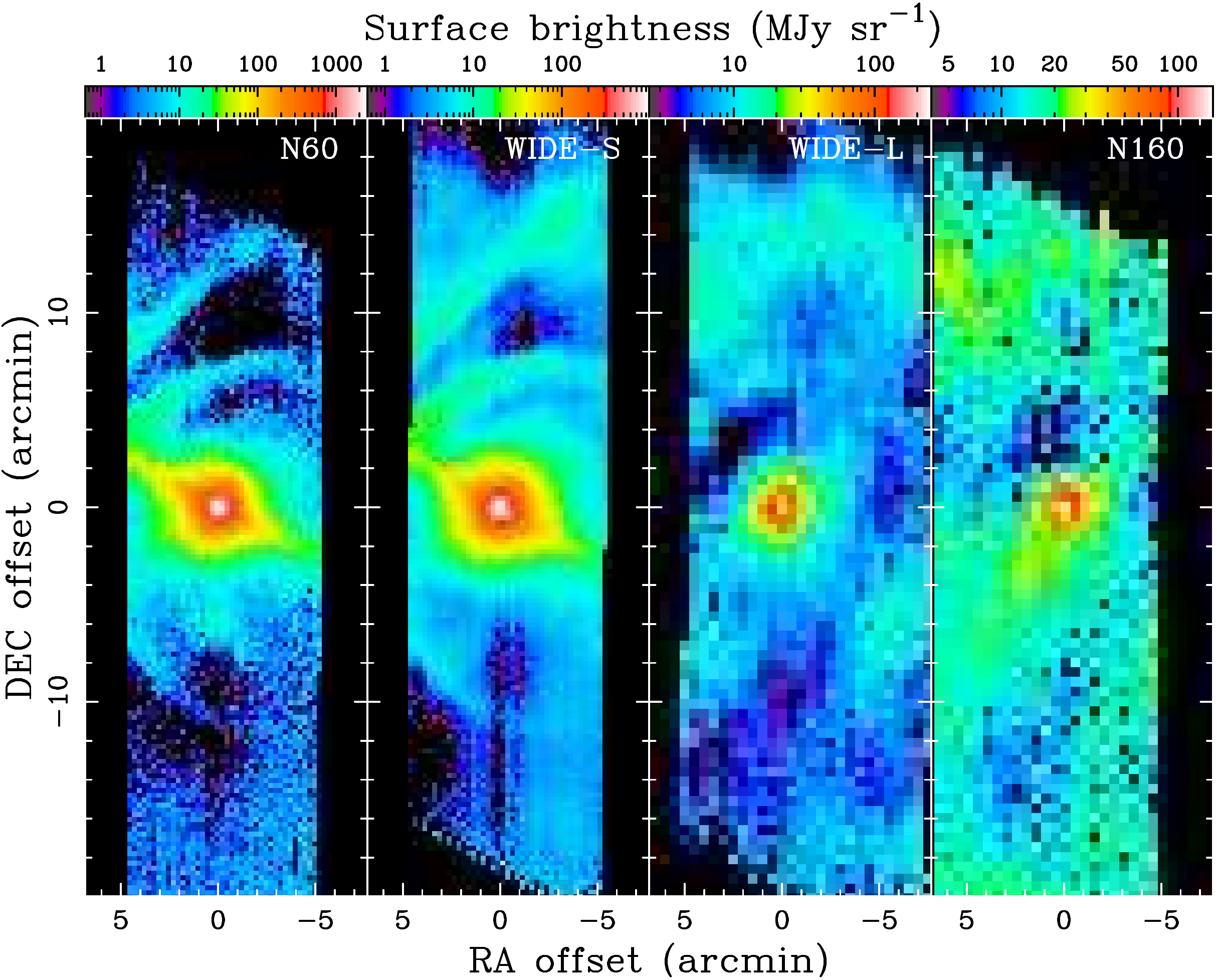}
\caption{AKARI/FIS false-colour maps of $\alpha$ Ori in the SW bands --- N60 (65\,$\mu$m) and WIDE-S (90\,$\mu$m) at 15\arcsec/pixel scale --- and in the LW bands --- WIDE-L (140\,$\mu$m) and N160 (160\,$\mu$m) at 30\arcsec/pixel scale --- from left to right, respectively \citep{Ueta2008PASJ...60S.407U}. Background emission has been subtracted by a combination
of temporal filters during data reduction. RA and Dec offsets (with respect to the stellar peak) are given in arcminutes. The wedges at the top indicate the
log scale of surface brightness in MJy sr$^{-1}$. North is up, and east to the left.}
\label{Fig:AKARI}
\end{figure}

For completeness, we show in this section the bow shock around Betelgeuse as observed with IRAS by \citet{Noriega1997AJ....114..837N} (Fig.~\ref{Fig:IRAS60}) and AKARI by \citet{Ueta2008PASJ...60S.407U} (Fig.~\ref{Fig:AKARI}).

\section{Additional figures}\label{App:additional_figures}
\begin{figure}[htp]
 \includegraphics[width=0.48\textwidth]{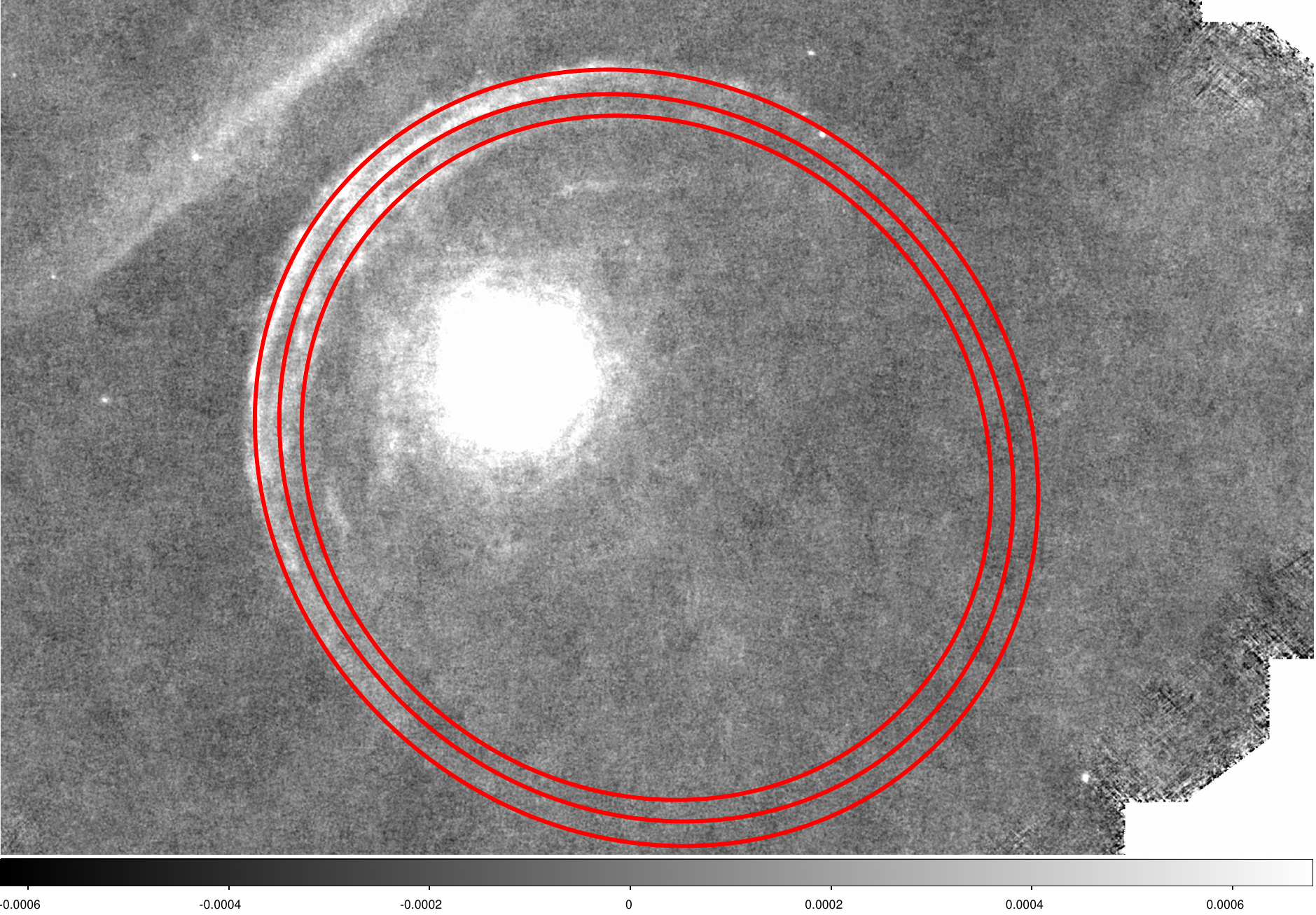}
\caption{PACS 70\,$\mu$m image overplotted in red with three concentric ellipses with the centre at $\alpha$\,=\,5h54m56.56s and $\delta$\,=\,7\deg22\arcmin16.18\arcsec\, radii of 622\arcsec, 585\arcsec, and 546\arcsec\ (as measured from the centre) and for a position angle of 47.7\deg.}
\label{FIG:PACS70_ellipse}
\end{figure}

Fig.~\ref{FIG:PACS70_ellipse} PACS 70\,$\mu$m image on which concentric ellipses were fitted to the different arcs.

\section{Herschel PACS image of R~Leo} \label{Sect:RLeo}
\begin{figure}[htp]
 \includegraphics[width=0.48\textwidth]{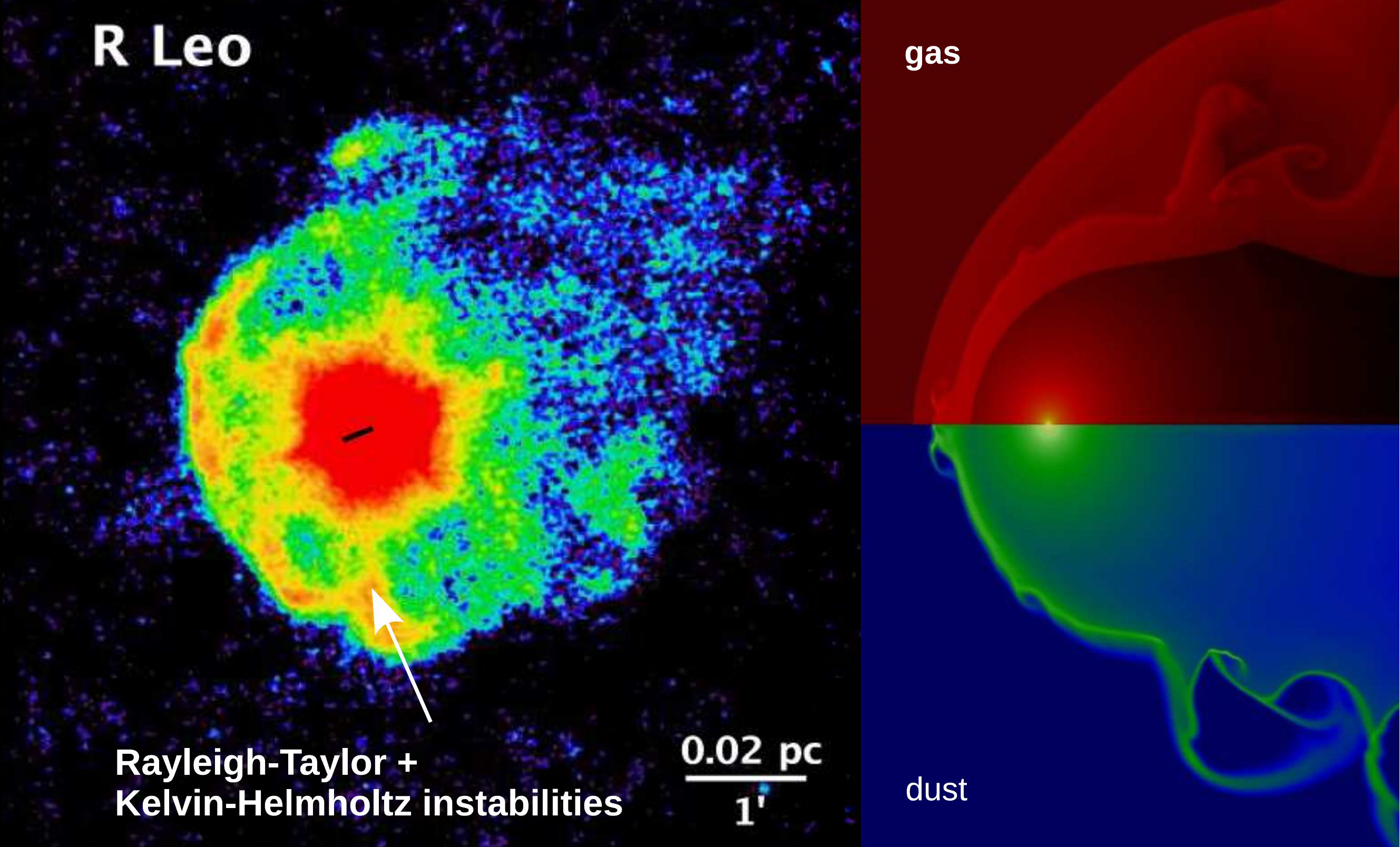}
\caption{PACS 70\,$\mu$m image of R~Leo \citep[left][]{Cox2012A&A...537A..35C} compared to a hydrodynamical simulation of the wind-ISM interaction computed with the {\tt AMRVAC} code \citep{vanMarle2011ApJ...734L..26V}. The parameters for this simulation are a stellar wind velocity of 15\,km/s, a constant gas mass-loss rate of $1 \times 10^{-6}$\,M$_\odot$/yr, a dust-to-gas mass ratio of 0.01, a space velocity of 25\,km/s, a local ISM density of 2 cm$^{-3}$, and an ISM temperature of 10\,K. The upper right figure shows the gas density, which ranges (on log scale) between $10^{-24}$ and $10^{-19}$\,g/cm$^{3}$; the lower right figure represents the dust grain particle density, ranging (on log scale) between $10^{-10}$ and $10^{-3.5}$\,cm$^{-3}$.}
\label{Fig:RLeo_hydro}
\end{figure}

The Herschel PACS 70\,$\mu$m image of the AGB star R~Leo are compared to a hydrodynamical simulation in Fig.~\ref{Fig:RLeo_hydro}. Rayleigh-Taylor instabilities, slightly deformed owing to the action of Kelvin-Helmholtz instabilities, are visible both in the data and in the simulations.

\section{Dust grain temperature in the envelope around Betelgeuse} \label{App:Tdust}
\begin{figure}[htp]
 \includegraphics[width=.48\textwidth]{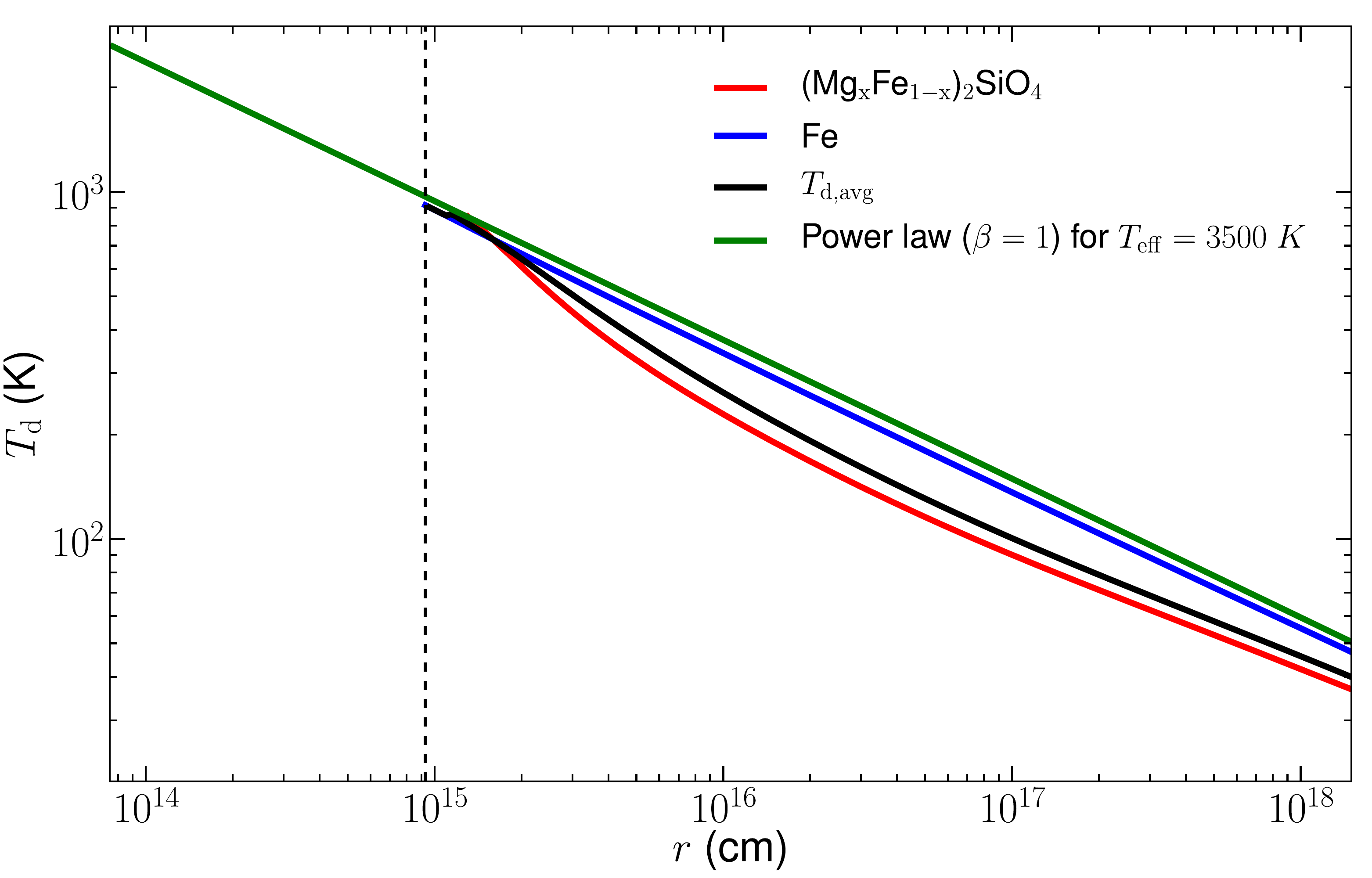}
\caption{The dust-temperature profiles in the envelope of Betelgeuse as modeled with {\sc mcmax} \citep{Min2009A&A...497..155M}. The different lines indicate the dust temperature of a  specific dust species: red for amorphous silicates and blue for metallic iron. The full black line gives the mean dust temperature profile, assuming thermal contact between iron and amorphous silicate grains. The green lines show a power-law distribution for the dust temperature for $\beta$\,=\,1. The vertical dashed line indicates the inner radius of the dust shell.}
\label{Fig:Robin}
\end{figure}

Under the assumption of a smooth continuous outflow and using the parameters as specified in Table~\ref{Table:input_AOri}, the dust grain temperatures for Fe and amorphous silicates in the envelope around Betelgeuse have been calculated using the {\sc mcmax} code \citep{Min2009A&A...497..155M}. Only the stellar radiation is taken into account as the energy source to compute the dust temperatures assuming radiative equilibrium. {The particle shape is represented by CDE \citep[continuous distribution of ellipsoids][]{Bohren1983asls.book.....B} for particles of which the equivalent spherical grain would have a grain size of 0.1\,$\mu$m.} 
The composition of the amorphous silicates is taken from \citet{deVries2010A&A...516A..86D}, who derived that the infrared spectrum of the AGB star Mira is best fitted with 65.7\% MgFeSiO$_4$ and 34.3\% Mg$_2$SiO$_4$, implying an average composition of Mg$_{1.36}$Fe$_{0.64}$SiO$_4$, having a $\beta$-index of $\sim$1.8. The resulting dust temperatures are shown in Fig.~\ref{Fig:Robin}. {The temperature for the amorphous silicate grains is lower than for the Fe-grains, owing to the higher opacity of Fe at the near-IR wavelengths, where the star emits most of its photons.}

\section{Numerical simulations of the bow shock} \label{App_numerics}

\subsection{Numerical method} 

For the simulations the {\tt MPI-AMRVAC} hydrodynamics code \citep{Keppens2012JCoPh.231..718K} was used. For a comparison to similar work, we refer to \citet{vanMarle2011ApJ...734L..26V}.
The  {\tt MPI-AMRVAC} code solves the conservation equations for mass, momentum, and energy on an adaptive mesh grid. It  includes optically thin radiative cooling, using the exact integration algorithm \citep{Townsend:2009}, and a cooling curve for gas at solar metallicity (Wang Ye, \textit{private communication}). For these  simulations, a two-dimensional approach was taken, assuming an inclination angle of 0\deg. As deduced by \citet{Cox2012A&A...537A..35C}, the inclination angle of the space motion with respect to the plane of the sky is  small ($\sim$8\deg).

 Dust was included in the simulations using a two-fluid approach, where the dust is treated as a pressure-less fluid \citep{Paardekoper2006A&A...453.1129P, Woitke2006A&A...452..537W, vanMarle2011ApJ...734L..26V, Cox2012A&A...537A..35C}. The dust is linked to the gas through the drag force \citep{Kwok1975ApJ...198..583K}:
\begin{equation}
f_d~=~(1-\alpha_T)\,\pi\,n_d\,\rho\,\,a^2\,\Delta v\,\sqrt{(\Delta v)^2+v_T^2},
\end{equation}
with $n_d$ the dust particle density, $\rho$ the gas density, $a$ the radius of the dust grain, $v_T=\sqrt{3kT/m_h}$ the thermal speed, and $\Delta v$ the velocity difference between dust and gas. For a given mass, the particle density decreases with $a^{-3}$. Therefore, the drag force is weaker for larger dust grains. 
The sticking coefficient for gas-dust collisions, $\alpha_T$, is estimated as 
\begin{equation}
\alpha_T~=~0.1~+~0.35\,e^{-\sqrt{(T_g+T_d)/500}}
\label{Eq_alphaT}
\end{equation}
\citep{Decin2006A&A...456..549D}, with $T_d$ the dust temperature and $T_g$ the gas temperature. 
This is an improvement over our earlier simulations \citep{vanMarle2011ApJ...734L..26V, Cox2012A&A...537A..35C}, where we kept the sticking coefficient constant. 
Because the dust temperature is typically much lower than the gas temperature {in the shocked CSM-ISM region}, we approximate the sticking coefficient by setting $T_d\,=\,0$\footnote{{This approximation is not valid in the region inside the termination shock where gas and dust temperatures might be similar far away from the star. As described by \citet{vanMarle2011ApJ...734L..26V}, this region is not simulated in our computations: the wind interaction is initialized by filling a spherical area of radius 0.1\,pc around the origin with wind material having equal gas and dust velocity. As seen in Fig.~\ref{FIG:info_Nick}, the observed dust temperature in the wind-ISM interaction region is $\sim$70\,K, which would yield an increase of the sticking coefficient of maximum 2.8\% compared to the results of Eq.~\ref{Eq_alphaT} for $T_d=0$. In the case of large dust grains  which penetrate a cold ISM (simulation~B), the change in the sticking coefficient in this ISM region would be $\sim$10\%. }}.

Unlike our previous models \citep{vanMarle2011ApJ...734L..26V, Cox2012A&A...537A..35C}, which used a cylindrical grid, these simulations are run on a two-dimensional spherical grid with a physical domain of $R=3$\,pc by $\theta=180\deg$. The grid has a minimal resolution of 160 radial by 80 azimuthal grid cells and  
 five additional levels of refinement are allowed, giving the simulations an effective resolution of $2\,560\times\,1\,280$ grid cells.

\subsection{Initial conditions}
At the inner radial boundary, the wind parameters are set according to the values in Table~\ref{Table:input_AOri}. Since the star is moving relative to the ambient medium,  the outer radial boundary is divided into two zones: (1)~for $\theta<90\deg$  an inflow boundary is set with the ambient gas entering the grid as a parallel stream with $v=28.3$\,km/s,  (2)~for $\theta>90\deg$ the gas is allowed to flow out of the grid. This is similar to the approach of \citet{Villaveretal:2012}.
The gas density in the ambient medium is set at $3 \times 10^{-24}$\,g cm$^{-3}$ (or $n_{\rm{H}} \approx 2$\,cm$^{-3}$) and a temperature of 100\,K (reflecting the temperature of a cold neutral medium). 
This temperature also serves as a lower limit for the gas temperature throughout the physical domain. We assume that there is no dust in the ambient medium. 

To avoid numerical instabilities we start the simulation without radiative cooling and let it run for 100\,000\,years. By that time the bow shock has reached its equilibrium position 
(where the ram pressure of the wind is balanced by the ram pressure of motion of the star through the ambient medium). From this point in time we include radiative cooling and let the simulation run for 200\,000\,years.

\Online

\section{Animations of bow shock simulations}\label{app-online}

\noindent\begin{minipage}{\textwidth}
\captionof{figure}{
Movie of Aori\_TISM100K showing gas density and dust particle density for the 200\,000\, year period from the introduction of radiative cooling to the end of the simulation 
for the model with a cold ambient medium. 
}
\label{fig:movie1}
\end{minipage}

\noindent\begin{minipage}{\textwidth}
\captionof{figure}{
Similar to Fig.~\ref{fig:movie1}, Aori\_TISM8000K showing the 200\,000 year period from the introduction of radiative cooling to the end of the simulation 
for the model with a warm ambient medium.
}
\label{fig:movie2}
\end{minipage}

\noindent\begin{minipage}{\textwidth}
\captionof{figure}{
Similar to Fig.~\ref{fig:movie1}, the third movie (Aori\_vardM) shows the 200\,000 year period from the introduction of radiative cooling to the end of the simulation 
for the model with a cold ambient medium and a varying mass-loss rate.
}
\label{fig:movie3}
\end{minipage}

\noindent\begin{minipage}{\textwidth}
\captionof{figure}{
Similar to Fig.~\ref{fig:movie2}, the fourth movie (Aori\_TISM8000K\_highvel) shows the 200\,000 year period from the introduction of radiative cooling to the end of the simulation 
for the model with a warm ambient medium and a high stellar velocity.
}
\label{fig:movie4}
\end{minipage}

\noindent\begin{minipage}{\textwidth}
\captionof{figure}{
Similar to Fig.~\ref{fig:movie1}, the fifth movie (Aori\_largegrain) shows the 200\,000 year period from the introduction of radiative cooling to the end of the simulation 
for the model with a cool ambient medium, but for large (100\,nm) dust grains.
}
\label{fig:movie5}
\end{minipage}

\end{appendix}

\end{document}